\documentclass[a4paper,fleqn,usenatbib]{mnras}

\usepackage{mathptmx}

\usepackage[T1]{fontenc}
\usepackage{ae,aecompl}

\usepackage{graphicx}	
\usepackage{amsmath}	
\usepackage{amssymb}	
\usepackage{wasysym}           
\usepackage{epstopdf}
\usepackage{mathrsfs}
\usepackage{natbib}
\usepackage{color}
\usepackage{hyperref}

\title[Tidal disruption events]{Tidal disruption events from supermassive black hole binaries}

\author[Coughlin et al.]{
Eric R. Coughlin,$^{1,2,3}$\thanks{email: eric\_coughlin@berkeley.edu}\thanks{Einstein fellow}
Philip J. Armitage,$^{2,3}$
Chris Nixon,$^{4}$
Mitchell C. Begelman,$^{2,3}$
\\
$^{1}$Astronomy Department and Theoretical Astrophysics Center, University of California, Berkeley, Berkeley, CA 94720 \\
$^{2}$JILA, University of Colorado and NIST, UCB 440, Boulder, CO 80309 \\
$^{3}$Department of Astrophysical and Planetary Sciences, University of Colorado, 391 UCB, Boulder, CO 80309-0391, USA\\
$^{4}$Theoretical Astrophysics Group, Department of Physics \& Astronomy, University of Leicester, Leicester LE1 7RH UK \\
}

\date{Accepted XXX. Received YYY; in original form ZZZ}

\pubyear{2016}

\begin{document}
\label{firstpage}
\pagerange{\pageref{firstpage}--\pageref{lastpage}}
\maketitle

\begin{abstract}
We investigate the pre-disruption gravitational dynamics and post-disruption hydrodynamics of the tidal disruption of stars by supermassive black hole (SMBH) binaries. We focus on binaries with relatively low mass primaries ($10^6M_{\odot}$), moderate mass ratios, and separations with reasonably long gravitational wave inspiral times (tens of Myr). First, we generate a large ensemble (between 1 and 10 million) of restricted three-body integrations to quantify the statistical properties of tidal disruptions by circular SMBH binaries of initially-unbound stars. Compared to the reference case of a disruption by a single SMBH, the binary potential induces significant variance into the specific energy and angular momentum of the star at the point of disruption. Second, we use Newtonian numerical hydrodynamics to study the detailed evolution of the fallback debris from 120 disruptions randomly selected from the three-body ensemble (excluding only the most deeply penetrating encounters). We find that the overall morphology of the debris is greatly altered by the presence of the second black hole, and the accretion rate histories display a wide range of behaviors, including order of magnitude dips and excesses relative to control simulations that include only one black hole. Complex evolution typically persists for many orbital periods of the binary. We find evidence for power in the accretion curves on timescales related to the binary orbital period, though there is no exact periodicity. We discuss our results in the context of future wide-field surveys, and comment on the prospects of identifying and characterizing the subset of events occurring in nuclei with binary SMBHs.
\end{abstract}




\section{Introduction}
Observations of the tidal disruption of stars by supermassive black holes {}{(SMBHs; }\citealt{hills75,rees88}) {}{can} yield unique 
information {}{about}
several, otherwise{}{-}inaccessible astrophysical phenomena. 
When a disruption occurs, {}{approximately half of the tidally-disrupted debris returns to the SMBH \citep{rees88}, circularizes through a sequence of dissipation processes \citep{evans89}, and generates a luminous, panchromatic accretion episode \citep{cannizzo90, ulmer99, lodato11}. Recent numerical investigations have shown that the material circularizes} 
on scales comparable to the tidal disruption radius, given 
by $r_t = \left( M_h / M_* \right)^{1/3} R_*$ for a star of mass $M_*$ and radius $R_*$ around a black 
hole of mass $M_h$ \citep{bonnerot16,hayasaki16,shiokawa15}. {}{If the black hole mass is less than $10^7M_{\odot}$, the rate at which material returns to the SMBH can lead to super-Eddington accretion rates lasting from days to years \citep{evans89}; in this case, radiation pressure can inflate the accretion disk into a quasi-spherical envelope extending to much larger radii \citep{loeb97, coughlin14}, strong winds can be driven from the surface of the disk \citep{strubbe09, strubbe11, shen16}, and relativistic jets can be launched from the system \citep{coughlin14, kelley14, coughlin15}, all of which lead to vastly different disk morphologies and appearances of the tidal disruption event (TDE)}. 

The rates {}{at which TDEs occur} 
are sensitive to the structure and relaxation processes taking place in galactic nuclei \citep{magorrian99,wang04}. Most disruptions are expected to occur around the lowest mass black holes, 
making TDEs particularly good probe{}{s} of the poorly{}{-}understood{}{,} low-mass end of the {}{SMBH} mass function \citep{stone16}.
Most promisingly, although 
much has {}{already} been {}{learned} from the first dozens of {}{observed} TDEs \citep{komossa15}, the rate of {}{TDE}
identifications is expected to increase by as much as two orders of magnitude as forthcoming 
wide-area surveys come online {}{(e.g., the Large Synoptic Survey Telescope; \citealt{ivezic08})}.

The presence of {}{SMBH}
binaries in galactic nuclei \citep{begelman80} may 
affect both the rate and {}{appearance}
 of TDEs. At early epochs, when the binary is still 
relatively wide, the secondary acts as a perturber that increases the rate at which stars 
diffuse into the loss cone of the primary \citep{polnarev94}. Later on, as the binary approaches 
coalescence, the rate of TDEs from {\em unbound} stars is predicted to be suppressed, as 
stars entering the loss cone and encountering the binary will most often be ejected rather 
than being tidally disrupted by either hole \citep{chen08}. Conversely, the rate of TDEs 
from {\em bound} stars may be greatly enhanced, as dynamics related to the Kozai-Lidov 
effect \citep{kozai62,lidov62} drive stars toward nearly{}{-}radial orbits \citep{ivanov05,chen09,chen11,li15}.

At still closer separations, the presence of a binary may alter not just the rate of events but also the 
evolution of individual TDEs. If the fallback time of the debris stream is on the order of the binary orbital time, then the fallback rate (and resultant disk formation) will be greatly modified from the single-SMBH case. For this to be relevant observationally requires a binary with an orbital period of years or less{; however,} provided this {}{additional}
criterion is satisfied{}{,} the {}{presence of the binary will} result {}{in a} 
modification of the fallback rate from the 
{}{predicted power-law} 
$\dot{M} \propto t^{-5/3}$ \citep{phinney89,evans89}{}{; thus, a binary companion could provide another means for the asymptotic fallback rate to deviate from the canonically-assumed $t^{-5/3}$ law (see \citealt{guillochon13} for a demonstration of how partial disruptions break from $t^{-5/3}$ and \citealt{hayasaki13}, who show that lower eccentricities provide another means of deviation from this behavior)}. In some subset of cases accretion 
on to the disrupting black hole can be interrupted \citep{liu11,ricarte16}{}, and in others 
accretion on to {\em both} black holes will ensue. \cite{liu14} have suggested that the 
observed lightcurve of a TDE in the galaxy SDSS J120136.02+300305.5 is {}{modified by a binary companion}
; irrespective of whether or not that claim is confirmed, it seems 
very probable that future samples of thousands of TDEs will contain some events {}{whose appearances have been altered by }
binary dynamics. 
Successfully isolating and modeling those 
events would not only {}{generate}
a sample of SMBH binaries at interestingly small separations{}{, it would also provide a new technique for finding SMBH binaries that is independent of emission line diagnostics that can be plagued by spurious correlations with gas dynamics \citep{muller15}}.

Our goal in this paper is to analyze the morphological evolution of the accretion disks generated from TDEs by SMBH binaries, and to quantify the accretion and fallback rates of the disrupted debris onto the SMBHs. Using restricted 3-body integrations, we 
first calculate encounter cross sections and geometries in the case where stars approach 
a circular binary on a parabolic trajectory {}{about the binary center of mass}. (In this paper we focus on unbound 
stars for reasons of simplicity, as modeling TDEs from bound stars cannot be disentangled 
from the history of the binary merger.) We then simulate the hydrodynamics of an unbiased sample of the resulting TDEs with smoothed-particle hydrodynamics (SPH) simulations, and we directly calculate the accretion rates onto the SMBHs. {}{In doing so }
we extend the work of \cite{ricarte16}, who studied TDEs in binaries using 
a test particle approach, and that of \cite{hayasaki16b}, who used hydrodynamic simulations 
to study the rare case 
of a TDE nearly 
coincident with the final coalescence of 
the binary. 

The plan of the paper is as follows: In \S\ref{sec:parameters} we justify our choice of 
binary parameters, which is driven by the requirements that the orbital 
period be short enough to affect the observable portion of the fallback, while being 
long enough that the {}{gravitational-wave} inspiral time not be unreasonably short. In \S\ref{sec:tbp} 
we describe the methods used for the 3-body integrations, and we analyze the statistics (i.e., the tidal disruption rate, the distribution of energies at the time of disruption, the time taken to be disrupted, etc.) determined from those three-body encounters. We use the results of the three-body encounters -- particularly the properties of the orbit of the star immediately before disruption -- to simulate a subset of those encounters with the smoothed-particle hydrodynamics (SPH) code {\sc phantom} \citep{price10, lodato10}, and we present the results of those simulations in section \S\ref{sec:moderate}. We discuss the implications of our findings in \S\ref{sec:discussion} before concluding and summarizing in \S\ref{sec:conclusions}.

\section{Selection of parameters}
\label{sec:parameters}
Binaries tight enough to affect the dynamics of TDE{}{s} 
are typically well into the 
regime where inspiral is dominated by gravitational wave losses. Since these losses increase 
steeply with decreasing separation {}{($\propto a^{-4}$, where $a$ is the binary semimajor axis; see equation \ref{tmerge}), we }
expect the population of TDEs 
noticeably perturbed by binary companions to be dominated by {}{systems with larger separations.} 

To estimate this 
relevant separation, consider a binary with masses $M_{h,1}$ and $M_{h,2}$ in  
a circular orbit of semimajor axis $a$. Ignoring stellar structure-dependent 
corrections \citep{diener95}, a star of mass $M_*$ and radius $R_*$ that interacts with the binary will be 
disrupted if it strays within a distance $r_{t,1} = (M_{1,h} / M_*)^{1/3} R_*$ 
of the primary, or $r_{t,2} = (M_{2,h} / M_*)^{1/3} R_*$ of the secondary.

After disruption, the bound debris 
is on high{}{ly-eccentric} 
orbits that span a range of semi-major axes around the disrupting black hole. 
For a disruption by the primary, for example, an orbit with semi-major axis 
$a_{\rm debris}$ will have an apocenter {}{of} 
 $\simeq 2 a_{\rm debris}$ and 
a period $P = 2 \pi ( a_{\rm debris}^3 / GM_{h,1} )^{1/2}$. At a time of $P$ 
after disruption, the binary will strongly perturb the fallback {}{of the bound material} if 
the apocenter distance of the debris is outside of the 
Roche lobe of the disrupting hole. Defining the mass ratio $q \equiv M_{h,2} / M_{h,1} < 1$, 
the spherical effective radius of the Roche lobe around the primary is approximately 
\citep{eggleton83},
\begin{equation}
 r_L \simeq \frac{0.49 q^{-2/3}}{0.6 q^{-2/3} + \ln( 1 + q^{-1/3})} a.
\end{equation}
The secondary's Roche lobe radius is given by the same expression with 
$q \rightarrow 1/q$. Imposing the condition that the apocenter of debris 
returning after time $P$ is $r_L$, the binary separation in units of the 
primary's tidal disruption radius is,
\begin{equation}
 \frac{a}{r_{t,1}} = \frac{0.6 q^{-2/3} + \ln (1 + q^{-1/3})}{0.245 q^{-2/3}} 
 \left( \frac{GM_*}{4 \pi^2 R_*^3} \right)^{1/3} P^{2/3}.
\label{TDE_a} 
\end{equation} 
There is no dependence on the primary mass (or on the secondary mass, 
for disruptions by the secondary). For a given type of star, binaries that 
can potentially perturb TDE fallback rates on a specified time scale $P$ have a fixed 
separation in units of tidal radii. For Solar type stars and $q=0.2$, for 
example, a binary with $a / r_{t,1} \approx 1.6 \times 10^2$ would 
perturb the fallback on to the primary on time scales of around one month.

For circular orbits, the gravitational wave coalescence time scale from 
separation $a$ is \citep{peters63},
\begin{equation}
 t_{\rm merge} = \frac{5 c^5}{256 G^3 M_{1,h} M_{2,h} (M_{1,h} + M_{2,h})} a^4. \label{tmerge}
\end{equation} 
With $a$ specified as in equation~(\ref{TDE_a}) and with a given orbital period, 
the merger timescale for binaries close enough to perturb TDEs scales with 
mass as $M^{-5/3}$. We therefore expect --- even more strongly than for 
TDEs as a whole --- that the population of TDEs with observable binary 
perturbations ought to be concentrated toward the low mass end of the 
SMBH mass function. At sufficiently low masses, however, the merger 
time scales are long enough that a non-zero number of binary-perturbed 
TDEs are expected in a large sample. Taking as an example a 
binary with a $10^6 \ M_\odot$ primary and a mass ratio $q=0.2$, 
the separation could be close enough to influence TDE dynamics on a  
time scale of a month while still having a merger time of $t_{\rm merge} \approx 
4 \times 10^7 \ {\rm yr}$. Ignoring for now any enhancement of the TDE rate, 
taking the ratio of this merger time to that of galactic mergers (of the order of 
a Gyr) would suggest, very roughly, that of the order of 1\% of TDEs in this 
mass range would take place in observationally interesting binaries. {}{From a more rigorous standpoint, Wegg \& Bode (2011) have calculated, by following the inspiral of the secondary SMBH during the disruption and ejection of stars in the vicinity of the primary, that 3\% of all TDEs should occur by binaries as they transition to a hardened state. } A substantially higher fraction is predicted if we relax the one month requirement 
by even a modest factor.

Finally, we ask whether the TDEs of observational interest in binaries are 
likely to arise primarily from bound {}{stars -- those that diffuse in energy space onto orbits that remain tightly bound to the binary  \citep{peebles72, bahcall76} --} or unbound stars{}{, which scatter primarily in angular momentum space and thus have effectively zero energy with respect to the black holes \citep{shapiro76, frank76, lightman77}; see \citet{chen09} and \citet{chen08}, respectively, for an analysis of each of these cases}.
The number of bound stars can be estimated if we assume that each black 
hole is initially surrounded by a cusp of stars with a density profile 
$\rho(r) \propto r^{-7/4}$ \citep{bahcall76}, normalized such that the 
enclosed stellar mass equals the black hole mass within the sphere of 
influence $r_{\rm BH} = GM_h / \sigma^2$, where $\sigma$ is the nuclear 
velocity dispersion. The nearest star then has an orbital radius 
$r_{*,1} \sim G M_*^{4/5} M_h^{1/5} \sigma^{-2}$. Adopting an 
$M$-$\sigma$ relation {}{of} $M_h = 10^6 ( \sigma / 60 \ {\rm km s^{-1}})^4 \ M_\odot$ 
(roughly consistent with \citealt{gultekin09}; though see \citealt{graham13}), the {}{distance of the }star nearest the primary 
black hole is

\begin{equation}
 \frac{r_{*,1}}{r_{t,1}} \sim 8 \left( \frac{M_{h,1}}{10^6 \ M_\odot} \right)^{-19/30},
\end{equation}
where $r_{t,1}$ is the tidal radius of the primary and we are assuming Solar-type stars. 
{}{This argument} suggests {}{that}, 
for the observationally 
interesting binaries (with $a / r_{t,1} \sim 10^2$){}{,} at most a handful of bound stars 
from the original cluster would be left to be 
tidally disrupted over the 
last few tens of Myr. {}{For this reason} we focus on TDEs from unbound stars in this paper. {}{However, we note that} bound 
stars could be more important if the stellar density on very small scales was 
enhanced {}{due to}, for example{}{,} 
the migration of resonantly captured stars \citep[in an 
analogous context,][]{yu01} or via the disruption of binaries \citep{hills88}.
\section{Three-body interactions}
\label{sec:tbp}
Here we describe the initial setup of both the binary and the stars for the three-body encounters (Section 3.1), before going on to our results concerning the statistics of those encounters (Section 3.2).  
\subsection{Three-body setup}
We consider the gravitational interaction between a star, treated as a test particle, and a circular binary with masses $M_1$ and $M_2$, with $M_1 \geq M_2$. The equations for the evolution of the star are then standard, and are given for completeness in Appendix \ref{sec:bineqs}. 

As noted in \S{\ref{sec:parameters}}, the interesting cases (from observational and physical standpoints alike) are when the orbital period of the binary is on the order of months and the masses are relatively small. Given these constraints, we will let the mass of the primary be $M_1 = 10^6M_{\odot}$ and the binary semimajor axis be $a = 100\,r_t = 100R_*(M_h/M_*)^{1/3}$, where $R_* = 1R_{\odot}$ and $M_* = 1M_{\odot}$ are the radius and mass of the star; these numbers then give $a = 2.26\times10^{-4}$ pc = $46.5$ AU and $T_{orb} = 106$ days, where $T_{orb}$ is the orbital period of the binary.

Our primary goal in this paper is to investigate the hydrodynamic interactions that take place after stellar disruption. We therefore focus on the single mass ratio $q = 0.2$, and run a large number (10 million) of realizations to gather good statistics. However, for comparative purposes, we will also integrate a smaller number (one million) of three-body interactions in which we vary the mass ratio between $q = 0.1$ and 1 in increments of 0.1.

To specify the initial positions and velocities of the star, we will assume that the galaxy nucleus hosting the SMBH binary is spherical, meaning that stars entering the sphere of influence of the binary do so isotropically. We assume that the stars of interest approach the binary on parabolic orbits about the center of mass (COM), and, as noted previously, we ignore any contribution from bound stars. 

For the initial conditions, we let the stars be uniformly distributed on a sphere of radius $r_0$, and we will choose $r_0 = 50$ in units of the binary semi-major axis. This large of a separation implies that the quadrupole correction to the potential relative to the monopole term is $\Phi_{quad} \simeq 50^{-2}$, so the star effectively sees the binary as a point mass at the COM. The square of the angular momentum of the star will be uniformly distributed over the range $\ell^2 = [0,4]$; we truncate the upper bound at 4 because the pericenter of the star is $r_p = \ell^2/2$, and the probability of interacting with the binary (and being tidally disrupted) is significantly reduced once $r_p \gtrsim 1$ (see  Figure \ref{fig:beta_pdf} for a validation of this notion). A uniform distribution is appropriate if the scattering events experienced by the tidally-disrupted star impart a large relative change in its angular momentum per orbit (the pinhole regime; \citealt{lightman77}). 

\subsection{Results}
We integrated ten million, restricted three body interactions between a star and a binary SMBH with a mass ratio of $0.2$ for a maximum of 1600 orbits; for comparative purposes, we also integrated one million encounters between a star and binary SMBHs with mass ratios in the range $q = 0.1 - 1$ in increments of 0.1, also for 1600 orbits. In every case the primary black hole had a mass of $10^6M_{\odot}$; while the explicit value of the mass of either hole is not needed to integrate the dynamical equations (which depend only on the mass ratio; see equations \ref{L} -- \ref{EL}), it is necessary for specifying when the star is tidally disrupted. If the star came within the tidal radius of either black hole, the star was considered ``tidally disrupted'' and contributed to the total number of encounters that ended in the destruction of the star{}{ (note that this criterion only accounts for total disruptions of most low-mass stars; see below)}; if the star receded to more than 100 semimajor axes from the binary, it was considered ``ejected'' and did not contribute to the total number of TDEs; and finally, if it stayed within 100 semimajor axes of the binary for more than 1600 orbits, the orbit was considered ``inconclusive'' and did not contribute to the total number of TDEs. Experimentation shows that these criteria, although approximate, suffice to yield reliable statistics for TDE dynamics in binaries. {}{We are also not considering partial disruptions, which occur for $\beta \gtrsim 0.5$ for a $\gamma = 5/3$ polytrope \citep{guillochon13}}. 

\begin{figure}
   \centering
   \includegraphics[width=0.47\textwidth]{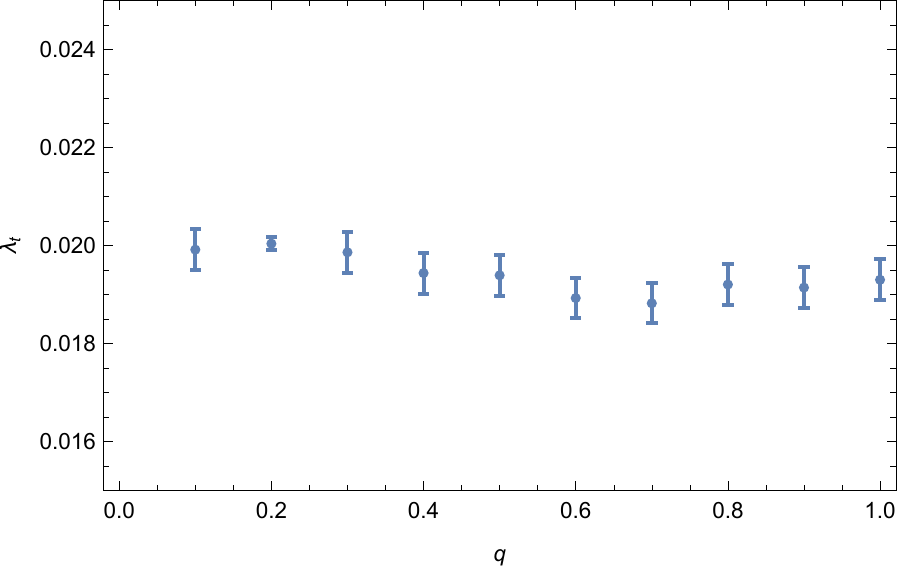} 
   \caption{The probability of being {}{fully} tidally disrupted, $N_t/N_e \equiv \lambda_t$ as a function of the mass ratio $q$, with the associated $3-\sigma$ error bars. The error bars for the case of $q = 0.2$ are smaller by a factor of $\sim3$ because of the larger number of three-body integrations done for that case.}
   \label{fig:tderate}
\end{figure}

Figure \ref{fig:tderate} shows the tidal disruption {}{fraction}, $N_t/N_e \equiv \lambda_t$, determined by simply taking the total number of tidal disruptions, $N_t$, and dividing by the total number of encounters, $N_e$; the $x$-axis is the mass ratio of the binary. The $3-\sigma$ error bars were calculated by assuming that the probability of being disrupted follows a Poisson distribution, from which the RMS value is $\sigma = \sqrt{\lambda_t/N_e}$. Interestingly, this figure demonstrates that the probability of being {}{fully} disrupted by the binary is nearly independent of the mass ratio; taking the average over all the mass ratios, we find that the average rate of disruption is $<\lambda_t> = 0.0194$. The exact values of the mean and $1-\sigma$ deviations for the different mass ratios are given in Table \ref{tab:1}.

\begin{figure} 
   \centering
   \includegraphics[width=0.47\textwidth]{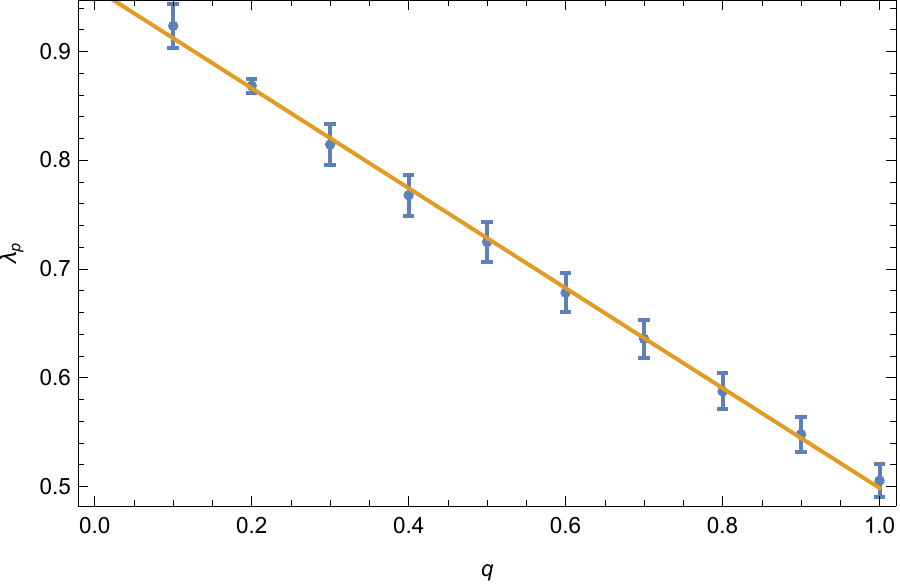} 
   \caption{The {}{fraction} of {}{full} disruptions contributed by the primary black hole, $\lambda_p$, as a function of the binary mass ratio, $q$. The error bars indicate the $3-\sigma$ deviations, and the yellow line shows the best-fit, linear relationship $\lambda_p = 0.96-0.46q$. As for Figure \ref{fig:tderate}, the smaller error bars for $q = 0.2$ are due to the fact that we integrated 10 million, as opposed to 1 million, three-body encounters.}
   \label{fig:Npofr}
\end{figure}

Figure \ref{fig:Npofr} shows the {}{fraction} of disruptions by the primary black hole, $N_p/N_t \equiv \lambda_p$, as a function of the binary mass ratio, $q$, calculated by simply taking the total number of disruptions by the primary, $N_p$, and dividing by the total number of TDEs, $N_t$. The ratio drops from 0.5 at $q=1$ to $\simeq 0.92$ at $q = 0.1$. Overall, we find that the {}{probability of being disrupted} by the primary as a function of $q$ is very well-fit by the linear relation

\begin{equation}
\lambda_p(q) = 0.96-0.46\,q, \label{bestfit}
\end{equation}
which is shown by the yellow line in Figure \ref{fig:Npofr}.

{}{The probability of being disrupted by the primary must satisfy $\lambda_p(q=0) = 1$ and $\lambda_p(q = 1) = 1/2$, and Equation \eqref{bestfit}, which is the best fit linear relationship that results from a least-squares regression, violates this requirement at $q = 0$. This lack of agreement could be due to the fact that the best fit line was only calculated with the mean values at each $q$, meaning that there is some uncertainty in the coefficients appearing in Equation \eqref{bestfit} due to the error bars in Figure \ref{fig:Npofr}. Alternatively, the ``true'' relationship for $\lambda_p$ as a function of $q$ may not be exactly linear, and a higher-order fit could cause the constant in Equation \eqref{bestfit} to change from 0.96 to 1. In either case, the fact that the best fit conforms almost exactly to the expected relationship based on the requirements that $\lambda_p(0) = 1$ and $\lambda_p(1/2) = 1/2$ likely implies that the true distribution is very nearly linear.}

The positive correlation between the number of TDEs caused by the primary SMBH and the mass ratio is predictable: the larger black hole has a larger tidal radius, and thus the probability of being tidally disrupted by that hole is correspondingly greater. However, if the increased rate of destruction by the primary were purely due to this geometric effect, then we would expect the probability of being disrupted by the primary to scale as $q^{2/3}$, i.e., as the ratio of the \emph{areas} of the black holes. It is apparent from Figure \ref{fig:Npofr}, though, that a linear relationship fits the rate of TDEs due to the primary black hole extremely well, meaning that there is an additional, dynamical effect that increases the probability of being disrupted by the primary. 

\begin{figure} 
   \centering
   \includegraphics[width=0.47\textwidth]{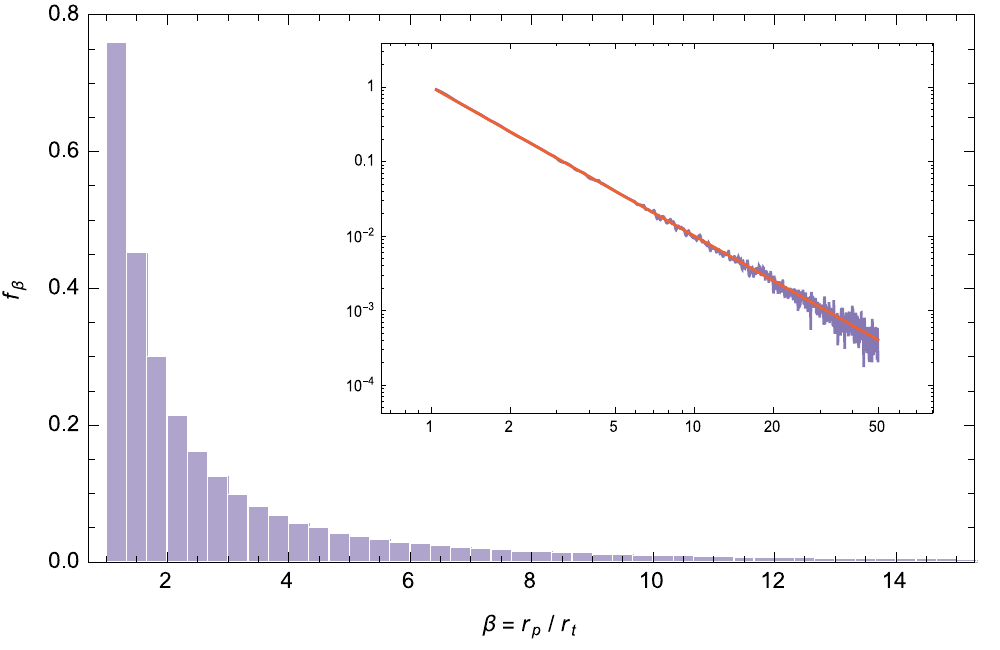} 
   \caption{The probability distribution of the impact parameter $\beta \equiv r_p/r_t$, where $r_p$ is the point of closest approach to the disrupting black hole and $r_t$ is the tidal radius of that hole, for a binary mass ratio of $q = 0.2$. The inset shows the value of the PDF at the midpoints of the bins (purple curve) and the analytic solution $f_{\beta} = 1/\beta^2$ (red curve) on a log-log scale. }
   \label{fig:beta_pdf}
\end{figure}

\begin{figure} 
   \centering
   \includegraphics[width=0.47\textwidth]{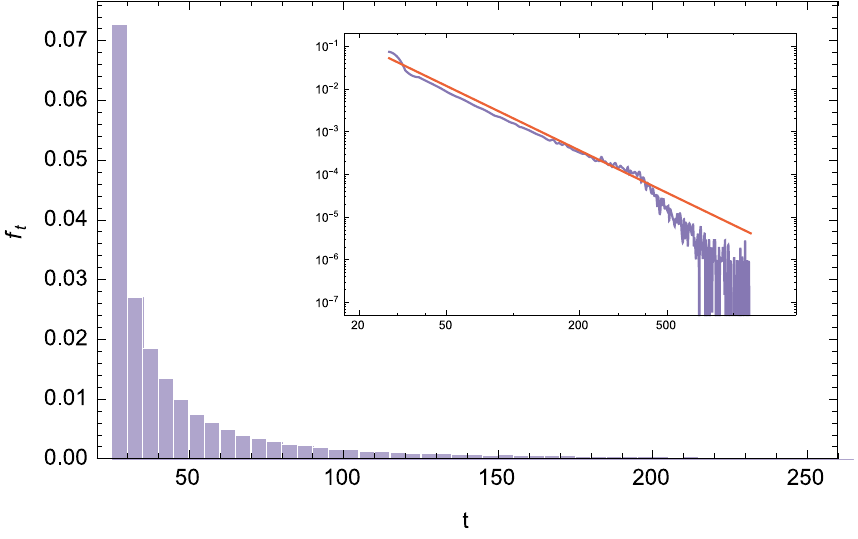} 
   \caption{The PDF for the number of binary orbits taken to be disrupted for the total sample of TDEs disrupted by a binary with mass ratio 0.2. The sharp cutoff for orbits less than roughly 25 is due to the fact that every star took approximately $50^{3/2}\sqrt{2}/(6\pi) \simeq 26.5$ orbits to just reach the binary. The inset shows the value of the PDF at the midpoints of the bins (purple curve), while the red curve shows the solution $f_t = 0.056\times(26.5/t)^{5/2}$.}
   \label{fig:t_pdf}
\end{figure}

\begin{figure*}
   \centering
   \includegraphics[width=0.325\textwidth]{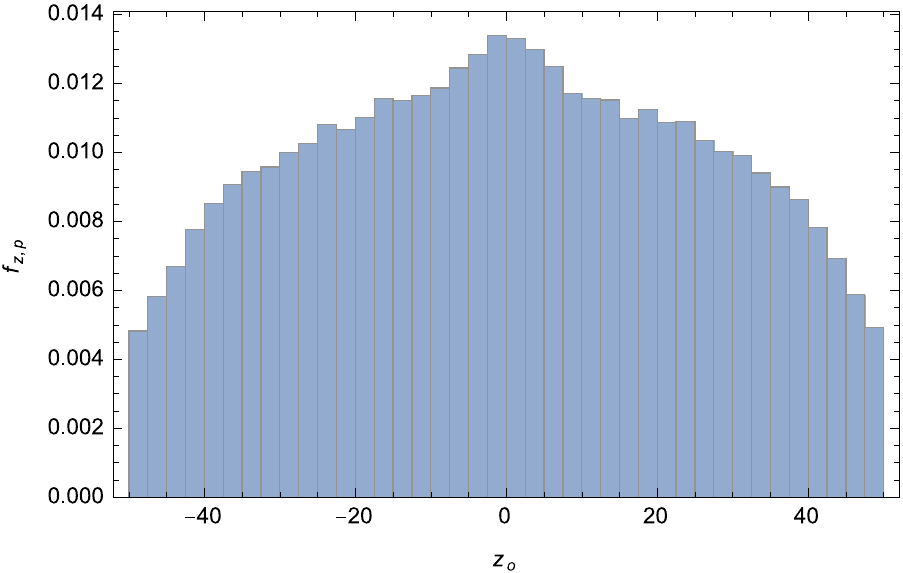} 
   \includegraphics[width=0.325\textwidth]{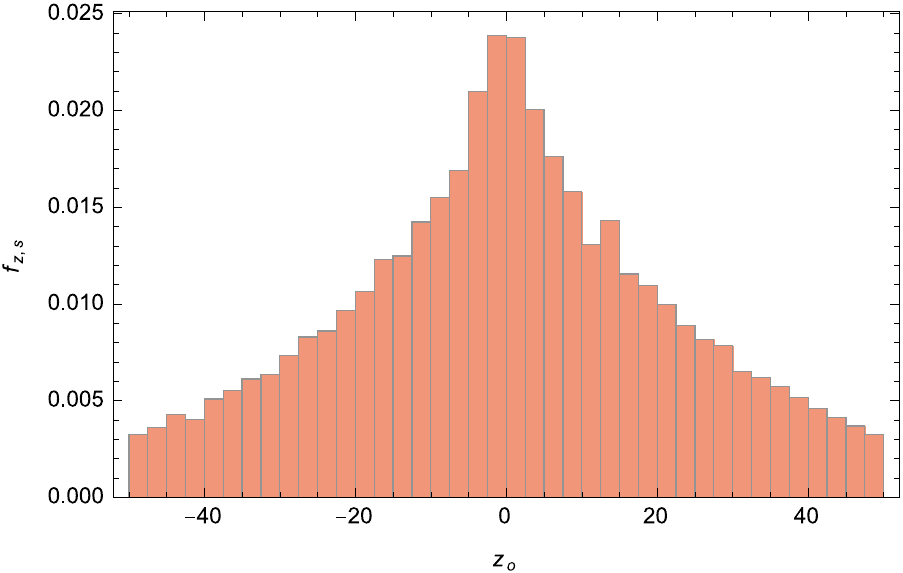} 
   \includegraphics[width=0.325\textwidth]{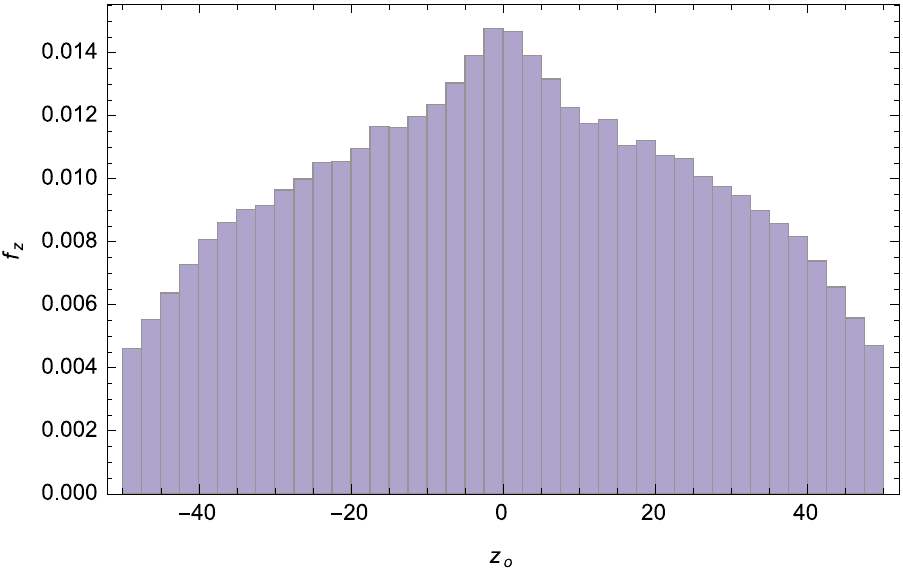} 
   \caption{The PDF for the initial $z$ coordinate, in units of the semimajor axis of the binary, of stars disrupted by the $q=0.2$ binary if the star is disrupted by the primary (left panel), the secondary (middle panel), and the composite PDF (right panel). This figure shows that disrupted stars are preferentially located in the plane of the binary, which arises from the fact that many stars are placed onto temporary ``bound'' orbits before being disrupted, which is also apparent from Figure \ref{fig:t_pdf}.}
   \label{fig:z_pdf}
\end{figure*}

\begin{figure*}
   \centering
   \includegraphics[width=0.325\textwidth]{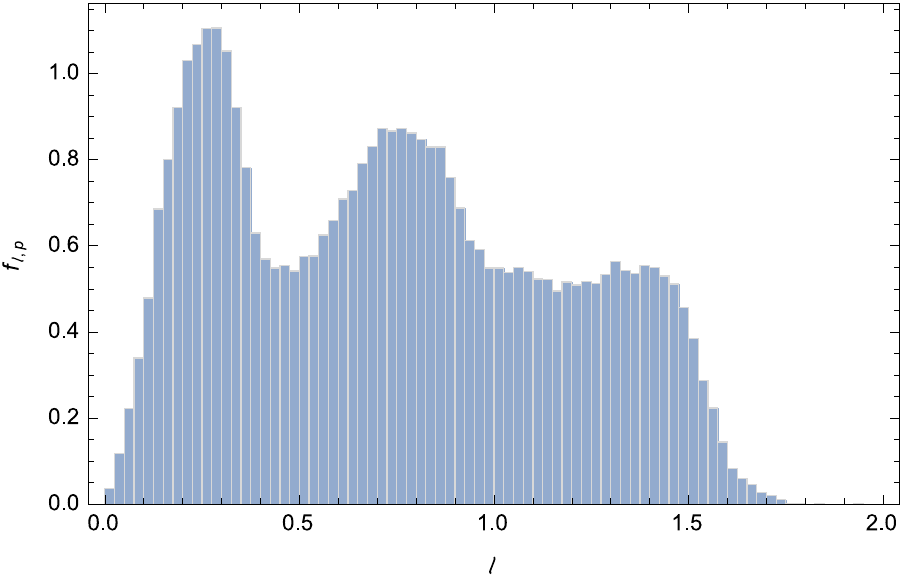} 
   \includegraphics[width=0.325\textwidth]{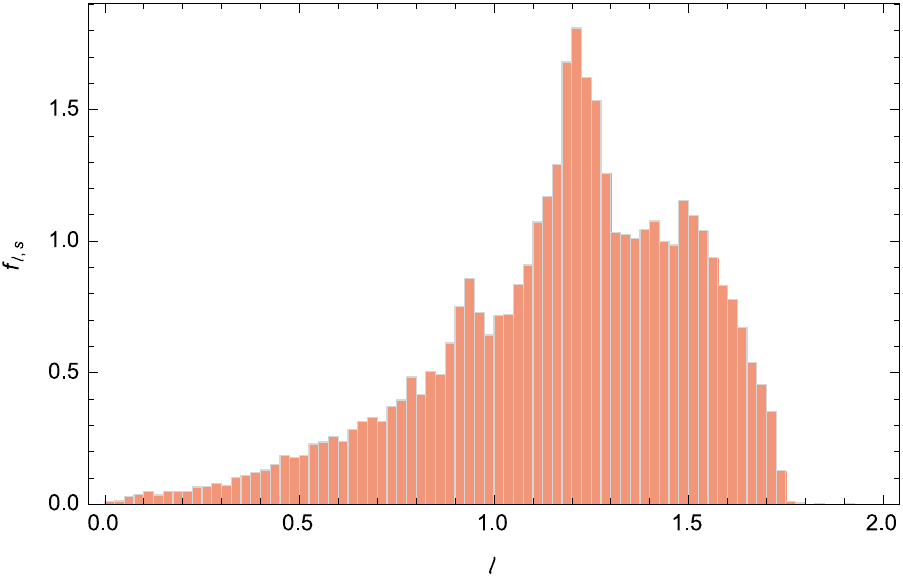} 
   \includegraphics[width=0.325\textwidth]{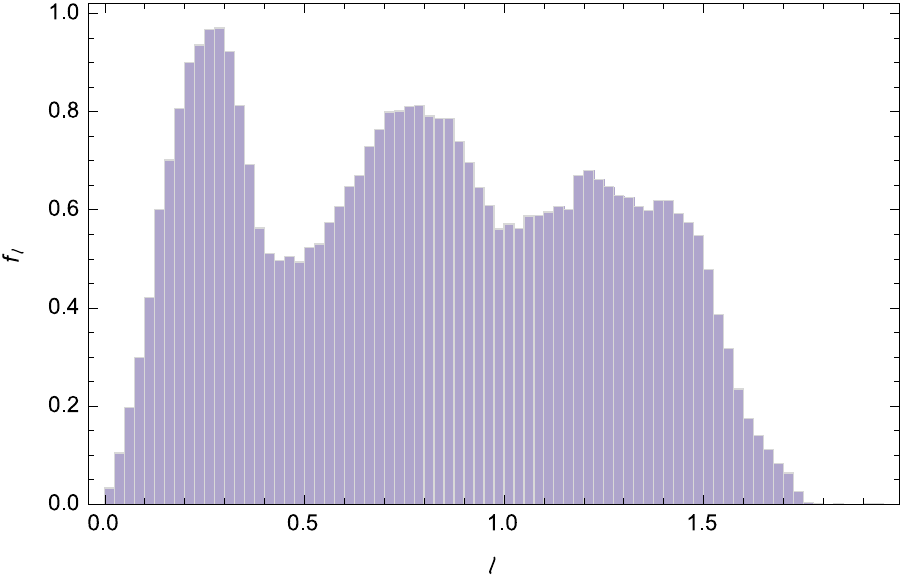} 
   \caption{The PDF for the original specific angular momentum, in units of $GMa$, of the stars disrupted by the binary with mass ratio 0.2. The left-hand panel represents stars disrupted by the primary, the middle panel by the secondary, and the right-hand panel is the total, composite distribution function. The sharp cutoff at specific angular momenta greater than about $\ell \simeq 1.8$ comes from the fact that the pericenter of the star is $r_p = \ell^2/2$, meaning that stars with angular momenta greater than about $\sqrt{2}$ will not interact with the binary.}
   \label{fig:ell_hist}
\end{figure*}

\begin{figure*}
   \centering
   \includegraphics[width=0.325\textwidth]{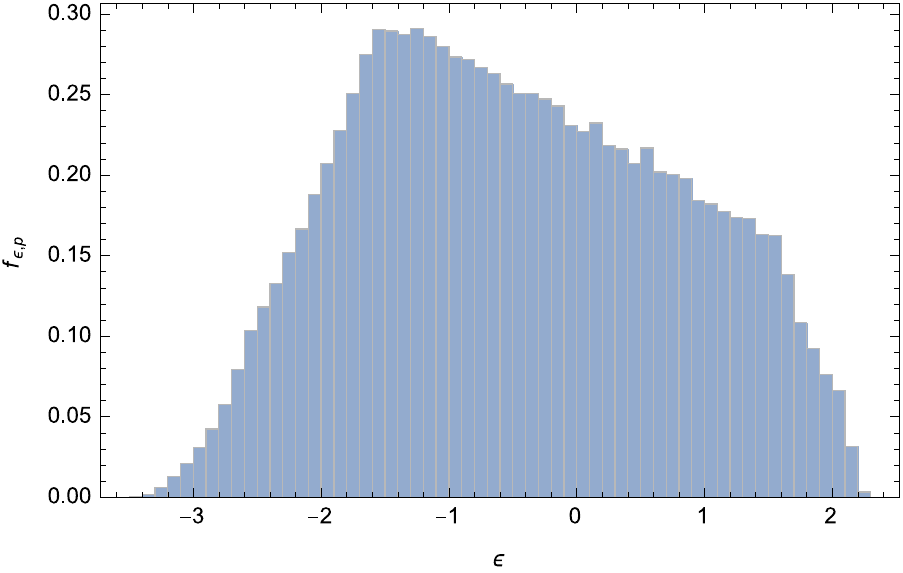} 
   \includegraphics[width=0.325\textwidth]{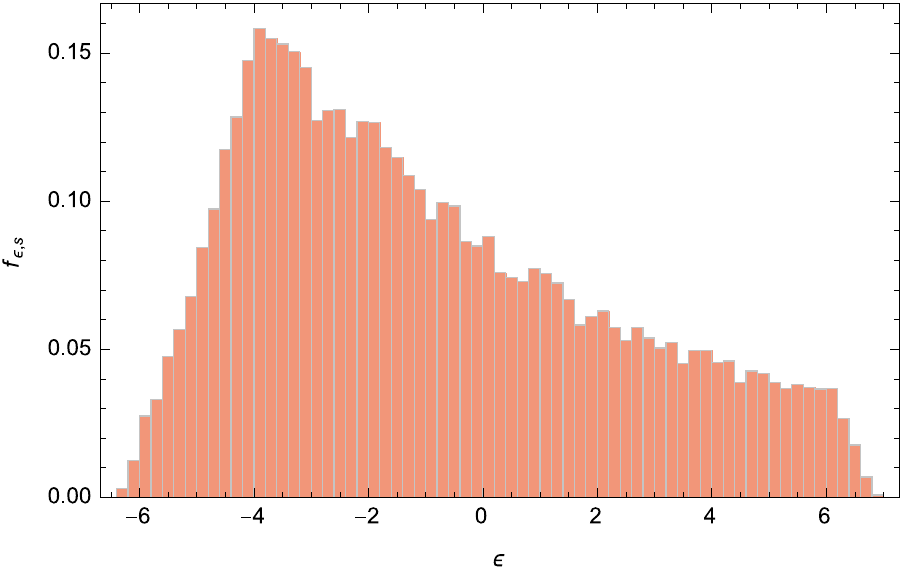} 
   \includegraphics[width=0.325\textwidth]{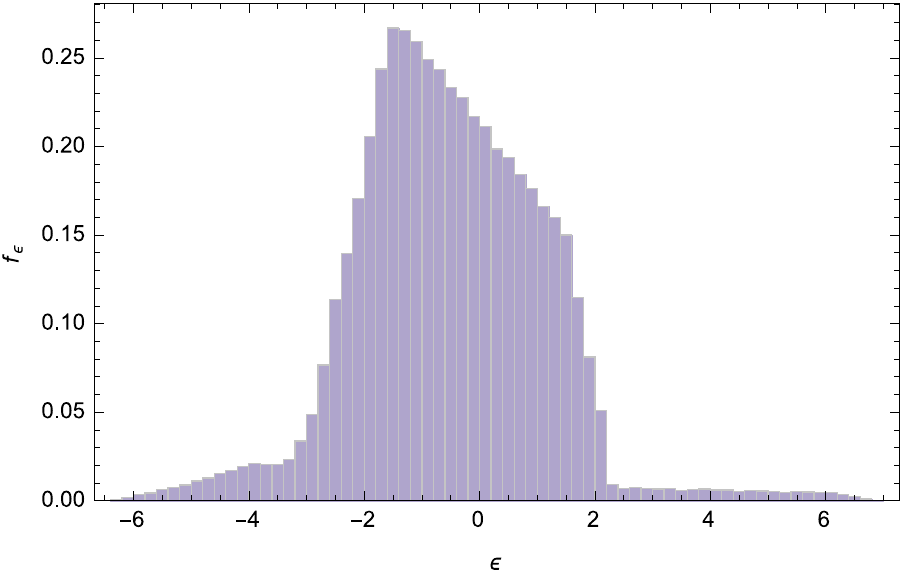} 
   \caption{The PDF for the specific energy at the time of disruption of stars disrupted by the $q=0.2$ binary, where the energy is in units of $GM/a$. The left-hand panel represents stars disrupted by the primary, the middle panel by the secondary, and the right-hand panel is the total, composite distribution function. The wider spread in the specific energy distribution for disruptions by the secondary arises from the higher relative velocities generated between the star and that hole (see equations \ref{R1eq} -- \ref{phieq}).}
   \label{fig:eps_hist}
\end{figure*}

Figure \ref{fig:beta_pdf} shows the probability distribution function (PDF), $f_{\beta}$, of the impact parameter $\beta = r_t/r_p$, where $r_p$ is the point of closest approach of the star to the disrupting black hole and $r_t$ is its tidal radius, for the full set of TDEs with a binary mass ratio of 0.2. The bin width for this figure is $\Delta{\beta} = 1/3$, which captures a large number of encounters (on the order of thousands) per bin. The inset in the figure shows, on a log-log scale, the PDF at the bin center (purple curve) alongside the power-law solution $1/\beta^2$ (red line), which demonstrates that the PDF for the impact parameter is very well approximated by the function $f_{\beta} = \beta^{-2}${}{; this demonstrates that the square of the angular momentum of the star about the disrupting SMBH is almost exactly characterized by a uniform distribution}. The $\beta$ that results in the star being swallowed whole by the primary is $\beta_{c,p} = 23.5$, while that for the secondary is $\beta_{c,s} = 118$ (these simply come from equating the tidal disruption radius to the Schwarzschild radius of the hole). The average $\beta$ for stars disrupted by the primary is then $<\beta_p> \simeq \ln(23.5) \simeq 3.2$, while that for the secondary is $<\beta_s> \simeq \ln(118) \simeq 4.8$. The average impact parameter $\mu_{\beta}$ (over both the primary and secondary, cut off at the Schwarzschild radius of the secondary) for all of the mass ratios is given in Table \ref{tab:1}.

Figure \ref{fig:t_pdf} illustrates the PDF for the time, in units of binary orbits, taken to be disrupted for the binary mass ratio of 0.2. The bin width for this figure was $\Delta{t} = 5$ orbits, and the sharp falloff for times less than roughly 25 orbits is due to the fact that the stars originated at a distance of 50 semimajor axes (and from such a distance each star takes approximately 27 orbits just to reach the binary). The inset shows the value of the PDF at the bin center (purple curve) and the solution $f_t \propto t^{-5/2}$ (red line) on a log-log scale. This inset therefore demonstrates that the PDF for the time taken to be disrupted, $f_t$, is well-matched by the power-law $f_t = 0.056\times(26.5/t)^{5/2}$ (the multiplicative constant comes from the normalization of the PDF) until the number of orbits surpasses roughly 500 orbits, at which point the PDF drops off much more rapidly.

Figure \ref{fig:z_pdf} shows the PDF for the initial $z$-coordinate of the disrupted stars for the mass ratio of $0.2$, where the left panel is for stars disrupted by the primary, the middle for the secondary, and the right-hand panel is the total PDF. Recalling that the original distribution of stars was isotropic (i.e., equal probability of being at any given $\{x_0,y_0,z_0\}$ constrained to a sphere of radius 50 in units of the binary semimajor axis), this figure reveals that disrupted stars are preferentially located in the plane of the orbit of the binary. The reason for this tendency is that the majority of the disrupted stars are temporarily captured by the binary before being disrupted (i.e., there is a significant fraction that are not disrupted on their ``first pass''), which is apparent from Figure \ref{fig:t_pdf}. To be placed on a bound orbit, however, the star must ``see'' one of the black holes with a lower velocity, meaning that it has effectively less kinetic energy with respect to the system.

Figure \ref{fig:ell_hist} displays the PDF for the original (i.e., when the center of mass of the star is at 50 binary separations) specific angular momentum (in units of $\sqrt{GMa}$) of the stars disrupted by the primary (left panel), the secondary (middle panel), and the composite (right panel) for a mass ratio of 0.2. The bin width for this figure is $\Delta{\ell} = 0.025$. This figure shows that stars with specific angular momenta greater than $\ell \gtrsim 1.8$, or pericenter distances greater than $r_p \gtrsim 1.6\,a$, are not tidally disrupted, which is a reasonable result: the star must either come close enough to be disrupted by one of the stars on the first pass or to be temporarily captured by the binary. Note, however, that there is a sizable fraction of stars that have pericenter distances greater than $r_p > 1$ ($\ell > \sqrt{2}$), which shows that the star does not actually have to pass \emph{through} the binary in order to be disrupted. The average angular momentum over the full set of disrupted stars for a mass ratio of 0.2 is $\mu_{\ell} = 0.81$, while the standard deviation is $\sigma_{\ell} = 0.44$; the averages and standard deviations for the other mass ratios are given in Table \ref{tab:1}. 

Figure \ref{fig:eps_hist} shows the PDF for the specific energy (in units of $GM/a$) of the disrupted stars at the time of disruption for a mass ratio of 0.2; the left-hand panel shows the distribution for the stars disrupted by the primary (the bin width for this panel is $\Delta\epsilon = 0.1$), while the middle and right-hand panels show the distribution for stars disrupted by the secondary and the composite PDF, respectively (the bin width for both of these panels is $\Delta\epsilon = 0.2$). The energy was calculated from the canonical Hamiltonian of the system evaluated at the time of disruption, i.e., $\epsilon = v^2/2+\Phi$, where $v$ is the speed of the star and $\Phi$ is the total potential of the binary, given by equation \eqref{Phieq}. Even though this energy is not conserved owing to the time-dependent nature of the potential, it should give some indication of the ``boundness'' of the disrupted debris to the binary. This figure demonstrates that, due to the interaction with the binary, the energies of the disrupted stars deviate from the original, parabolic value of $\epsilon = 0$. Therefore, the specifics of the evolution of the tidally-disrupted debris may differ significantly from the disruption by a single SMBH -- not only because of the presence of the secondary, but also because of this change in the specific energy of the center of mass (see Section \ref{sec:moderate}). The average energy for the mass ratio of 0.2 is $\mu_{\epsilon} = -0.52$, with a standard deviation of $\sigma_{\epsilon} = 1.63$; those quantities for the other mass ratios are given in Table \ref{tab:1}.

\begin{figure}
   \centering
   \includegraphics[width=0.47\textwidth]{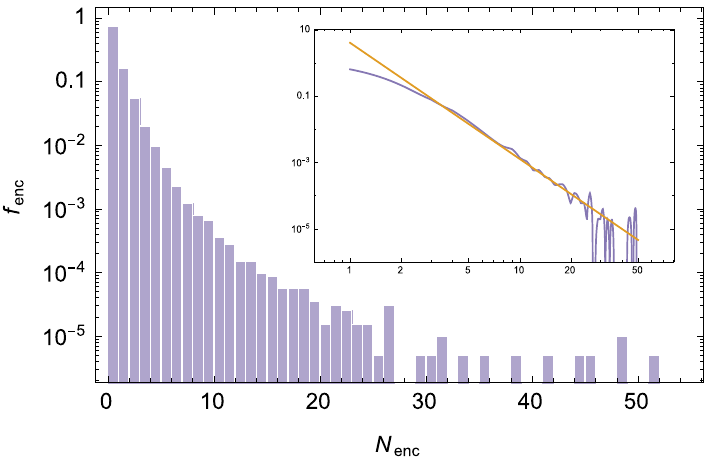} 
   \caption{The log of the PDF for the number of additional times that the disrupted stars come within three tidal radii of either SMBH, $N_{enc}$ (i.e., disrupted stars that only come within three tidal radii at the time of disruption have $N_{enc} = 0$), for the mass ratio of 0.2. The inset plots the values of the PDF at the bin center (purple curve) and the power-law $f_{enc} \propto N_{enc}^{-3.5}$, which approximates the PDF reasonably well.}
   \label{fig:Nenc}
\end{figure}

It is also apparent from the second panel in Figure \ref{fig:eps_hist} that the ``wings'' at larger energies in the composite PDF arise primarily from the disruptions by the secondary black hole. This feature is due to the fact that the change in the energy of the center of mass of the star originates from the relative motion between the star and the disrupting black hole: if the SMBH is moving away from the star at the time of disruption, then the effective kinetic energy of the star is decreased, leading to a bound state (and conversely for motion away from the star at the time of disruption). Since the speed of the secondary black hole is greater than that of the primary, the spread in energies at the time of disruption is greater for stars disrupted by the secondary. 

We have found that the positions of the stars at the time of disruption were isotropically distributed about the disrupting SMBH. This result may seem surprising given Figure \ref{fig:z_pdf}, which shows that the tidally-disrupted stars are preferentially located within the plane of the binary (i.e., small initial z-position), and we therefore might expect this preference to be reflected in the position of the star at the time of disruption. However, because the tidal disruption radius is a small fraction of the binary separation, even a very small initial $z_0$ correlates to a large deviation from the binary midplane (in units of tidal radii) when the star is tidally disrupted; for this reason the initial preference to be located near the midplane is negligible when the stars reach the disrupting SMBH. 

Figure \ref{fig:Nenc} shows the PDF for the number of ``additional encounters,'' $N_{enc}$, that each tidally disrupted star experiences; an additional encounter occurs any time that the star comes within 3 tidal radii of either SMBH (the encounter experienced when the star is actually disrupted, i.e., comes within one tidal radius of a SMBH, is not counted), for the mass ratio of 0.2. The inset shows the values of the PDF at the midpoints of the bins (purple curve) alongside the curve $f_{enc} \propto N_{enc}^{-3.5}$, which fits the solution fairly accurately and shows that the PDF falls off quite rapidly for larger values of $N_{enc}$. This figure demonstrates that, while the falloff of the PDF with $N_{enc}$ is rather steep, there is a non-zero fraction of stars that are ``nearly disrupted'' a number of times before actually crossing the tidal radius of either hole. In these cases, the star may be mildly to significantly distorted by the tidal interaction with the SMBHs, which could conceivably heat and rotate the star, cause it to inflate (rendering it more easily tidally disrupted), and change the stellar orbit in a manner not captured by our point particle treatment {}{(see also \citealt{antonini11}, who investigated some of these effects for the close passage of binary stars near isolated SMBHs)}. 

{}{We note that our criterion for disruption, that the pericenter distance of the star to either black hole be less than $r_t(M_{1,2}/M_*)^{1/3}$, only accounts for \emph{full} disruptions of $\gamma = 5/3$ polytropes (and stars with similar central densities; see below), in which there is no surviving core of the star. However, the distance characterizing complete disruption does depend on the stellar composition, with higher central densities generally requiring larger $\beta$'s \citep{guillochon13, manukian13}. Accounting for these composition-dependent differences would then result in an increase or decrease in the rate of disruption, though averaging over different stellar types may have little effect.}

{}{Perhaps more importantly, partial stellar disruptions do still generate debris streams that are bound to the SMBHs. In these cases, the overall amount of accreted material is less than that for a full disruption of the same star, and the asymptotic fallback rate differs from $t^{-5/3}$ \citep{guillochon13}. Nevertheless, in our analysis here we only consider full disruptions, the hydrodynamics of which we now consider, and leave the effects of partial disruptions to a future investigation.}

\begin{table*}
\fontsize{5.0pt}{10pt}
\selectfont
\center
\begin{tabular}{|c | c |c |c |c |c |c |c |c |c |c|}
\hline
$q$ & 0.1 & 0.2 & 0.3 & 0.4 & 0.5 & 0.6 & 0.7 & 0.8 & 0.9 & 1 \\
\hline
$\lambda_t\times10^{2}\,\,(\sigma_t\times10^{4})$ & 1.96 (9.4) & 2.00 (7.2) & 1.97 (3.4) & 1.96 (4.4) & 1.95 (4.8) & 1.91 (37) & 1.88 (2.6) & 1.89 (4.5) & 1.91 (9.5) & 1.93 (2.8)\\
\hline
$\lambda_p\times10\,\,(\sigma_p\times10^2)$ & 9.2 (1.0) & 8.7 (0.28) & 8.1 (1.5) & 7.7 (1.4) & 7.2 (1.0) & 6.7 (1.2) & 6.4 (1.1) & 5.9 (1.1) & 5.5 (1.2) & 5.0 (0.68)\\
\hline
$\mu_{\beta,p}$ ($\mu_{\beta,s}$) & 3.7 (5.5) & 3.8 (5.2) & 3.9 (4.9) & 3.9 (4.5) & 3.9 (4.3) & 3.8 (3.7) & 3.9 (4.0) & 4.0 (3.9) & 4.0 (3.7) & 3.9 (3.7) \\
\hline
$\mu_t$ ($\sigma_t$) & 80 (93) & 57 (63) & 50 (53) & 48 (49) & 46 (49) & 44 (43) & 43 (41) & 44 (44) & 43 (42) & 43 (40) \\
\hline
$\mu_{\ell}$ ($\sigma_{\ell}$) & 0.70 (0.45) & 0.81 (0.44) & 0.86 (0.43) & 0.91 (0.41) & 0.95 (0.40) & 0.97 (0.39) & 0.99 (0.39) & 1.0 (0.38) & 1.0 (0.37) & 1.0 (0.37)\\
\hline
$\mu_{\epsilon}$ ($\sigma_{\epsilon}$) & -0.42 (1.0) & -0.52 (1.6) & -0.52 (2.0) & -0.56 (2.3) & -0.60 (2.5) & -0.59 (2.6) & -0.59 (2.7) & -0.60 (2.7) & -0.65 (2.7) & -0.64 (2.7) \\
\hline
$\mu_{z}$ ($\sigma_z$) & -0.086 (27) & 0.10 (25) & 0.026 (24) & 0.094 (24) & -0.17 (23) & 0.097 (23) & -0.13 (23) & 0.16 (23) & 0.086 (23) & -0.045 (23) \\
\hline
$\mu_{N} (\sigma_N)$ & 0.75 (2.4) & 0.45 (1.4) & 0.36 (0.94) & 0.34 (0.74) & 0.33 (0.74) & 0.32 (0.72) & 0.31 (0.70) & 0.30 (0.63) & 0.30 (0.66) & 0.30 (0.68) \\
\hline
\end{tabular}
\caption{The mean values of parameters describing the orbits of the disrupted stars for the full set of simulated mass ratios, $q$; the numbers in parentheses are the $1-\sigma$ deviations. The quantities tabulated here are $\lambda_t$: the {}{probability} of {}{full} disruption; $\lambda_p$: the {}{probability} of {}{being fully disrupted} by the primary SMBH; $\mu_{\beta,p(s)}$: the mean impact parameter for {}{full} disruptions by the primary (secondary); $\mu_t$: the expectation value of the amount of time taken to be disrupted; $\mu_{\ell}$: the mean original angular momentum of the disrupted stars; $\mu_{\epsilon}$: the mean energy of the center of mass of the star at the time of disruption; $\mu_z$: the mean of the original $z$-coordinate of the disrupted stars; $\mu_N$: the mean number of additional ``close encounters,'' one such encounter occurring any time a star passes within 3 tidal radii of either black hole without being disrupted. }
\label{tab:1}
\end{table*}

\section{Hydrodynamic simulations of TDEs by binary SMBHs}
\label{sec:moderate}
In the previous section we found that the energy and angular momentum of the center of mass of the star at the time of disruption differ from those of a parabolic orbit. The evolution of the tidally-disrupted debris will therefore exhibit characteristics that are not predicted from the standard picture of TDEs in which the COM is assumed to follow a parabolic orbit; for example, the fallback time of the most bound debris may be earlier or later than otherwise expected, and more (or less) of the tidally-disrupted debris may be bound to the binary. In addition, we expect the motion of the disrupting black hole and the presence of the second SMBH to induce other variations in the fallback of the debris and the formation of the accretion disk (or disks). To investigate these effects on the hydrodynamic evolution of the tidally-disrupted debris, in this section we use numerical methods to simulate the tidal disruption of stars by a SMBH binary.

\subsection{Simulation setup}
\label{sec:setup}
We used the smoothed-particle hydrodynamics (SPH) code {\sc phantom} \citep{price10, lodato10} to simulate the tidal disruption of a solar-like star (i.e., one with a solar mass and a solar radius) by a SMBH binary. {\sc phantom} has been compared to grid-based algorithms for accuracy \citep{price10}. This code has been highly effective for simulating complex fluid geometries \citep{nixon12, nixon13, martin14a, martin14b, nealon15, dogan15} and has also been used to study TDEs, including the disruption process itself \citep{coughlin15}, the evolution of the disrupted debris \citep{coughlin16b}, and the formation of the accretion disk \citep{bonnerot16}, with the results being consistent with expectations from analytic estimates \citep{rees88, coughlin16b}.

We employ an artificial viscosity to maintain particle order and to capture shocks. Specifically, we adopt a variable $\alpha$ viscosity that evolves according to the prescription outlined in \citet{cullen10}, with $\alpha^{AV}_{min} = 0.01$ and $\alpha^{AV}_{max} = 1$, while we set $\beta^{AV} = 2$. Past investigations have shown \citep{price10} that this choice of $\beta^{AV}$ accurately captures shocks when the Mach number is not too extreme. 

For the parameters of the binary, we let the primary mass be $M_1 = 10^6M_{\odot}$, the secondary mass be $M_2 = 2\times10^5M_{\odot}$ (i.e., a mass ratio of 0.2), and the binary separation be $a = 100r_t$, where $r_t$ is the tidal radius of the primary SMBH. As mentioned in \S\ref{sec:parameters}, a separation this large guarantees that the binary orbital time is on the order of the fallback time of the tidally-disrupted debris. Also, this combination of masses and separation yields a fairly long gravitational-wave inspiral time (a few million years), meaning that there is a high probability that there will be at least one TDE before the coalescence of the binary. In particular, using the fact that the rate of disruption for a mass ratio of 0.2 is $\lambda_t = 0.02$ (Figure \ref{fig:tderate} and Table \ref{tab:1}) and the rate at which stars are scattered into the loss cone of an isolated SMBH is $\lambda_s = 10^{-4}$ -- $10^{-5}$ per galaxy per year \citep{frank76, stone16}, we expect the number of TDEs over the lifetime of the binary to be roughly $N_{tde} \simeq \lambda_t\lambda_s\tau_{bin}a/r_t$, where $\tau_{bin}$ is the lifetime of the binary. Adopting $\tau_{bin} \simeq 4\times10^{7}$ yr (Section \ref{sec:parameters}), we find that roughly $N_{tde} \simeq $ 800 -- 8000 stars should be disrupted before the binary coalesces. (We note that this number is actually an upper limit, as the fraction $a/r_t$ is a decreasing function of time; however, since the binary inspiral time is very sensitive to the separation (Equation \ref{tmerge}), the binary spends the majority of its lifetime at larger separations, and hence the true number should be close to the one quoted here.)

The disrupted star was constructed as an $n = 3/2$ polytrope \citep{hansen04} with a Solar mass and a Solar radius. We generated this polytrope by first placing the particles on a close-packed sphere and then stretching that sphere to closely resemble a polytropic distribution. The resulting configuration was then relaxed for ten sound-crossing times to achieve a static initial state, which closely matches the analytical polytrope solution. While a $\gamma = 5/3$ polytrope does not yield as high a density contrast as is appropriate for the true Sun \citep{guillochon13}, it approximates well the density distribution of real stars, especially at the low-mass end \citep{hansen04} where the IMF is thought to peak \citep{kroupa01}.

As we noted above, the properties of the stellar orbit at the time of disruption differ from those of the original parabolic encounter because of the interaction with the binary. To accommodate these deviations and to investigate their effects on the evolution of the tidally-disrupted debris, we initialized our SPH simulations with randomly selected encounters from Section \ref{sec:tbp} that ended in the disruption of the star, subject only to a cut to avoid the most deeply penetrating encounters (see below for details). To mitigate the computational cost, however, we did not simulate the entire three-body problem before the disruption. Instead, we opted to start the relaxed polytrope at 5 tidal radii from the disrupting black hole with the velocity of every SPH particle equal to that of the center of mass of the star, the orbit of the center of mass determined from the restricted three-body integrations; this approach is similar to what has been done in past investigations (e.g., \citealt{lodato09}). The choice of 5 tidal radii was made so that the star was sufficiently far from the disrupting black hole to be virtually unaffected by the tidal force, but close enough to not waste an excessive amount of computational resources in retracing the restricted three-body orbit. We ran 120 such simulations (i.e., of randomly selected orbits from Section \ref{sec:tbp} that ended in the disruption of the star) and we also ran 120 ``control'' simulations with identical initial conditions but without the presence of the other SMBH (i.e., if the disrupting black hole was the primary, then the control simulation was initiated with the same positions and velocities as the binary disruption but boosted into the rest frame of the primary; the evolution then proceeded with only the point mass potential of the primary). The control simulations then isolate the dynamical effects introduced by the presence of the second SMBH from the alterations to the energy and angular momentum of the center of mass of the star.

The gas was assumed to follow an adiabatic equation of state throughout the evolution of the TDE, with the adiabatic index equal to the polytropic index of $\gamma = 5/3$. These simulations therefore assume that the gas cools very efficiently after being shocked, and that the temperature remains low enough such that the contribution of radiation to the gas dynamics is small. Both the shock heating and the contribution of radiation would serve to ``inflate'' the gas and extend the resulting accretion regions around the binary and individual SMBHs to larger radii. While these effects have obvious important astrophysical implications, we leave their study to future investigations and focus primarily on the (hydro)dynamics in this paper. 

At all stages of the simulation the self-gravity of the gas was included. It has been demonstrated, both analytically \citep{kochanek94, coughlin16b} and numerically \citep{guillochon14, coughlin15, coughlin16a}, that self-gravity can be very important for modifying the structure of the disrupted debris, especially when the pericenter distance of the disrupted star is larger ($\beta \simeq 1$) and the adiabatic index of the gas is stiffer ($\gamma \ge 5/3$; \citealt{lodato09, coughlin16b}). The self-gravity of the debris was calculated by using a tree algorithm alongside an opening angle of 0.5 \citep{barnes86, gafton11}.

In every simulation we used $5\times10^5$ SPH particles. To assess the impact of resolution on our results, we also ran a lower-resolution ($2.5\times10^5$ particles) and a higher-resolution (1 million) simulation of one particular run. We discuss the convergence properties of the simulations based on these additional runs in Section \ref{sec:resolution}.

The gravitational field of the binary was modeled by two Newtonian point masses, and each point mass acted as a sink particle with an accretion radius of 0.5 times the tidal radius of the respective SMBH. Any particle passing within the accretion radius of either black hole is ``accreted'' and removed from the simulation. 

Finally, because the probability distribution of the impact parameter of disrupted stars falls off somewhat weakly with $\beta$ ($f_\beta = 1/\beta^2$), there are some orbits that have very small pericenter distances. For these cases, where general relativistic effects become very important even on the initial encounter, treating the potential as Newtonian likely introduces very large inaccuracies for the resulting distribution of debris. Furthermore, the star is intensely shocked in these instances, meaning that the ensuing expansion of the debris will be very sensitive to the thermodynamic properties of the gas and resolution. Because we feel that our present approach could not accurately capture these effects, we opted not to simulate any TDEs that had impact parameters greater than 5. {}{We note that this is still a very deep encounter, as the amount of vertical compression -- assuming that the envelope of the star undergoes effective freefall as it enters the tidal sphere of the disrupting hole -- is roughly $\delta{R}/R_* \simeq 1/\beta^3 \simeq 1/125$ (\citealt{carter83}; see also \citealt{rosswog09}, who found that one could ignite nuclear burning in white dwarfs compressed by similar amounts). Overall, our results for higher $\beta$ are in agreement with past investigations that considered similar scenarios (e.g., \citealt{ramirez-ruiz09}, who considered TDEs with $\beta = 3$ by intermediate mass black holes).}

\subsection{Results}
\label{sec:results}
\subsubsection{Morphological evolution}
\label{sec:morphology}
\begin{figure*}
   \centering
   \includegraphics[width=0.975\textwidth]{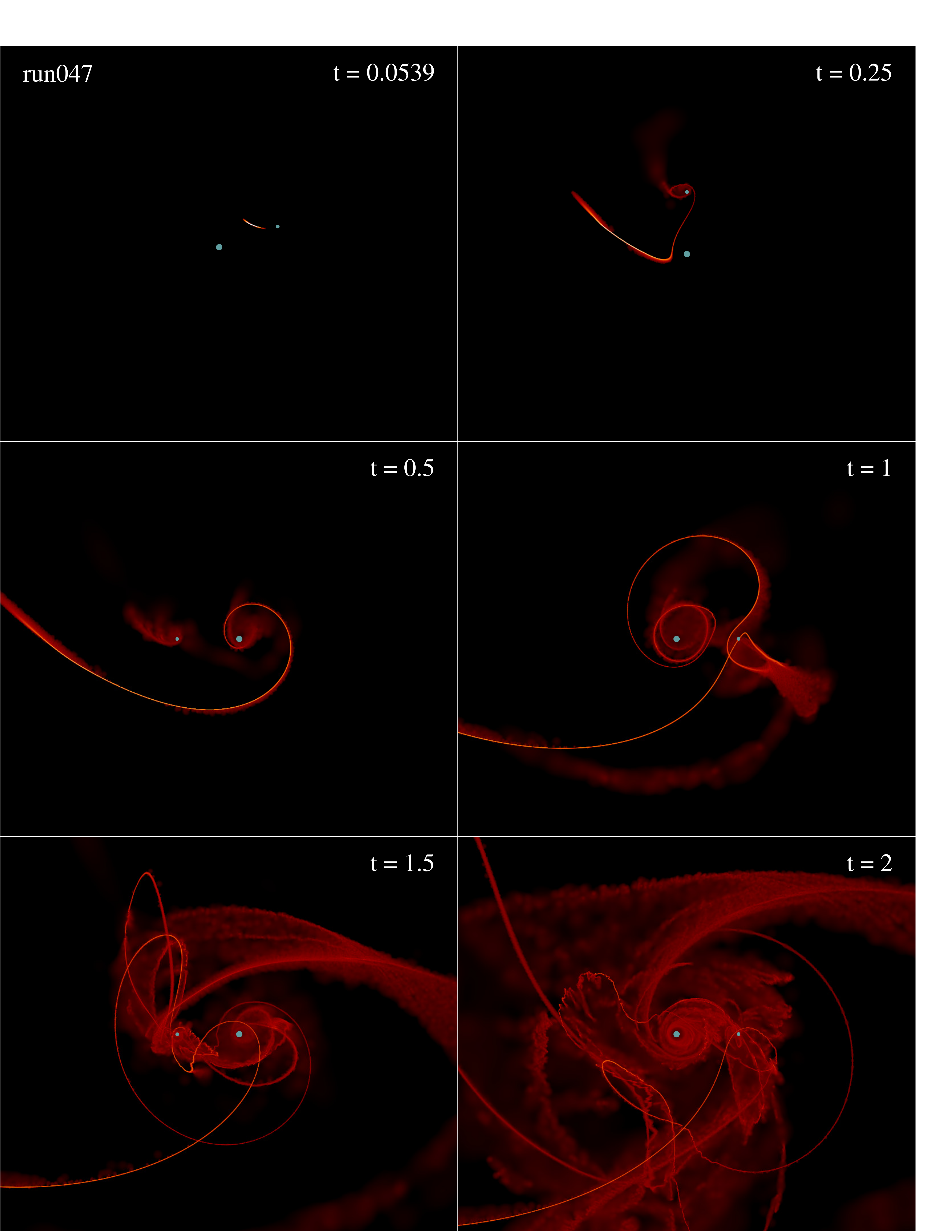} 
   \caption{The $x$-$y$ projection of the evolution of the debris generated from a TDE, with the colors indicating the value of the column density (brighter colors correspond to denser material); the blue circles represent the SMBHs, with the size of each circle being ten times the size of the tidal radius of the respective hole. The $x$-axis covers roughly 7 binary semimajor axes, while the $y$-axis covers approximately 6. In this case the star was disrupted by the secondary, and the time since the start of the simulation (when the star was at $5$ $r_t$) is shown in the top right corner in units of binary orbits. A movie of this simulation can be found \href{http://w.astro.berkeley.edu/~eric_coughlin/movies.html}{here}.}
   \label{fig:inplane_tile}
\end{figure*}

\begin{figure*} 
   \centering
   \includegraphics[width=0.975\textwidth]{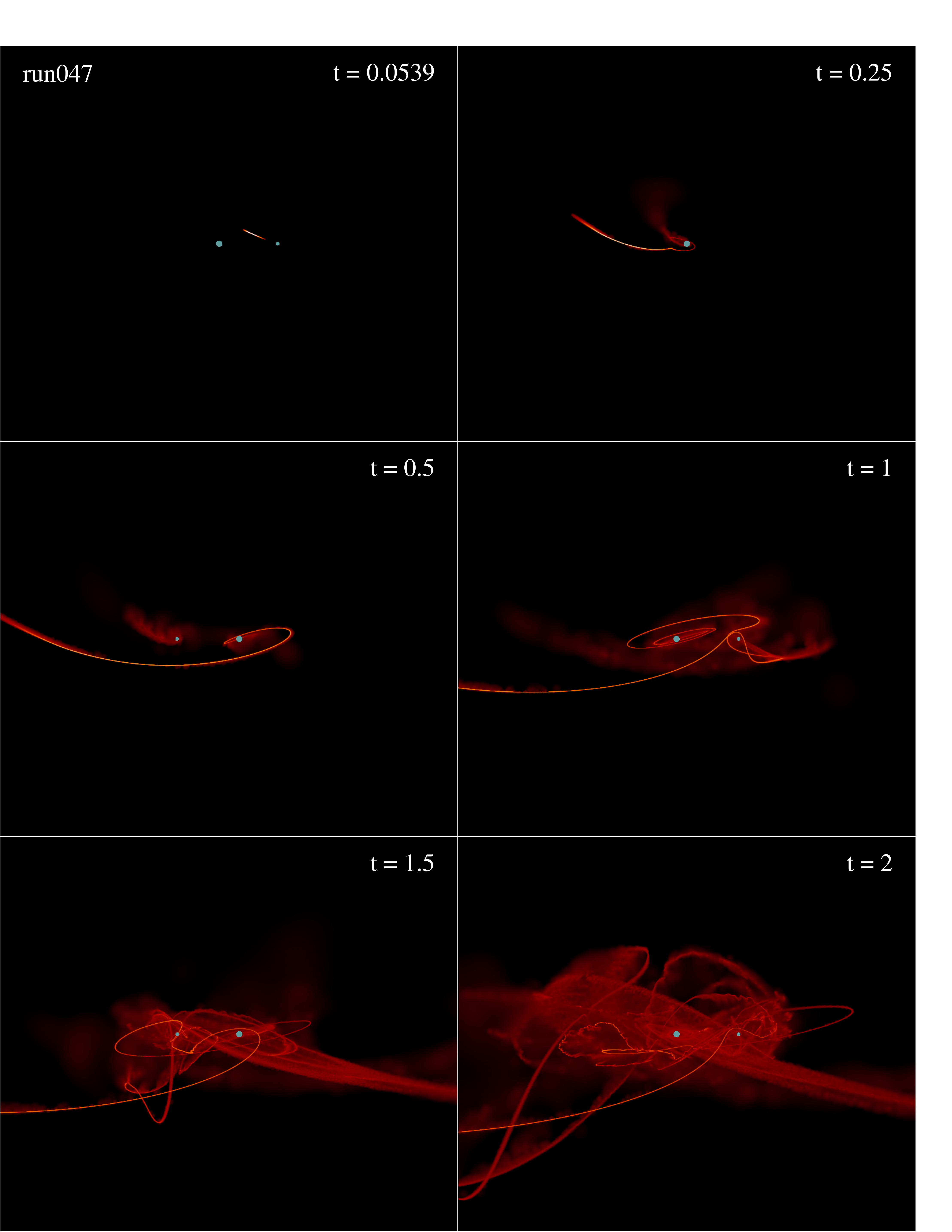} 
   \caption{The $x$-$z$ projection of the evolution of the debris generated from the same TDE shown in Figure \ref{fig:inplane_tile}. The times, colors, etc., correspond to those in Figure \ref{fig:inplane_tile}.}
   \label{fig:outofplane_tile}
\end{figure*}

Figures \ref{fig:inplane_tile} and \ref{fig:outofplane_tile} show the $x$-$y$ (in-plane) and $x$-$z$ (out-of-plane) projections, respectively, of the evolution of the debris generated from one of the 120 simulated TDEs; in this particular case the star was disrupted by the secondary. In these panels the colors correlate with the projected column density of the material, with the brightest colors corresponding to the densest regions, and the times are just after disruption (top left), 0.25 (top right), 0.5 (middle left), 1 (middle right), 1.5 (bottom left), and 2 binary orbits (bottom right). The two blue circles represent the SMBHs, with the radius of each circle set, for visualization purposes, to ten times the tidal radius of the respective black hole. 

Originally the TDE proceeds as if the disrupting SMBH were isolated (the stream from the control case appears identical to the stream in the top-left panel of Figure \ref{fig:inplane_tile} at that time). However, once the tidally-disrupted debris stream leaves the Roche Lobe of the secondary, the primary SMBH begins to significantly alter the morphology of the flow following the disruption of the star. These effects from the primary start to become apparent in the top-right panel of Figure \ref{fig:inplane_tile}, where the debris stream accreting onto the secondary is bent by the gravitational field of the companion hole. By the middle-left panel, we see that the primary has ``stolen'' the incoming debris stream, leaving a small, eccentric accretion disk around the secondary and feeding material onto the primary in an extended disk. When the binary makes a complete orbit (middle-right panel), the secondary comes close enough to the stream to significantly alter its dynamics, generating a large debris ``loop'' that becomes disconnected from the system. Continued perturbations from the secondary SMBH result in a highly chaotic distribution of material by the bottom two panels (1.5 and 2 orbits from left to right), with small-scale (i.e., within the Roche Lobe of either hole) accretion disks around each hole and a large-scale cloud of debris that surrounds the binary and extends to roughly ten binary separations.

The out-of-plane projection, Figure \ref{fig:outofplane_tile}, demonstrates a qualitatively similar evolution to the in-plane projection: while the TDE originally proceeds as one would expect from an isolated SMBH, the system rapidly becomes highly disorganized due to the perturbing effects of both the primary and the secondary. This figure also shows that the original stellar orbit was fairly inclined with respect to the orbital plane of the binary. Furthermore, while the stream initially stays relatively confined to the binary plane (note that the orientation of the accretion disk around the primary is quite tilted with respect to the original plane of the disrupted debris), by the bottom two panels the large-scale distribution of material is roughly spherical and surrounds the entire binary.

\begin{figure*}
   \centering
   \includegraphics[width=0.975\textwidth]{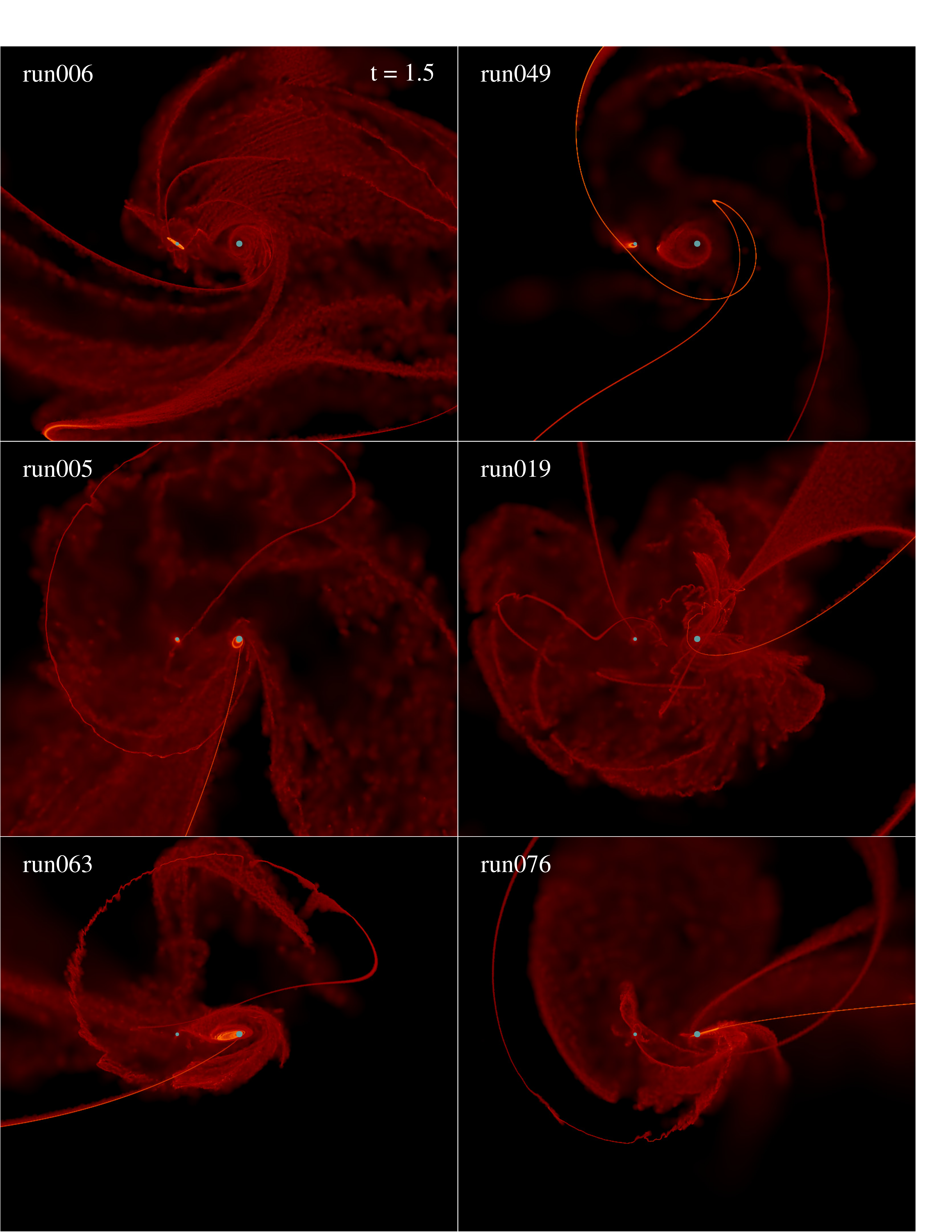} 
   \caption{The $x$-$y$ projection of the debris produced from six different simulated TDEs at a time of 1.5 binary orbits. The top two plots were disruptions by the secondary SMBH, while the bottom four were disrupted by the primary. The colors, plot ranges, etc.~are all identical to those of Figure \ref{fig:inplane_tile}.}
   \label{fig:panels_multi}
\end{figure*}

Figure \ref{fig:panels_multi} shows the in-plane ($x$-$y$) projection of the distribution of the debris from six different simulated TDEs (none of which corresponds to that shown in Figure \ref{fig:inplane_tile}), all at a time of 1.5 binary orbits (roughly 5 months post-disruption). The colors scale with the column density of the gas, with lighter colors corresponding to denser areas, and the plot range and black hole sizes are the same as those for Figure \ref{fig:inplane_tile}. The top two panels resulted from stars disrupted by the secondary SMBH, while the bottom four were disrupted by the primary. 

As is apparent from Figure \ref{fig:panels_multi}, the qualitative appearance and dynamical evolution of a TDE generated by a SMBH binary depends very sensitively on the properties of the star at the time of disruption. Figure \ref{fig:inplane_tile} represents somewhat of a ``Goldilocks'' situation in which the star is disrupted by the secondary but its trajectory is toward the primary. In this case, the primary SMBH is guaranteed to have a very large effect on the evolution of the disrupted debris early on (indeed, by half of an orbit the stream is significantly deflected by its gravitational influence). However, in all of the cases illustrated in Figure \ref{fig:panels_multi}, the configuration of the debris has been transformed into a very disordered state. 

We note that each TDE generates a series of ``loops'' of debris that tend to recede away from the binary, and these features are apparent in all of the panels of Figure \ref{fig:panels_multi} (as well as the panels in Figure \ref{fig:inplane_tile}). These loops originate when the bound debris starts to leave the Roche lobe of the disrupting hole, causing its new pericenter position to deviate significantly from the position of the hole. 

We also see from Figure \ref{fig:panels_multi} that there are ``gaps'' in these debris loops (one particularly obvious gap can be found near the top-left corner of the middle-left panel in Figure \ref{fig:panels_multi}), which occur when the debris stream self-intersects, generating a shock that blasts material from the less-dense portion of the stream. {}{These self-intersections are reminiscent of those usually invoked to explain the formation of post-disruption accretion disks around isolated SMBHs, with the self-intersection in those cases induced by the general relativistic advance of periapsis (though nodal precession, if the SMBH is spinning, can reduce the likelihood and efficiency of self-intersection; \citealt{rees88, kim99, stone12, dai13, dai15, guillochon15, hayasaki16, jiang16, bonnerot16b}).} 

{}{In our simulations, however, the black holes are modeled with Newtonian gravity, meaning that the self-intersections found here are induced by a combination of the differences in the orbital properties of the debris and the motion of the SMBHs (i.e., the time dependence of the binary system and its gravitational potential). Furthermore, while apsidal precession generally causes stream intersections that occur primarily in the plane of the motion of the stream itself, resulting in a large dissipation of kinetic energy, here the material can rapidly swing out of the original plane of the disrupted star (as evidenced from Figure \ref{fig:outofplane_tile}). In these instances, stream collisions can occur over a wide range in angles that do not necessarily coincide with a single plane, thereby confining the intersections to narrow portions of the stream (creating the gaps) and resulting in only small portions of the stream being affected. Nevertheless, there are some instances, for example the disruption in Figures \ref{fig:inplane_tile} and \ref{fig:outofplane_tile}, where the stream self-intersection does lead to the formation of the accretion disk (see the movie \href{http://w.astro.berkeley.edu/~eric_coughlin/movies.html}{here}).}

Interestingly, in some cases the debris generated from the TDE was completely ejected from the system, generating no fallback whatsoever. Of the 120 simulations performed, there were four such cases, and all of them were associated with disruptions by the secondary SMBH (there were, however, also some instances in which only a small amount of material was bound to the system, and these corresponded to disruptions by both the secondary and the primary). In these completely-unbound instances, the specific energy of the center of mass of the star at the time of disruption happened to be so large that the entire debris stream had positive energy with respect to the binary. 

To obtain a rough idea of the energy of the center of mass required to eject the entire stream, we can assume that the energies of the gas parcels comprising the star are ``frozen in'' at the time that the center of mass passes through the tidal radius of the disrupting hole. This assumption then gives

\begin{equation}
\epsilon_i = \frac{1}{2}v_i^2-\frac{GM_{1,2}}{r_i} \simeq \epsilon_c+\epsilon_*m_{1,2}^{1/3}\eta_i,
\end{equation}
where $v_i$ and $r_i$ are the velocity and position of gas parcel $i$, respectively, and $M_{1,2}$ is the mass of the disrupting hole. The last line of this expression follows from assuming that the entire star moves with the center of mass prior to disruption, and we have set $\epsilon_c = v_c^2/2-GM_{1,2}/r_c$ as the energy of the center of mass ($v_c$ and $r_c$ are the center of mass speed and position, respectively), $m_{1,2} \equiv M_{1,2}/M_*$, $\epsilon_* \equiv GM_*/R_*$, and $\eta_iR_* \equiv r_i-r_t$. Since $|\eta_i| \le 1$, the entire stream will be ejected if

\begin{equation}
\epsilon_c-\epsilon_*m_{1,2}^{1/3} > 0.
\end{equation}
From Figure \ref{fig:eps_hist}, we see that $\epsilon_c = N\epsilon_{b}$, where $\epsilon_{b} = G(M_1+M_2)/(2a)$ is the binding energy of the binary and $N$ is a pure number, and we will set $a = k\,r_{t,1}$ (for our simulations $k = 100$). Using these relations and rearranging, this inequality then yields

\begin{equation}
N_i > \frac{2\,m_1^{1/3}m_i^{1/3}}{m_1+m_2}k. \label{Ncrit}
\end{equation}
This equation is valid for any binary mass ratio and any value of $k$; if we now plug in parameters specific to our set of simulations, this inequality becomes

\begin{equation}
N_i \gtrsim \frac{2\,m_i^{1/3}}{1.2\times100}.
\end{equation}
Thus, for disruptions by the primary, the energy of the center of mass at the time of disruption must satisfy $\epsilon_c \gtrsim 1.6\times\epsilon_b$, while disruptions by the secondary require $\epsilon_c \gtrsim 1.0\times\epsilon_b$. 

This scaling argument demonstrates that it is easier for the secondary SMBH to entirely eject streams than the primary for two reasons, the first being, as we just showed, that disruptions by the secondary result in a smaller limit on the energy of the center of mass required to achieve complete ejection. The second reason arises from the fact that the specific energy distribution for stars disrupted by the secondary has an inherently wider spread than that for stars disrupted by the primary, as can be seen in Figure \ref{fig:eps_hist}. Noting from this figure that there are very few stars disrupted by the primary that satisfy $\epsilon_c > 1.6 \,\epsilon_b$, we expect, and indeed the simulations confirm, that only a small fraction of disruptions by the primary will result in completely unbound streams.

A precisely analogous argument holds when we consider entirely-bound streams: if even the most energetic portion of the stream is to remain bound to the binary, then we require $\epsilon_c \lesssim -1.6\,\epsilon_b$ for disruptions by the primary, and $\epsilon_c \lesssim -1.0\,\epsilon_b$ for the secondary. As we can see from Figure \ref{fig:eps_hist}, this scenario is much more likely for either SMBH. 

We will return to a further discussion of these points and their ramifications in Section \ref{sec:discussion}. 

\subsubsection{Accretion and fallback rates}
\label{sec:rates}
As was mentioned in Section \ref{sec:setup}, the accretion radius of each SMBH was set to half of its tidal radius, and particles entering within the accretion radius of either hole were ``accreted'' and removed from the simulation. The rate at which particles are accreted then gives the raw accretion rate of either SMBH.

An accretion radius of $r_t/2$, where $r_t$ is the tidal radius of the relevant SMBH, was chosen to offset two competing goals, the first being to minimize the computational cost of the simulations: particles orbiting very close to the SMBH require very short timesteps which can dominate the computational cost of the simulation. On the other hand, maintaining a smaller accretion radius generally results in a more realistic distribution of tidally-disrupted debris. Indeed, if the accretion radii were made too large, every incoming debris stream would be swallowed entirely upon reentry, and much of the interesting hydrodynamics that generated Figures \ref{fig:inplane_tile} -- \ref{fig:panels_multi} would have been lost. From these two competing effects, it was decided that $r_t/2$ was a reasonable compromise. 

It may seem as though a smaller accretion radius always results in a more accurate distribution of debris (up until one reaches the Schwarzschild radius, which corresponds to accretion radii of $r_t/24$ and $r_t/69$ for the primary and secondary, respectively). However, with the number of particles used here ($5\times10^5$), it is not possible to resolve the flow at such small radii. Furthermore, an accretion radius of $r_t/2$ already corresponds to only $24$ Schwarzschild radii of the primary SMBH, meaning that decreasing the accretion radius only slightly more would result in general relativistic effects becoming very important. Using a smaller accretion radius would therefore require a large increase in particle number and the inclusion of general relativity, which is outside the scope of this paper. 

\begin{figure*} 
   \centering
   \includegraphics[width=\textwidth]{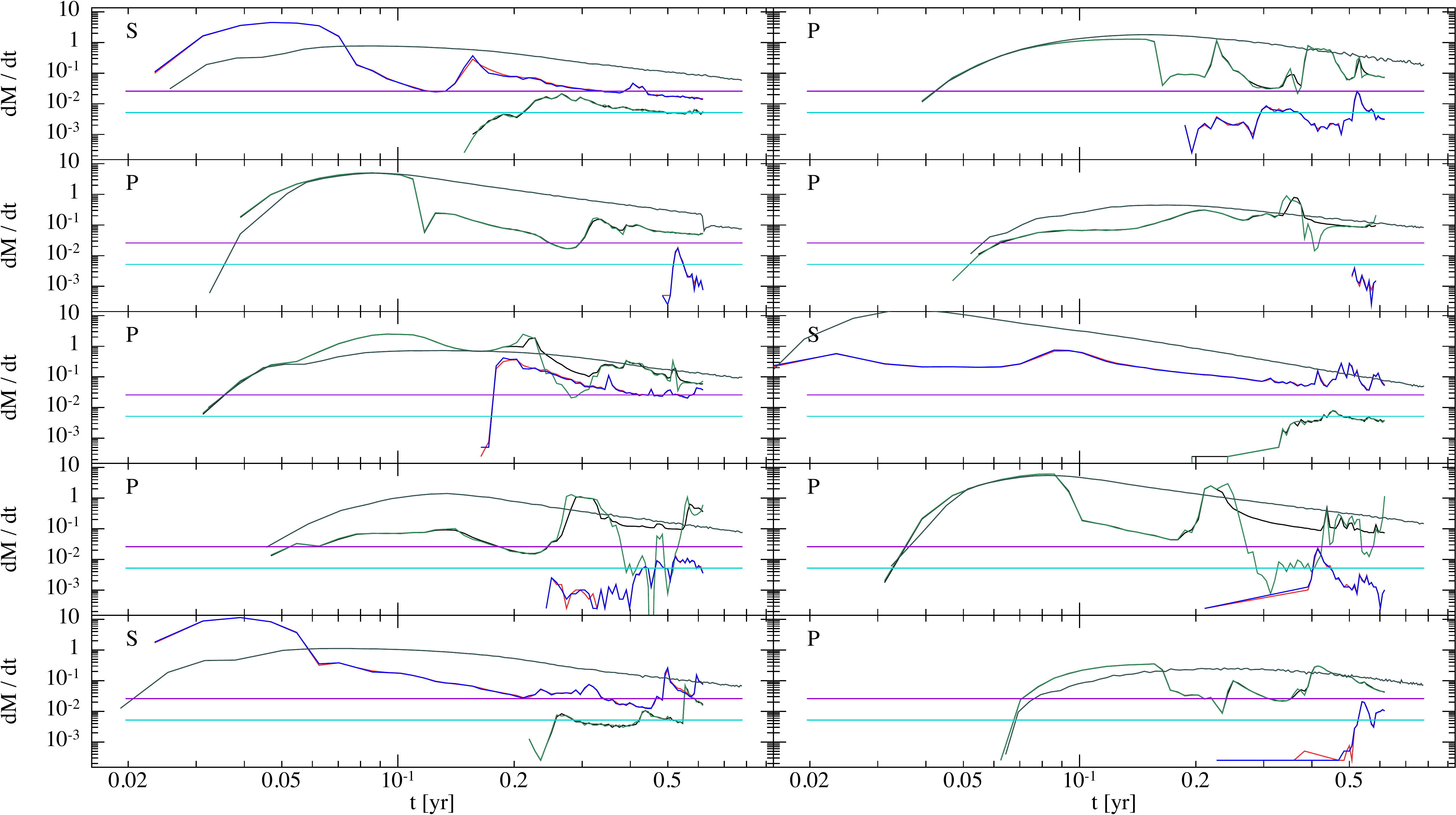} 
   \includegraphics[width=\textwidth]{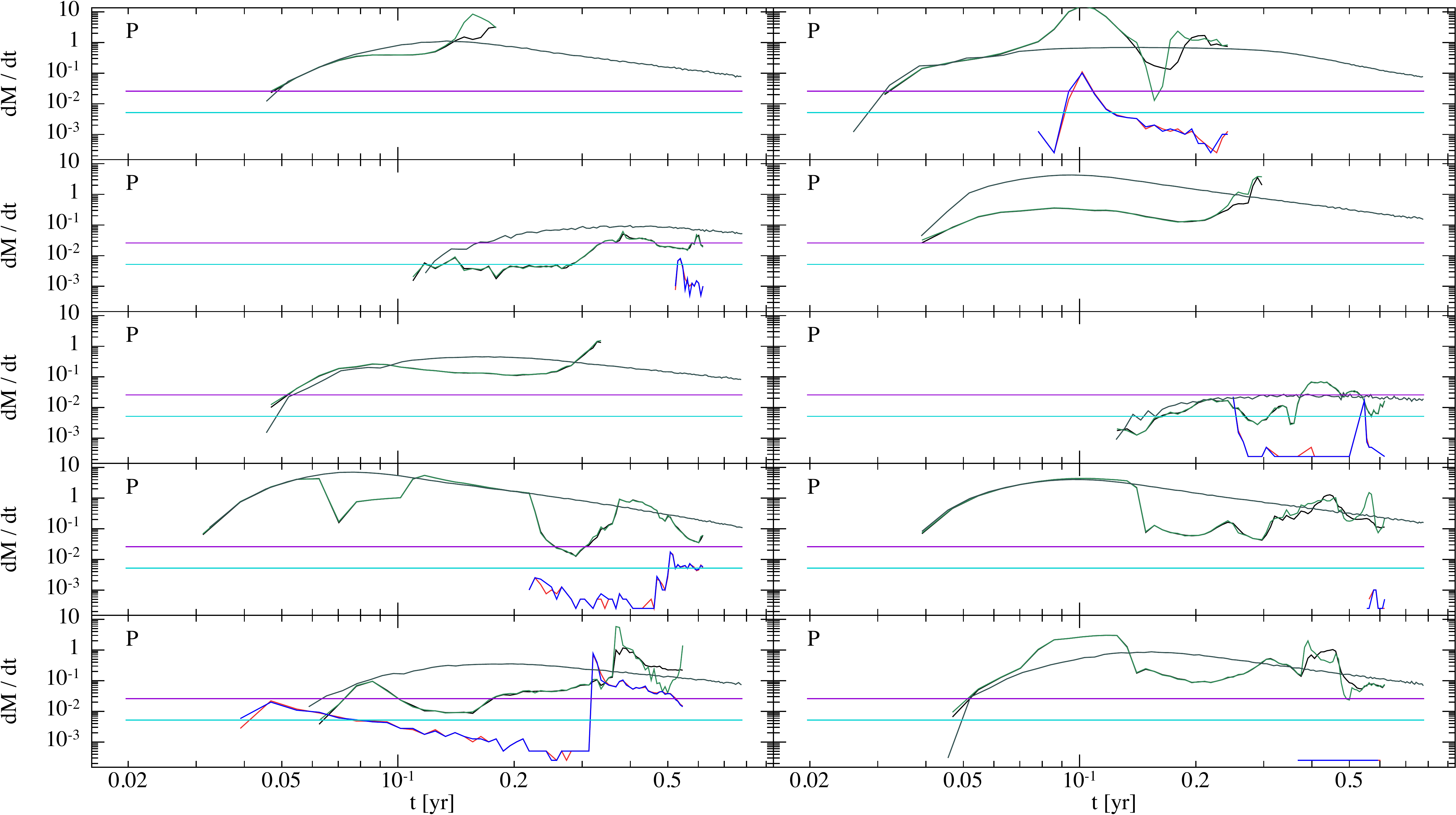} 
   \caption{A log-log plot of the accretion rate of the primary (black curve), the fallback rate of the primary (green curve), the accretion rate of the secondary (red curve), the fallback rate of the secondary (blue curve), and the accretion rate of the control (grey curve) for 20 different simulations. The disrupting black hole is indicated in the top-left corner, with a P indicating disruptions by the primary, an S the secondary. Accretion and fallback rates are measured in units of Solar masses per year, while time is measured in years (one binary orbit corresponds to 0.3 years). The magenta and cyan lines represent the Eddington limit for the primary and secondary SMBHs, respectively, assuming an accretion efficiency of 0.1. Simulations with no accretion whatsoever were complete ejections of the stream. The accretion and fallback rates for the other 80 simulations can be found in Appendix \ref{sec:sup}.}
   \label{fig:mdots_1}
\end{figure*}

\begin{figure*} 
   \centering
   \includegraphics[width=\textwidth]{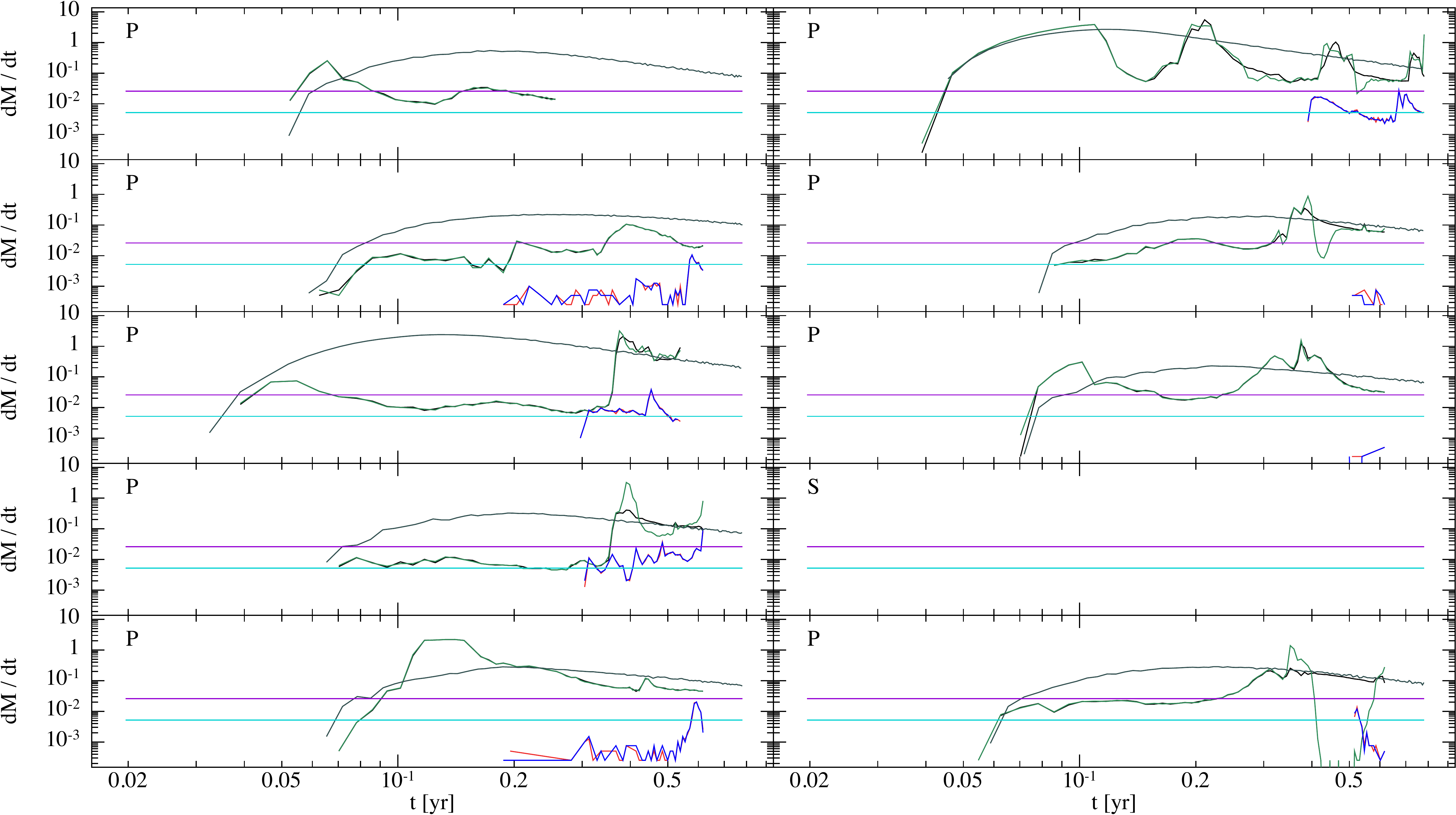} 
   \includegraphics[width=\textwidth]{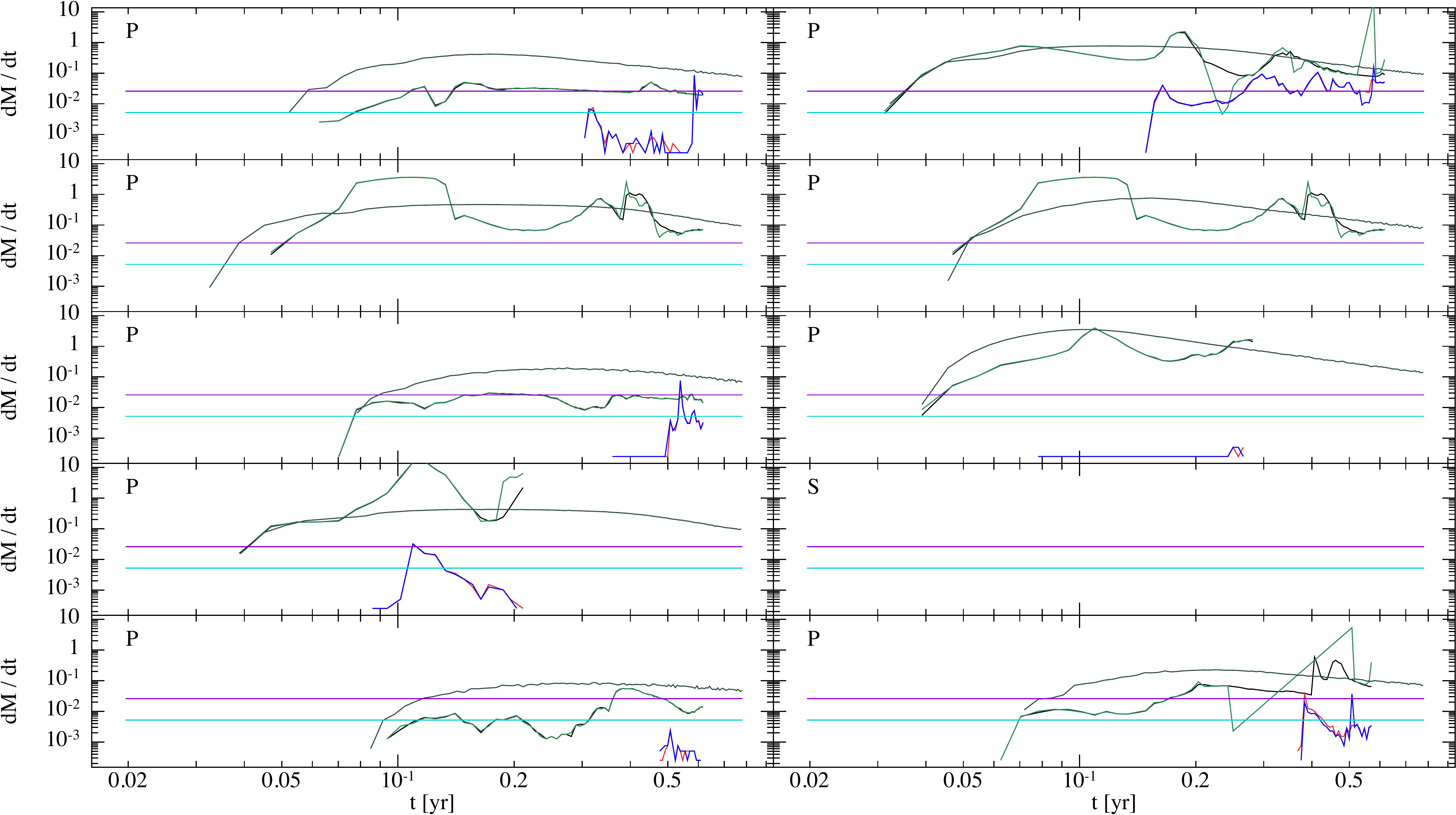} 
   \caption{See the caption for Figure \ref{fig:mdots_1}.}
   \label{fig:mdots_2}
\end{figure*}

As stated in Section \ref{sec:setup}, the heat generated from the shocks caused by stream-stream collisions does not contribute to the thermodynamics{}{; the physical interpretation of this restriction is that the heat injected into the gas through shock heating is radiated very efficiently, and thus the polytropic constant relating the pressure to the gas density remains unaffected.} Because of this restriction, the accretion disks around the individual SMBHs generally stay confined to within a few tidal radii in radial extent and are very thin (see Section \ref{sec:disks} for a verification of this point). However, the inclusion of shock heating could cause these disks to ``puff up'' significantly{}{, similarly to what was seen in the simulations of \citet{hayasaki16} when they included the heat generated by shocks in the thermodynamic properties of the gas}; furthermore, we find that the accretion rates from our simulations can be super-Eddington by two to three orders of magnitude in some instances (see Figures \ref{fig:mdots_1} and \ref{fig:mdots_2}), meaning that radiation pressure should be very important for expelling gas and inflating the accretion disks \citep{strubbe09, coughlin14}. It is therefore also reasonable to consider the effective accretion rate as the rate at which disrupted debris enters into some larger region around either SMBH, as this would mimic the modulation of the black hole accretion rate by a surrounding accretion flow.

Given the aforementioned uncertainties in relating the numerical ``accretion rate'' to observational quantities -- such as a luminosity -- we also calculate a ``fallback rate'' by counting all of the particles that cross a radius of $3\,r_t$, the radius of $3\,r_t$ chosen for consistency with \citet{coughlin15} who also calculated a fallback rate by this method. Furthermore, to successfully add to the fallback rate, the particle must have a negative specific energy with respect to the SMBH onto which it is falling (i.e., we do not count unbound particles that happen to stray too closely to one or the other SMBH). 

Figures \ref{fig:mdots_1} -- \ref{fig:mdots_2} show, on a log-log scale, the accretion and fallback rates of both SMBHs in Solar masses per year (the black, green, red, and blue curves represent the primary accretion rate, primary fallback rate, secondary accretion rate, and secondary fallback rate, respectively) as a function of time in years, and the control accretion rate (grey curve) for 40 different simulations{}{; accretion and fallback rates from the other 80 simulations can be found in Appendix \ref{sec:sup}. The disrupting black hole is indicated in the top-left corner of each plot, $P$ being disruptions by the primary, $S$ by the secondary. The straight lines give the Eddington limits of the SMBHs assuming an accretion efficiency of 10\%, with the top, magenta line corresponding to that of the primary (being $\dot{M}_{Edd,p} \simeq 0.026\,M_{\odot}$ yr$^{-1}$), the bottom, cyan line representing that of the secondary (being $\dot{M}_{Edd,s} \simeq 0.0052\,M_{\odot}$ yr$^{-1}$). There are a few panels that have no accretion, and these are the instances in which the stream was completely ejected from the binary system. The accretion rates from the other 80 simulations can be found in Appendix \ref{sec:sup}.

Overall, it is obvious that the presence of the second SMBH has dramatic, sporadic effects on the evolution of the accretion rates (from here onward we will refer to both the raw accretion rates and the fallback rates by simply accretion rates), and those effects are quite pronounced over many binary orbits (i.e., long after disruption; one binary orbit corresponds to roughly 0.3 years). In many cases, the accretion proceeds in a fairly regular fashion before abruptly changing by an order of magnitude (or more) on very short timescales relative to the binary orbit. It is also apparent that the canonical fallback rate of $\dot{M}_{fb} \propto t^{-5/3}$ \citep{phinney89} is hardly ever actualized despite the fact that the control cases, reassuringly, do display this behavior. We also see that the accretion rates, which count particles that cross $0.5\,r_t$, and the fallback rates, which count bound particles that enter within $3\,r_t$, follow each other closely for most simulations, indicating that the precise choice of accretion radius does not have a major influence on the results. The mapping between the accretion rate and the luminosity in a specific band involves several additional physical considerations \citep{lodato11}, but these are likely to be the same between TDEs generated by isolated and binary SMBHs.

Disruptions by the secondary SMBH display the most deviation from the controls, \emph{even at early times}. The reason for this behavior is that the Roche lobe of the secondary is a small fraction of the binary separation, and the tidally-disrupted debris (even the most bound debris) recedes to large distances from the disrupting hole before returning. Thus, the primary can significantly deflect the path of the debris stream (and change the energies and angular momenta of the gas parcels comprising the stream) soon after the star is disrupted. Furthermore, the accretion rates can either significantly exceed or fall well below the controls, which arises from the direction of the disrupted debris stream: in the cases where the accretion rate exceeds that of the control, the motion of the ejected stream is approximately antiparallel to the direction pointing from the secondary SMBH to the center of mass of the binary. Thus, as the stream recedes from the secondary, it sees more mass \emph{in the direction of the secondary}, causing the material to more rapidly accelerate back towards the disrupting hole and increasing the accretion rate. Contrarily, when the stream is ejected towards the primary SMBH the accretion rate falls below the control, because it experiences an additional gravitational force in the direction \emph{away from the direction of the disrupting hole}. This counteralignment then decreases the infall velocity of material onto the secondary and correspondingly diminishes the accretion rate.

For disruptions by the primary, the accretion rate onto the primary SMBH, in most cases, approximately follows the control rate at early times. This result is reasonable, as the Roche radius of the primary is a fairly large fraction of the separation; therefore, the material only ``sees'' the primary black hole for a fairly long time post-disruption, i.e., until the apocenter distances of the disrupted gas parcels start to exceed the Roche radius of the primary. This interpretation is also consistent with the rapid falloffs exhibited by the accretion rates of the primary: the sudden decline marks the point where the returning debris had, at its apocenter distance, left the Roche lobe of the primary and is therefore deflected from its original trajectory in the direction of the center of mass. We will return to a further discussion of this point and its implications in Section \ref{sec:discussion}.

The control simulations show the isolated effects caused by the alteration in the specific energy and angular momentum of the center of mass of the star at the time of disruption. We see that, even though all of the curves tend to follow the $t^{-5/3}$ law at late times, there are large differences in the return time of the most bound debris and the duration and magnitude of the peak in the fallback rate. 

\section{Discussion}
\label{sec:discussion}

In this section we discuss in more detail some of the most interesting findings presented in the previous section. In particular, we more closely analyze the accretion disks generated from the TDEs, we investigate the potential periodicities in the accretion rates, we consider the effects of resolution on the outcomes of our simulations, we make some remarks on the long term (i.e., after many binary orbits) evolution of the accretion morphologies and rates, and we consider the implications of totally-bound and unbound debris streams.

\subsection{Accretion flows}
\label{sec:disks}
It is apparent from Figures \ref{fig:inplane_tile} -- \ref{fig:panels_multi} that, following the disruption of the star, much of the shredded stellar material recedes to very large distances from the binary. Depending on the specific energy of the material and the deviations induced to that energy from the binary potential, that material either leaves the system on hyperbolic trajectories (as is the case for roughly half of the debris when a star is disrupted by an isolated SMBH) or returns to the sphere of influence of the binary on very weakly-bound orbits. As shown in the last panels of Figures \ref{fig:inplane_tile} and \ref{fig:outofplane_tile}, those orbits are distributed about the binary in a quasi-spherical fashion. 

\begin{figure*} 
   \centering
   \includegraphics[width=\textwidth]{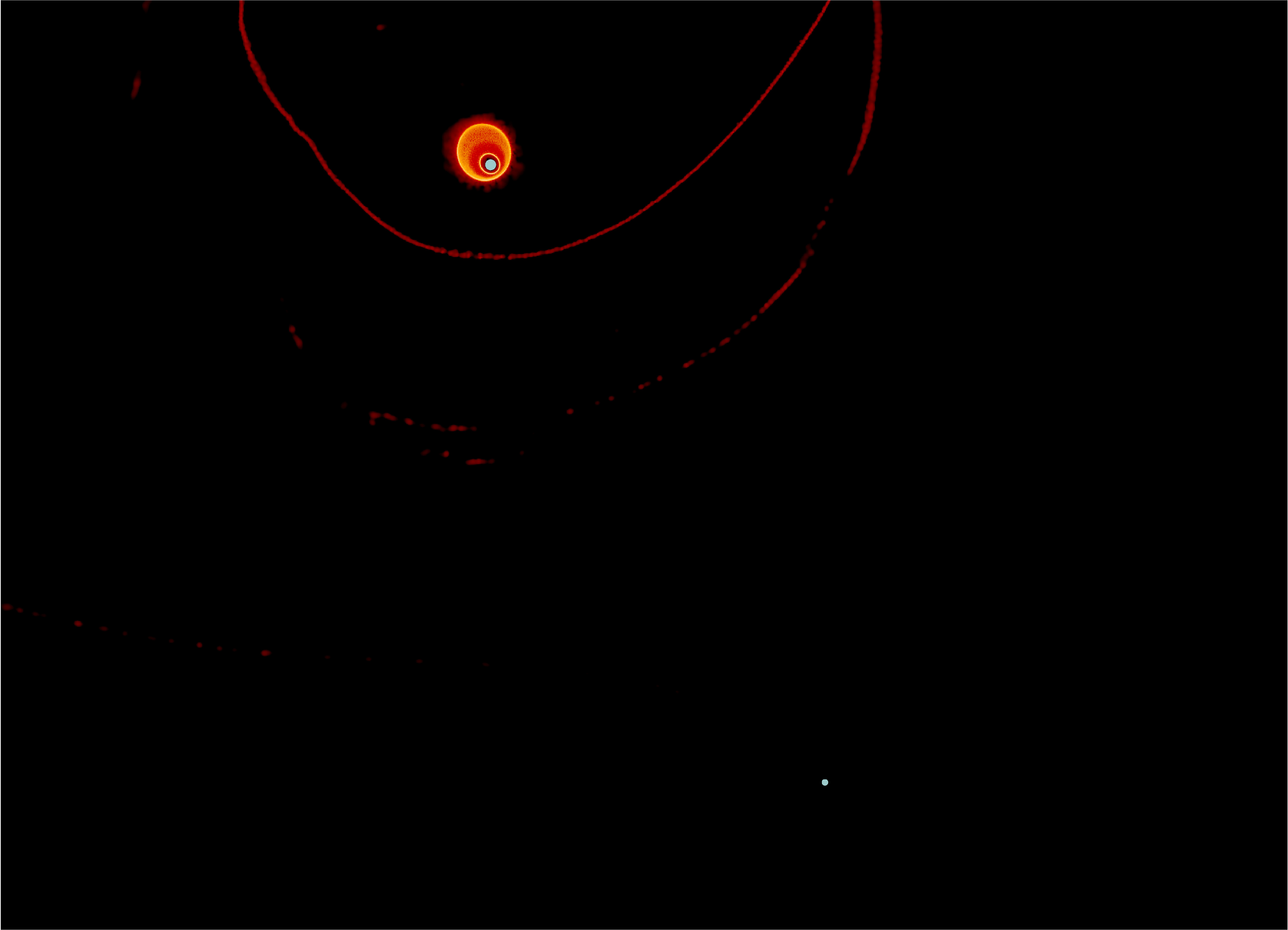} 
   \caption{The $x$-$y$ projection of the accretion flow generated from one simulation at a time of approximately 1.75 binary orbits; in this case the star was disrupted by the primary. The scale here shows the size of the accretion disk around the primary, and the sizes of the blue circles are set to two times the accretion radius of the respective SMBH.}
   \label{fig:disk_scale}
\end{figure*}

\begin{figure*} 
   \centering
   \includegraphics[width=0.50\textwidth]{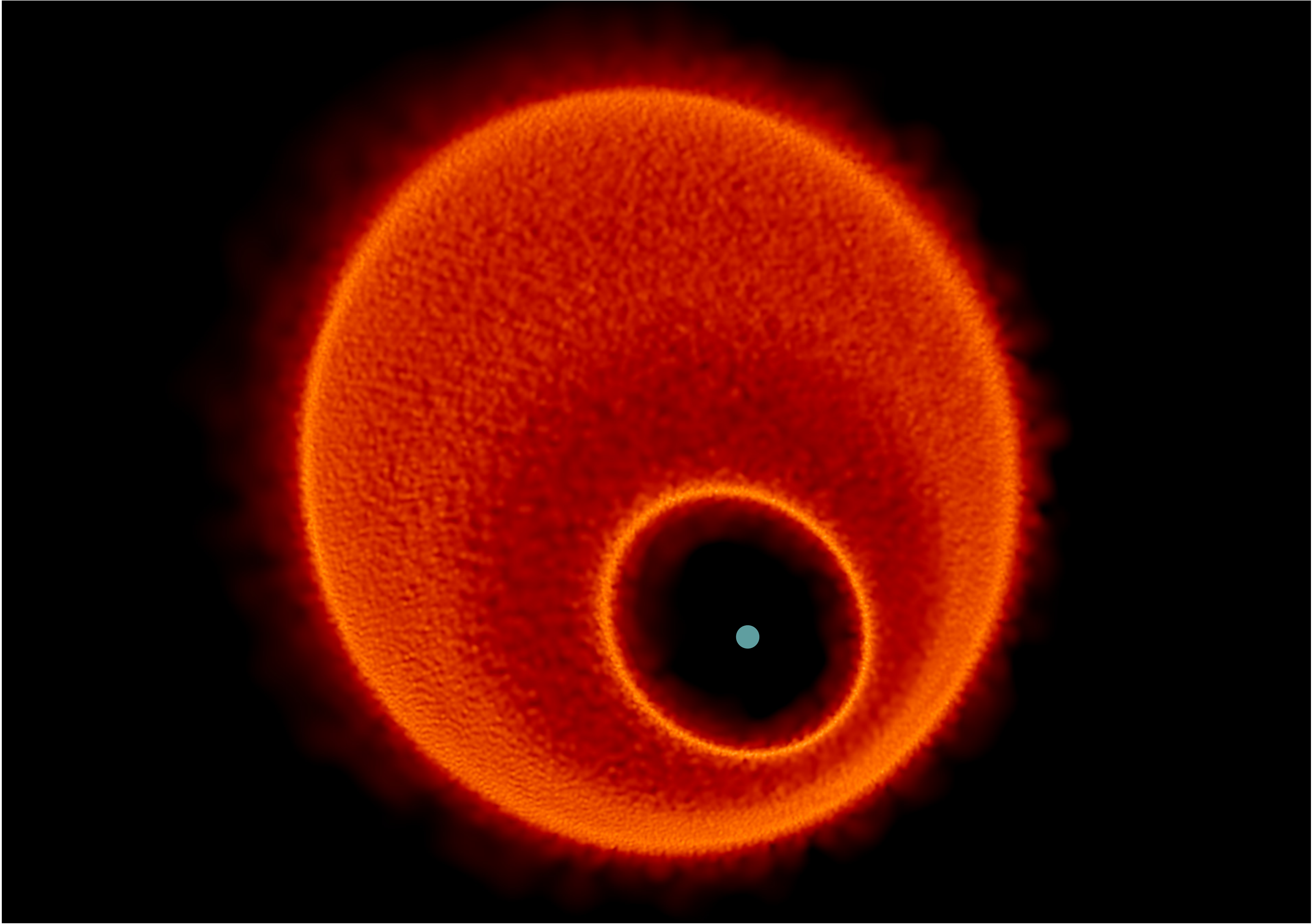} 
   \includegraphics[width=0.4875\textwidth]{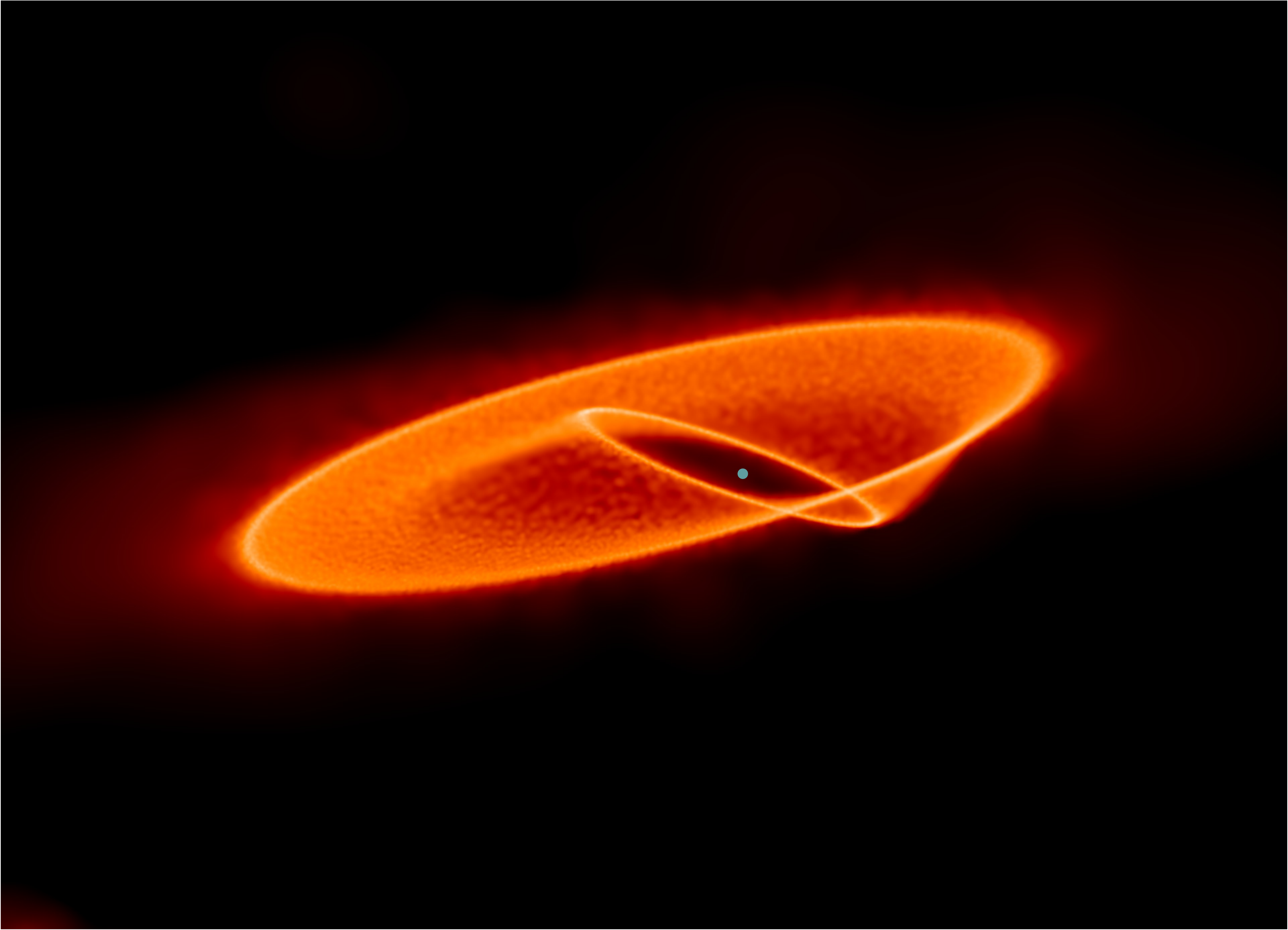} 
   \caption{Left-hand panel: a closeup of Figure \ref{fig:disk_scale} showing the detailed structure of the accretion disk in the plane of the binary. Right-hand: same as the left-hand panel, but rotated by 60$^{\circ}$ in a counterclockwise sense about the $x$-axis (i.e., the accretion disk is rotated by $60^{\circ}$ into the plane). This panel shows that the inner accretion flow possesses a significant tilt angle with respect to the larger accretion flow, and this geometric effect causes the apparent column density to appear to decrease in regions closer to the SMBH. A movie of the formation and evolution of the accretion disk can be found \href{w.astro.berkeley.edu/~eric_coughlin/movies.html}{here}.}
   \label{fig:disk_closeup}
\end{figure*}

However, some of the tidally-disrupted debris also stays confined to one or both of the SMBHs in the form of actively-accreting disks, morphologically evident from Figure \ref{fig:panels_multi}. Figure \ref{fig:disk_scale} shows the result of one particular simulation at a time of roughly 1.75 orbits in which a disk has only formed around the primary -- the disrupting black hole for this case (in this figure, the size of each blue circle was set to two times the accretion radius of the respective SMBH for the purpose of visualization). This was also found to be the general trend: disruptions by the primary \emph{usually} resulted in an accretion disk only formed around the primary, though Figure \ref{fig:panels_multi} shows that disks also formed around the secondary in some of those instances; Figures \ref{fig:mdots_1} and \ref{fig:mdots_2} also show that a substantial amount of material is sometimes accreted by the secondary for disruptions by the primary. On the other hand, disruptions by the secondary generally resulted in accretion onto both SMBHs.

From Figure \ref{fig:disk_scale} we see that the accretion disk around the primary extends to approximately 10\% of the semimajor axis of the binary (approximately 5-10 AU in physical units). Furthermore, it is already evident from this scale that the disk itself has quite a bit of substructure, being somewhat elongated in the direction of the semimajor axis of the binary, having a sharp, well-defined inner and outer radius, and possessing a radial gradient in its column density profile. The left-hand panel of Figure \ref{fig:disk_closeup} shows a closeup of the $x$-$y$ projection of the disk (in this figure the size of the SMBH has been set to 0.1 times its accretion radius simply for the sake of presentation). Here it is evident that the location of the inner boundary is somewhat ellipsoidal, while the outer edge of the accretion flow remains more circular.

The right-hand panel of this figure is the same as the left-hand panel, but we have rotated the coordinates by $60^{\circ}$ in a counterclockwise sense about the $x$-axis (in other words, the top half of the left-hand panel is rotated by $60^{\circ}$ into the page, the bottom half correspondingly by $60^{\circ}$ out of the page). This panel demonstrates that the inner and outer radii of the accretion disk are significantly inclined with respect to one another. Furthermore, it is apparent that \emph{neither} disk has its angular momentum vector aligned to that of the binary.

We have found that the orientation and tilts of the accretion disks are relics of the orientation of the stream at the time that it began to actively accrete onto the binary. Moreover, the inner and outer flows are actually comprised of two \emph{distinct disks} that formed at two different times: the outer accretion disk forms after the first passage of the primary through the returning debris stream, occurring at roughly  0.5 binary orbits. The second accretion disk, on the other hand, is generated when the primary passes through the debris stream a second time at 1.5 binary orbits. 

It is because the disks formed from discrete portions of the returning debris stream that they have well-defined radii, truncated at the approximate circularization radius of the material out of which they formed (which, because of the time-dependent nature of the potential of the binary, is not necessarily conserved from the disruption). Furthermore, the relative tilt between the inner and outer disks is observed to change over time, with the warp connecting the two viscously torquing the inner one and causing it to align with the outer one. At the instant that the inner accretion disk forms (at roughly 1.5 orbits) its relative tilt to the outer disk is close to $90^{\circ}$, while the tilt angle decreases and becomes nearly coplanar by a time of 2 orbits.

In addition to the continued impact of the stream onto the accretion flow, the binary itself should, in a more general sense, also have some effect on the evolution of the accretion disks formed around the primary and the secondary. In particular, we expect the misalignment between the angular momentum vector of the disk(s) and the binary to induce a precession of the accretion disk itself, and that precession will have some spatial dependence. Thus, over a long enough timescale, the binary should induce a warped structure in the accretion disk, with the magnitude of the warp depending on the local viscosity within the disc (see \citealt{nixon16} for a review of warped disc physics, and e.g. \citealt{larwood96, fragner10, dogan15} for simulations of misaligned discs in binary systems). These discs may also be susceptible to Kozai-Lidov oscillations on longer timescales \citep{martin14a, martin14b, martin16}.

Finally, we see that the disks formed in our simulations are very thin, and this thinness is a direct consequence of our neglect shock heating -- if the material retained the heat generated when incoming debris self-intersects to form the original accretion disks or when the debris impacts the pre-existing accretion disks, the resulting flows would likely be considerably more extended in the radial and vertical directions. Additionally, the back reaction of the generated accretion luminosity was entirely ignored in our analysis, even though it is apparent from Figures \ref{fig:mdots_1} and \ref{fig:mdots_2} that the accretion rates often exceed the Eddington luminosity of one or both holes. Therefore, these disks could be much more puffed up in the vertical direction and possess a combination of winds and outflows, such as those suggested by \citet{strubbe09} and \citet{coughlin14}.

For a more complete view of the dynamics discussed in this subsection, we refer the reader to \href{http://w.astro.berkeley.edu/~eric_coughlin/movies.html}{this link} for a movie that demonstrates the formation and evolution of the accretion disk in the simulation shown in Figures \ref{fig:disk_scale} and \ref{fig:disk_closeup}.

\subsection{Fallback periodicity} 
\label{sec:periodicity}
It is evident from Figures \ref{fig:mdots_1} and \ref{fig:mdots_2} that the accretion and fallback rates onto one or both SMBHs are highly variable from simulation to simulation. In particular, the time taken for the return of the most-bound material varies by more than an order of magnitude in time, even between disruptions by the same SMBH; the peak(s) in the fallback rates occur orders of magnitude below and above the estimates from the control cases and, as mentioned in Section \ref{sec:morphology}, depend sensitively on the trajectory of the disrupted debris; at later times, the fallback rates only seem to conform to the $t^{-5/3}$ rate in a very small subset of cases; and the rates themselves are highly variable on very short timescales (i.e., on timescales that amount to small fractions of a single binary orbit) at both early and late times.

Nevertheless, and in spite of all these obvious differences, the disruptions by the primary tend to generate accretion rates that exhibit a quasi-periodic behavior, with dips and spikes below and above the control rate on timescales that are comparable to the binary orbit. At first, this finding seems consistent with the restricted three-body integrations of \citet{liu14} and \citet{ricarte16}, who argued that the secondary SMBH could intersect (or come sufficiently close to) the tidally-disrupted debris stream and interrupt the accretion onto the primary. Under this interpretation, then, the accretion rate onto the primary should drastically decrease at some discrete time following the disruption of the star, quickly resume the previous accretion rate following the passage of the secondary, and this behavior should be repeated once per binary orbit.

We find, however, that the quasi-periodic behavior exhibited in our accretion and fallback rates is not entirely consistent with this picture. For one, it is apparent that while the accretion rate of the primary does fall below the control value at times, at others the binary rate actually \emph{exceeds} that of the control. In addition, if we recall that a binary orbit amounts to approximately 0.3 years, we see that the timescale(s) associated with the periodicity are generally only a fraction of one orbit. The changes in the accretion rate onto the primary are also, at least in some instances, uncorrelated with the position of the secondary, i.e., decreases in the accretion rate onto the primary occur at times when the secondary is nowhere near the incoming debris stream. Finally, the ability of the secondary to intersect the stream is clearly maximized when the disruption occurs in a manner that is coplanar with the binary, but the periodic nature of our accretion rates seem uncorrelated with the angle the debris stream makes with binary orbital plane.

We suspect that the quasi-periodic behavior of our accretion rates likely arises more from the \emph{motion of the primary} instead of the direct presence of the secondary. Specifically, when the star is originally disrupted by the primary, the event proceeds as if the SMBH were isolated (as is apparent, in most cases, from the fact that the accretion rate onto the primary originally matches that of the control in Figures \ref{fig:mdots_1} and \ref{fig:mdots_2}) and the stream returns approximately to the point of disruption. However, as the apocenter distances of the bits of the returning stream start to extend beyond the Roche Lobe of the primary, the stream no longer ``sees'' the primary but is, instead, influenced approximately by the monopole moment of the gravitational field of the binary. The pericenter distances of the returning gas parcels then start to deviate from the original site of the disruption and are deflected toward the COM of the binary. 

Once the stream starts to be deflected in this way, the returning debris will, in general, miss the primary and generate a dip in its accretion rate, \emph{unless} the position of the primary, the COM, and the point at which the stream reenters the Roche Lobe of the primary are approximately colinear. When this alignment occurs, accretion will resume onto the primary at a rate that could either fall below or exceed the original fallback rate. Geometrically this alignment will happen twice per orbit, meaning that the frequency over which we expect these variations in the accretion rate to occur is approximately equal to half the binary orbital period.

This argument ignores additional aspects of the problem that could complicate and potentially confound the ``clean'' periodicity we would otherwise expect. For example, the gravitational potential of the binary does not instantaneously change from a point mass at the location of the primary to a point mass at the location of the COM as the material leaves the Roche Lobe. Furthermore, the quadrupole moment of the potential will still induce additional, time-dependent variations in the orbits of the incoming gas parcels, which could introduce additional periodicities in the accretion rate. Finally, the secondary SMBH does accrete a sizable amount of material in a subset of disruptions by the primary, as can be seen in Figures \ref{fig:mdots_1} and \ref{fig:mdots_2}, demonstrating that the secondary is capable of physically impacting the stream. In these instances, then, there may be an additional periodic variation induced in the accretion rate of the primary with a periodicity on the order of one binary orbital period, in line with the expectations of \citet{liu14} and \citet{ricarte16}.

\begin{figure} 
   \centering
   \includegraphics[width=0.47\textwidth]{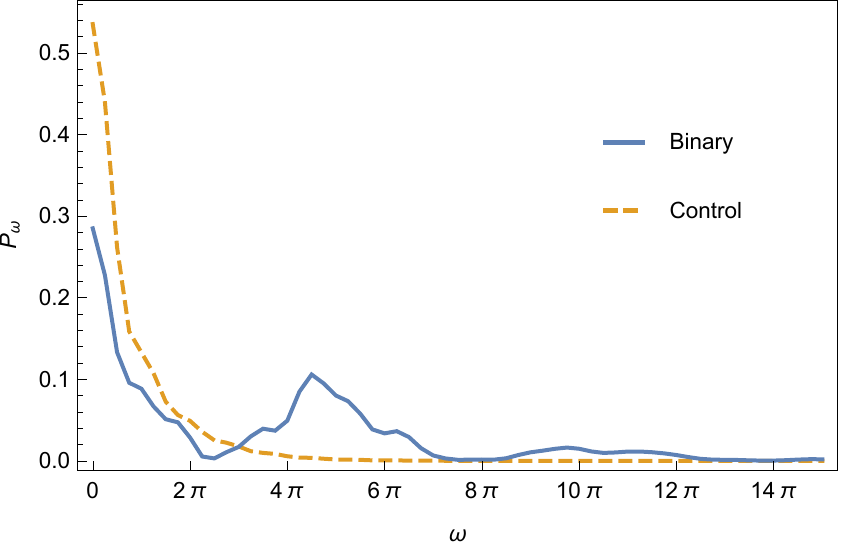} 
   \caption{The power spectrum of the primary accretion rate corresponding to the top-right panel of Figure \ref{fig:mdots_2}, with the solid, blue curve being the binary simulation, the dashed, orange curve the control. Here $\omega = 2\pi$ corresponds to variations on one orbital period, and the prominent spike -- showing that there is indeed a periodicity within the temporal accretion rate -- occurs at a frequency of $4.5\pi$, or a timescale of 0.44 times the binary period.}
   \label{fig:power}
\end{figure}

To more concretely assess the potential periodic nature of the accretion rates, one can analyze the power-spectra of the binary runs in comparison to the controls. Figure \ref{fig:power} shows one specific power spectrum of the accretion rate of the primary, the temporal variation of which is shown in the top-right panel of Figure \ref{fig:mdots_2}. In this figure, $\omega = 2\pi$ corresponds to the angular frequency of one binary period, the solid blue curve is from the simulation with the binary, while the dashed orange curve is the control case. There is a very prominent peak that occurs at $\omega \simeq 4.5\pi$, yielding a periodic timescale of $\simeq 0.44$ binary orbits -- a timescale that is consistent with our proposed interpretation of the motion of the primary generating the variation. There are also smaller peaks at frequencies near $\omega = 10\pi$, which could be indicative of higher order resonances caused by the effects of the binary orbital motion on the debris stream.

The top-right panel of Figure \ref{fig:mdots_2} has an apparent, by-eye, quasi-periodic variability, and it is therefore not overly surprising that the power spectrum of such a lightcurve possesses a well-defined peak. However, we have also found that many of the other simulations -- even those that do not possess obvious patters of variation -- also exhibit peaks at frequencies that correspond to approximately integral numbers of the binary orbital frequency, the largest of those peaks occurring at either one or two times the binary period. This finding is then indicative of the notion that TDEs generated by SMBH binaries generally exhibit variations on timescales that are comparable to the orbital period of the binary.

The frequency corresponding to half a binary orbital period, as we argued above, ultimately comes from the motion of the primary. Thus, as the binary mass ratio starts to decrease, this feature will start to become less prominent in the power spectrum. When $q \ll 1$, we would then expect only the motion of the secondary to potentially alter the distribution of tidally-disrupted debris, resulting in a peak at a frequency corresponding to one, as opposed to one half, binary orbital period. However, as noted by \citet{ricarte16}, the original orbit of the star must be nearly coplanar with the binary orbital plane in order for there to be a large effect.

\subsection{Resolution}
\label{sec:resolution}
In this paper, all of the simulations we used to analyze the accretion morphologies and rates were performed with $5\times10^5$ SPH particles (for more details of the numerical methods, see Section \ref{sec:setup}). An important question, then, is how much those morphologies and rates depend on the number of particles used. 

To obtain a rough idea of the dependence of our results on resolution, we ran two additional simulations employing the same initial setup for the star that produced Figures \ref{fig:inplane_tile} and \ref{fig:outofplane_tile}. However, one simulation was characterized by having a low resolution and only used $2.5\times10^5$ particles, while the other had a comparatively high resolution and used $10^6$ particles.

\begin{figure*} 
   \centering
   \includegraphics[width=0.325\textwidth]{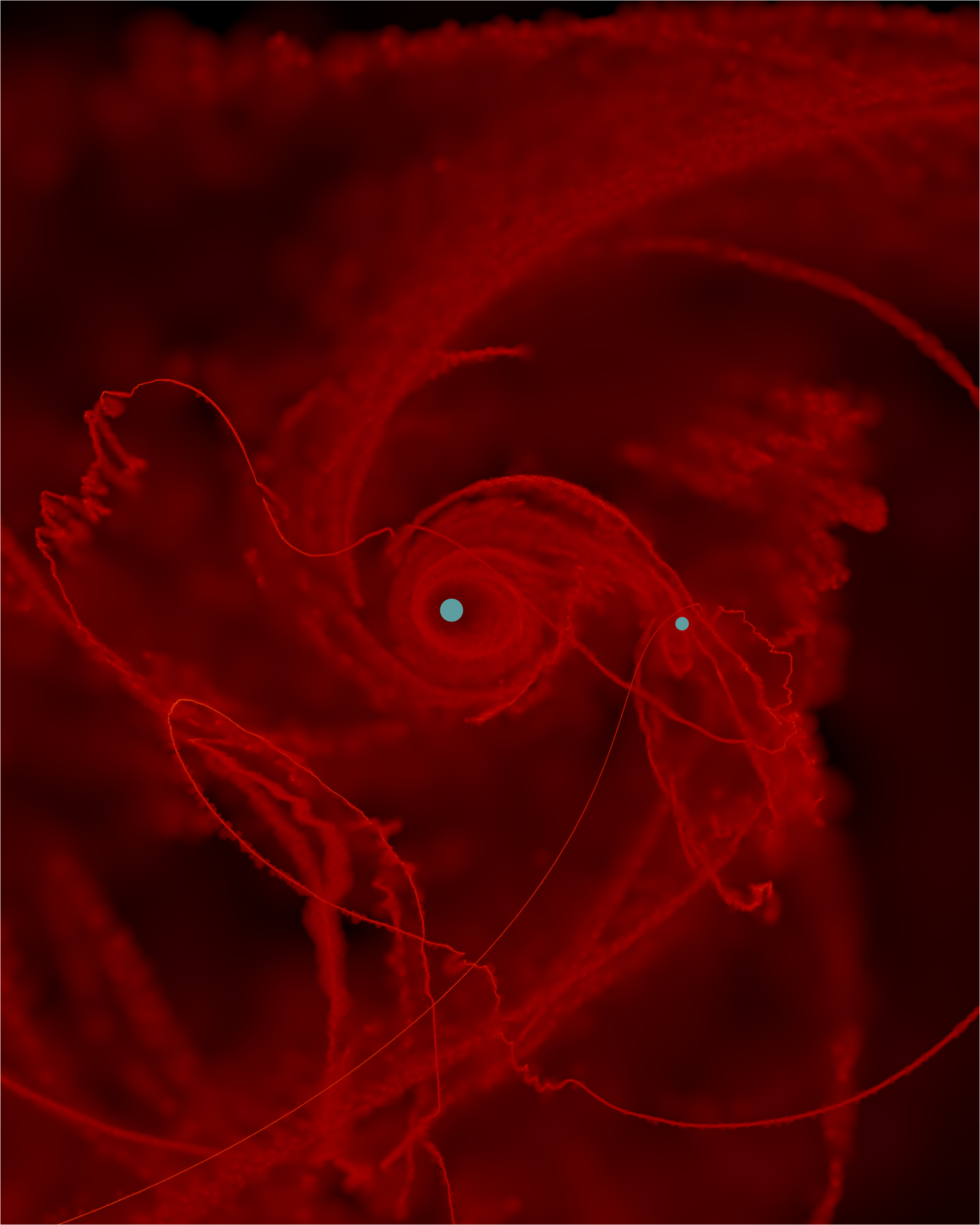}
   \includegraphics[width=0.325\textwidth]{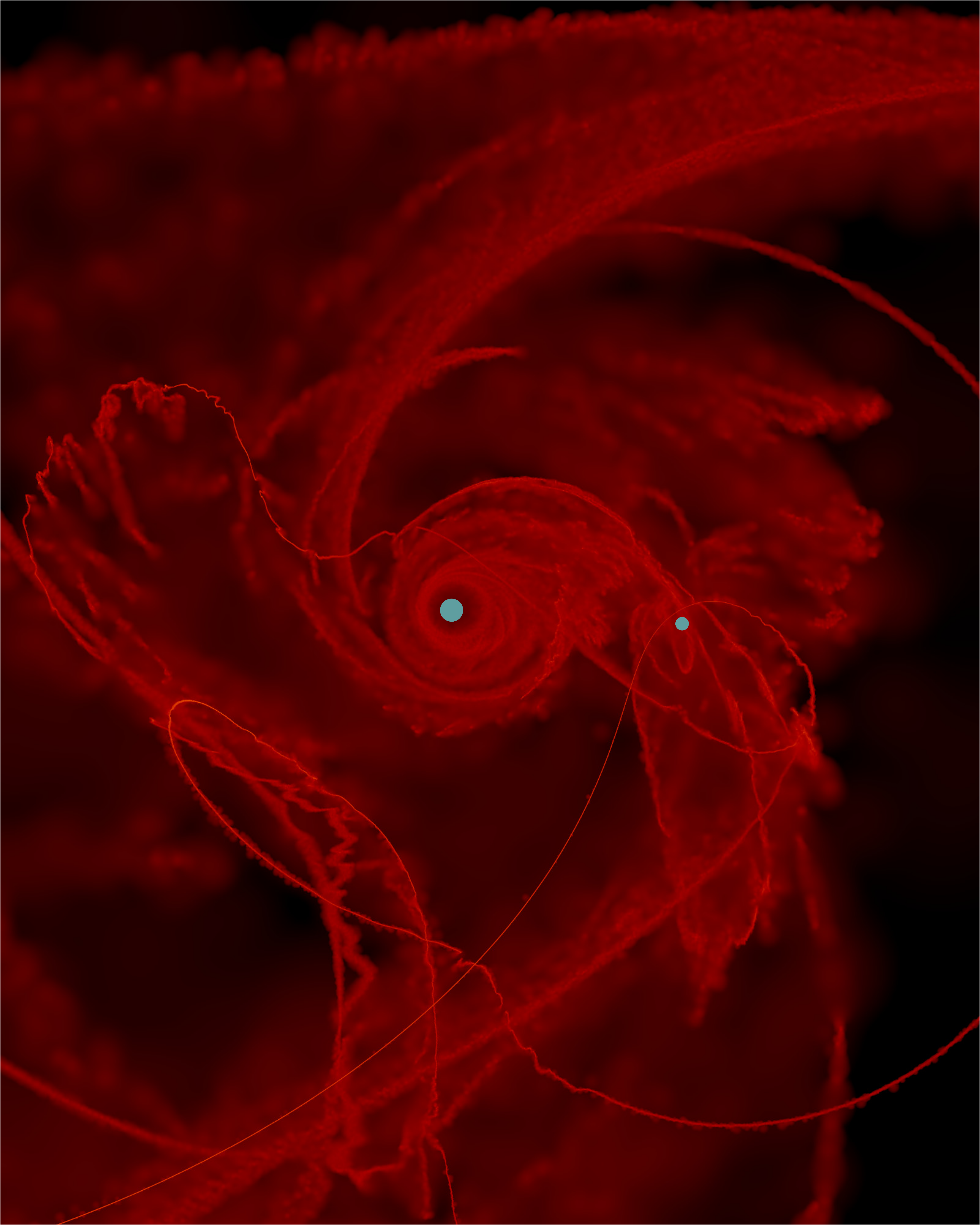}
   \includegraphics[width=0.325\textwidth]{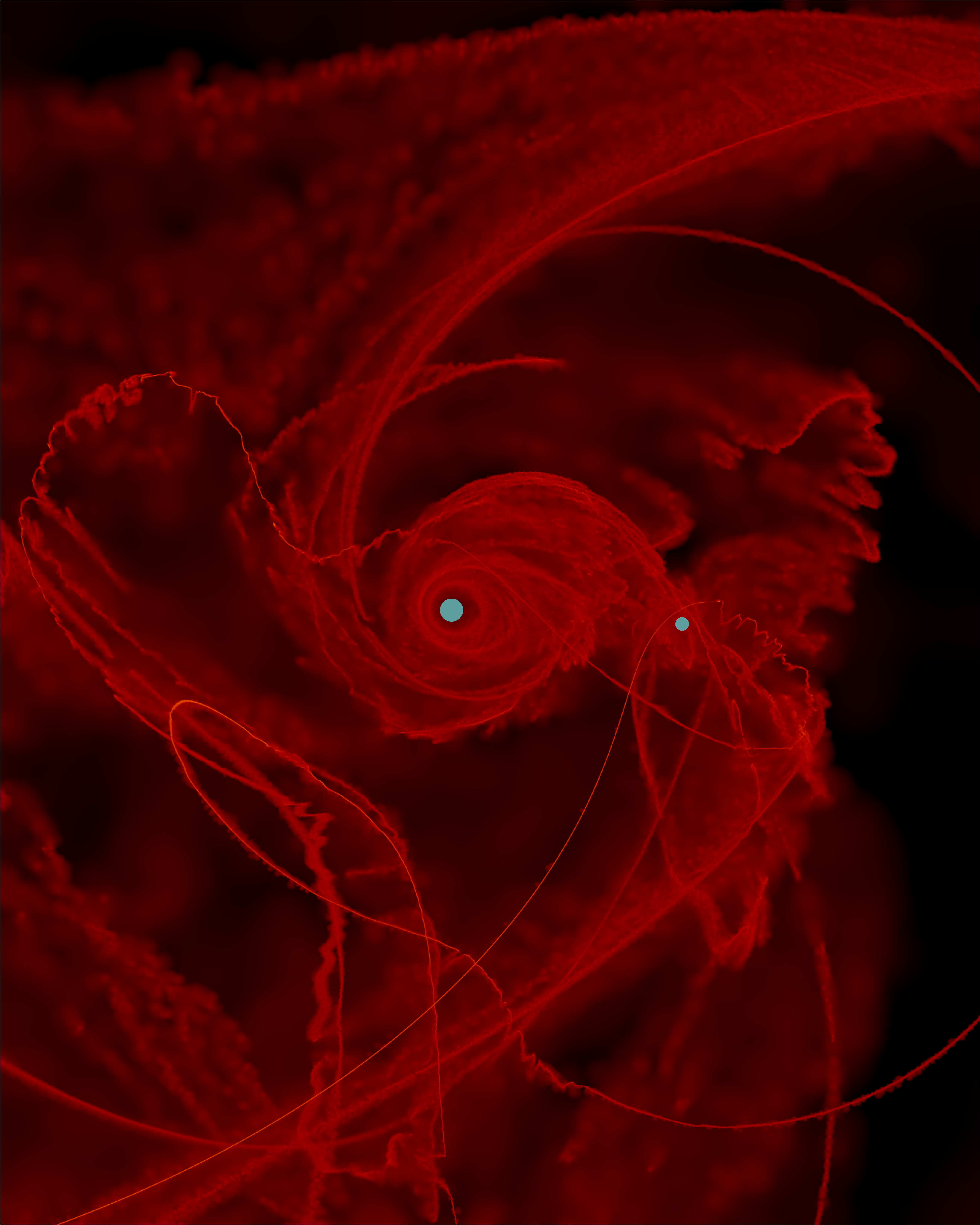} 
   \caption{The $x$-$y$ projection of run 47 at a time of 2 orbits; the left, middle, and right panels show the low ($2.5\times10^5$ particles), medium ($5\times10^5$ particles), and high ($10^6$ particles) resolution runs, respectively. This figure demonstrates that the overall structure of the accretion flow is largely unaffected by resolution.}
   \label{fig:res_flows}
\end{figure*}

Figure \ref{fig:res_flows} compares the morphology of the accretion and fallback flows generated from the three simulations, with low-resolution on the left, medium-resolution (with $5\times10^5$ particles) in the middle, and high resolution on the right. Qualitatively we see that there is very little difference in the overall, bulk structure and organization of the fluid around the binary, and it is only until one investigates the smallest scales that one finds any deviations between the panels. This behavior is, of course, expected: on large scales the dynamics is primarily ballistic, with the main hydrodynamic effects occurring intermittently when streams intersect. On small scales, shocks, disk circularization, and possible hydrodynamic instabilities are all more challenging to resolve.

\begin{figure} 
   \centering
   \includegraphics[width=0.47\textwidth]{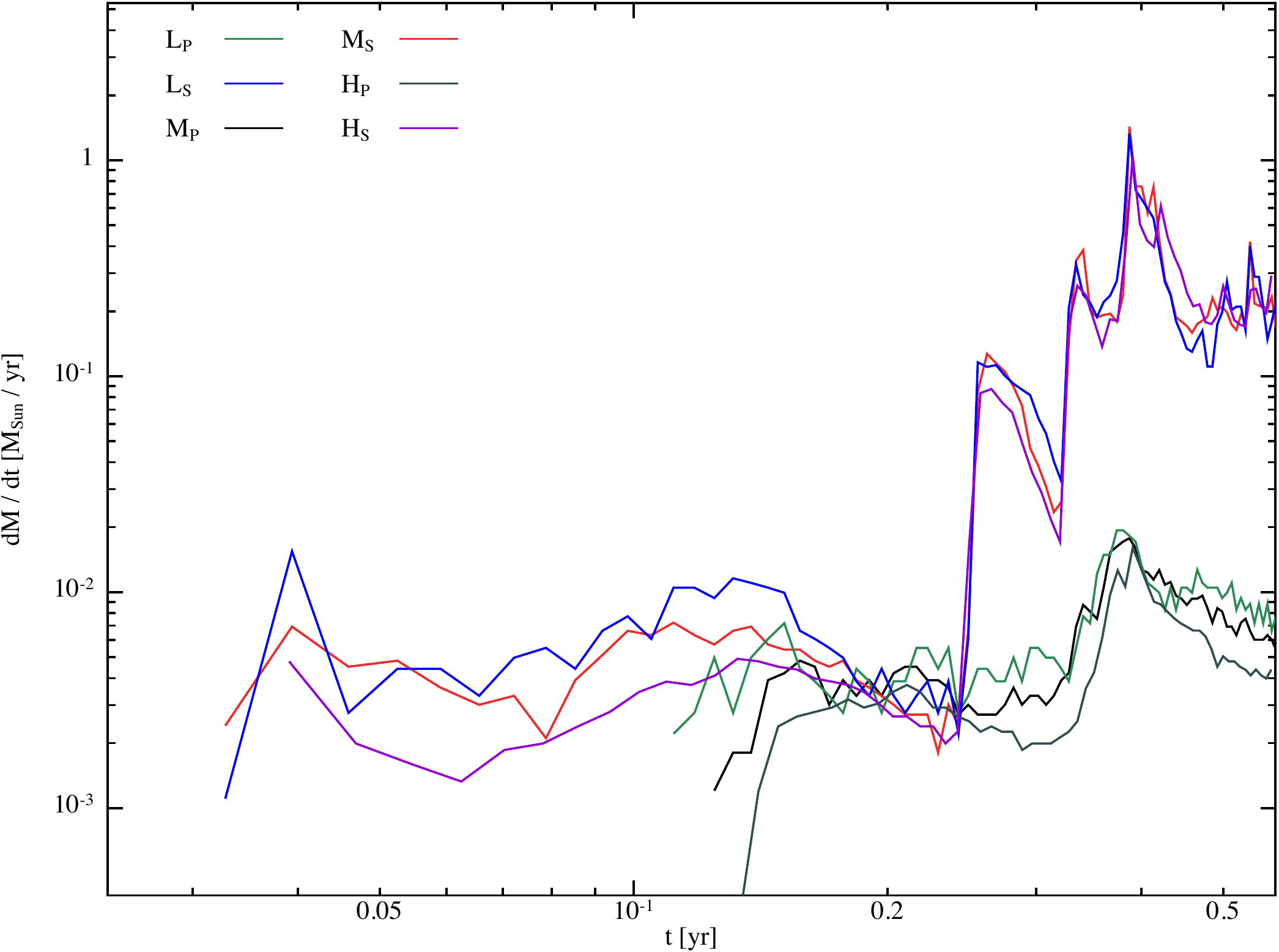} 
   \caption{The accretion rate onto the primary for the low-resolution ($\textrm{L}_\textrm{P}$), medium-resolution ($\textrm{M}_\textrm{P}$), and high-resolution ($\textrm{H}_\textrm{P}$) simulations of run 47, and the accretion rate onto the secondary for the low-resolution ($\textrm{L}_\textrm{S}$), medium-resolution ($\textrm{M}_\textrm{S}$), and high-resolution ($\textrm{H}_\textrm{S}$) runs. }
   \label{fig:res_mdots}
\end{figure}

Figure \ref{fig:res_mdots} shows the accretion rates onto both the primary and secondary for the low, middle, and high-resolution simulations; in the legend, L, M, and H stand for low, medium, and high resolution, while the subscripts P and S stand for accretion onto the primary and secondary (e.g., $\textrm{L}_\textrm{P}$ is the accretion rate onto the primary for the low-resolution test). This figure demonstrates that the accretion rates are reasonably well-resolved, with discrepancies between simulations only amounting to a few tens of percent, at most. 

The one consistent trend apparent from Figure \ref{fig:res_mdots} is that higher resolution results in a shift toward lower accretion rates, with the most significant decreases occurring when the accretion rates themselves are relatively low (e.g., compare the accretion rates onto the primary at a time of roughly 0.3 years). This trend likely signifies that the regions in the immediate vicinity of the accretion radii of either black hole at these times are not completely resolved: a smaller number of particles generates a correspondingly higher, effective numerical viscosity. This larger viscosity then increases the ability of the gas to transport angular momentum through the disk, thereby artificially augmenting the accretion rate onto the hole. However, we note that this trend is only particularly apparent when the accretion rates are low, with larger accretion rates -- which have a greater number of particles in the inner regions of the disks -- showing very little deviation between the resolution tests. 

\subsection{Long term evolution}
It is especially apparent from the accretion curves (Figures \ref{fig:mdots_1} -- \ref{fig:mdots_2}) and, to more or less of an extent, from the morphologies of the surrounding flows themselves (Figures \ref{fig:inplane_tile} -- \ref{fig:panels_multi}) that, even after two complete binary orbits (or nearly six months in time following the original disruption of the star), the binary system has not settled into an asymptotic, ``steady state'' following the disruption of the original star. In particular, most, if not all, of the simulations show rapid variation in the fallback and accretion curves for months after the disruption of the star, and most have not exhibited any type of a power-law decline. To investigate how long it could conceivably take for such a state to be reached, we evolved the run shown in Figures \ref{fig:inplane_tile} and \ref{fig:outofplane_tile} for a total of six binary orbits, or 1.7 years, and analyzed the long-term accretion rates and flow morphologies.

\begin{figure*} 
   \centering
   \includegraphics[width=0.495\textwidth]{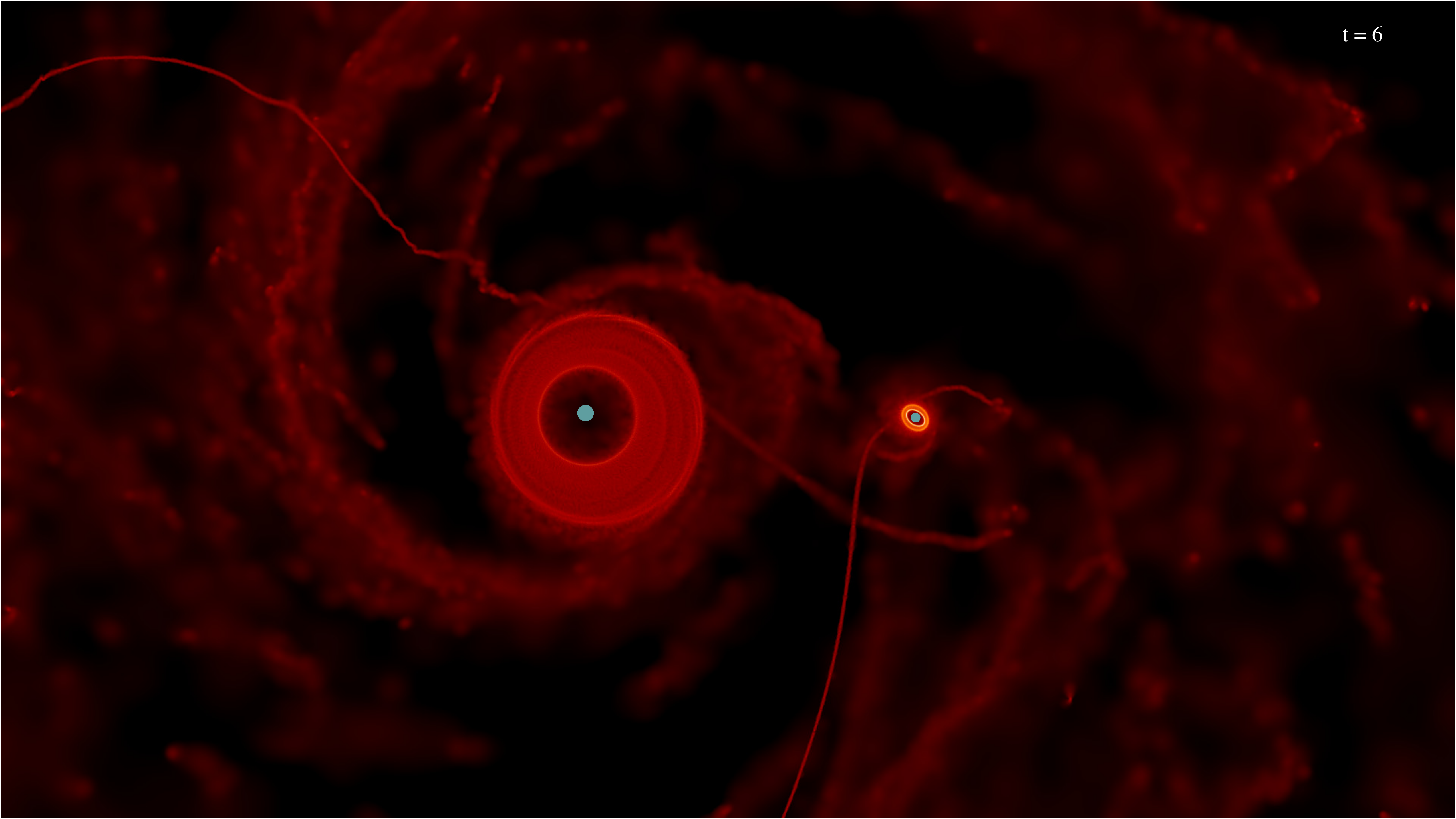} 
   \includegraphics[width=0.495\textwidth]{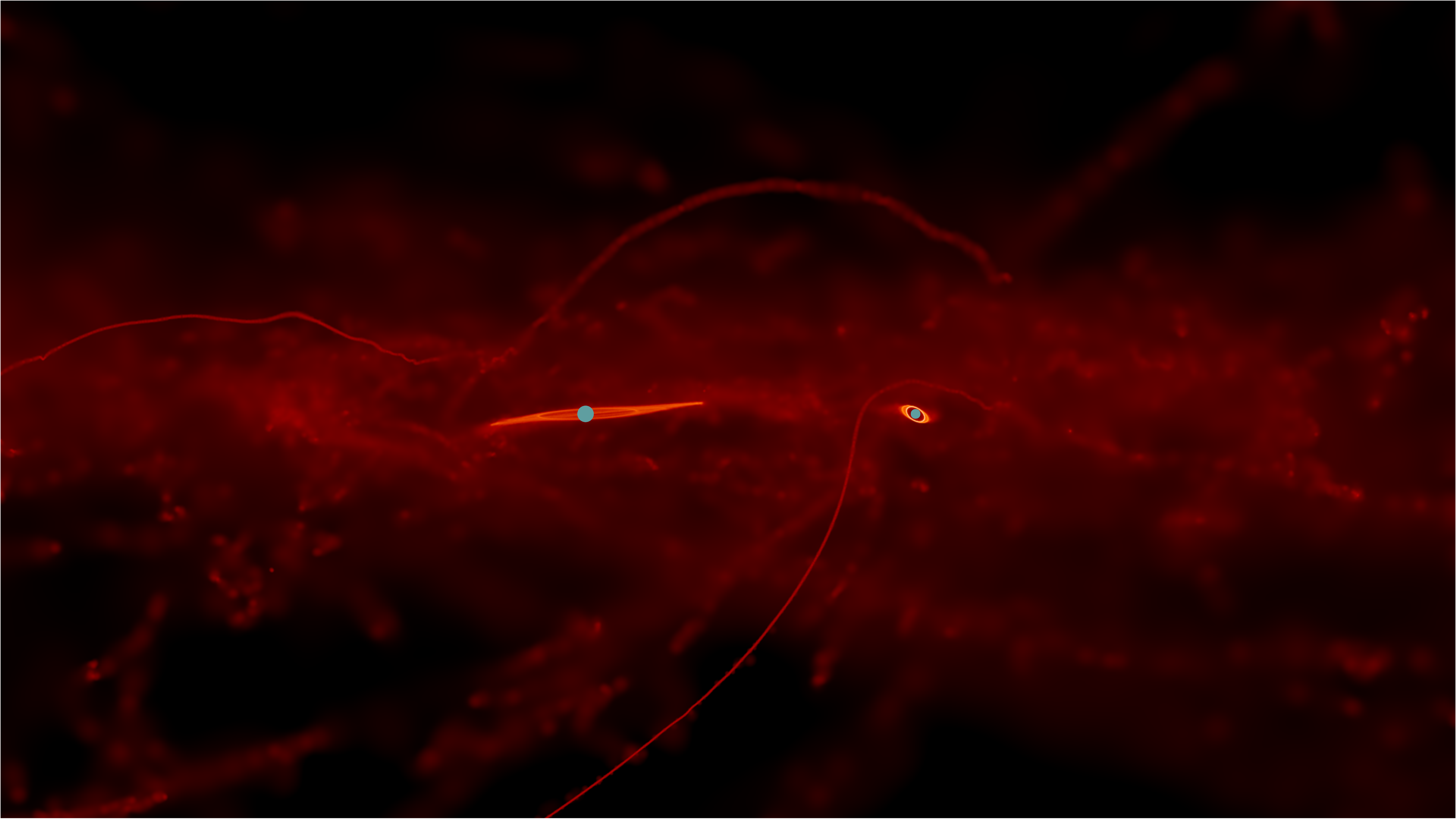}
   \caption{The morphology of the debris produced from run 047 at a time of 6 binary orbits, or approximately 1.7 years after the disruption of the star. The left-hand panel shows the projection in the plane of the binary, while the right-hand panel shows the out-of-plane projection.}
   \label{fig:lt_panels}
\end{figure*}

\begin{figure} 
   \centering
   \includegraphics[width=0.47\textwidth]{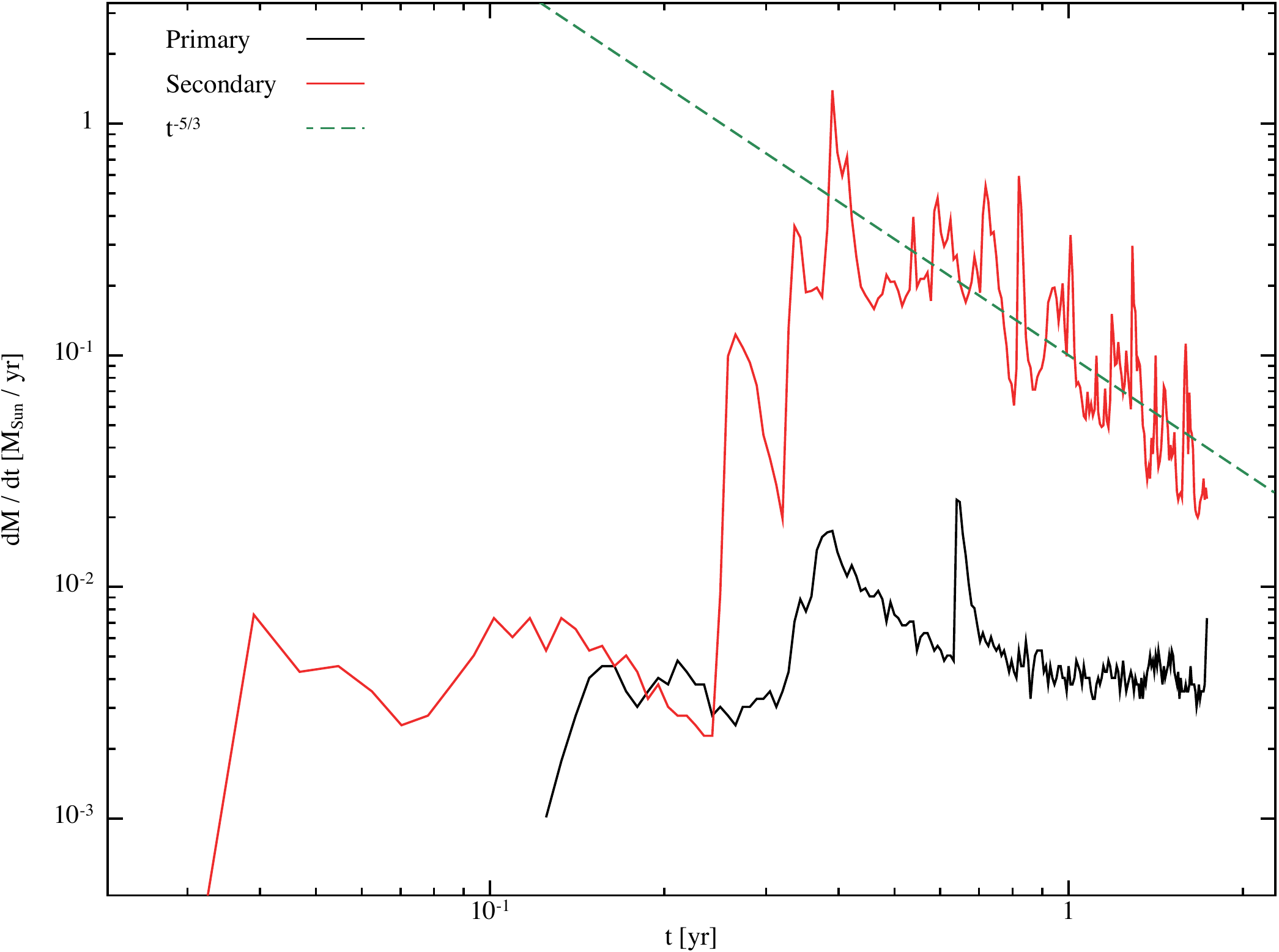} 
   \caption{The accretion rates onto the primary (black curve) and secondary (red curve) as functions of time, and the $\propto t^{-5/3}$ power-law (green, dashed line) for reference, calculated for a total of 6 binary orbits (or 1.7 years) for run 047. }
   \label{fig:lt_mdots}
\end{figure}

Figure \ref{fig:lt_panels} shows the distribution of disrupted debris after six binary orbits, the left-hand panel being the projection onto the binary plane, the right-hand panel the projection out of the plane. As is apparent, much of the material has conformed to two, well-defined disks surrounding each individual SMBH. Nevertheless, there is still a large amount of debris that lies outside of these two disks, with a very pronounced, spiral structure emanating from the disk surrounding the primary. The incoming debris stream is also still evident, and the right-hand panel shows that it makes a significant angle with respect to the binary plane. We also see, from the right-hand panel, that the debris has become more localized to the plane of the binary (compare this panel to the bottom-right panel of Figure \ref{fig:outofplane_tile}). 

Figure \ref{fig:lt_mdots} illustrates the accretion rate onto the primary (black curve), the accretion rate onto the secondary (red curve), and the power-law $t^{-5/3}$ (green, dashed curve) for reference. We see that the accretion rate onto the secondary has, on average, started to follow an approximate, $t^{-5/3}$ decay. However, there is still a large degree of variability, with changes ranging by roughly a factor of five above and below the $t^{-5/3}$ decay. The accretion rate onto the primary, on the other hand, remains approximately constant following a particularly large outburst around 0.7 years.

These figures demonstrate that, even after 6 binary orbits, or nearly two years following the disruption of the original star, the accretion rates onto the SMBHs have not settled into something that is easily identifiable as a steady state. Surprisingly, the accretion rate onto the primary SMBH does not resemble anything that would be obviously indicative of an accretion episode following a TDE. This finding seems to imply that, even though the primary is considerably more massive than the secondary, many orbits are required in order to deflect the stream to the point where it returns to the center of mass (approximately coincident with the location of the primary). 

As we noted above, this particular TDE was exceptionally dynamic and resulted in an extremely disordered distribution of material. Some of the other simulations, specifically those in which the primary was the disrupting SMBH, may start to exhibit less chaotic variability by six (or fewer) orbits following the initial disruption. Indeed, this is already somewhat apparent from some of the accretion curves at two orbits. 

\subsection{Entirely unbound/bound streams}
We noted in Section \ref{sec:results} that some debris streams were completely ejected from the binary system, which resulted from the fact that the energy of the stellar center of mass (COM) at the time of disruption was so high that even the most-bound debris had positive energy. As we showed at the end of Section \ref{sec:rates}, the energy of the stellar COM, $\epsilon_c$, must satisfy $\epsilon_c/\epsilon_b > N_i$ in order for the stream to be completely ejected, where $\epsilon_b$ is the binding energy of the binary and

\begin{equation}
N_i = \frac{2\,m_1^{1/3}m_i^{1/3}}{m_1+m_2}k. \label{Ncrit1}
\end{equation}
In this expression, $m_{1,2} \equiv M_{1,2}/M_*$ is the ratio of the primary or secondary mass to the stellar mass, $m_i$ is the mass of the disrupting hole (divided by the stellar mass), and $k$ is the ratio of the binary separation to the tidal radius of the primary.

For our specific set of simulations we adopted $k = 100$, which gave $N_i \simeq 1$. Investigating the histograms that show the distribution of the energy of the COM at the time of disruption (Figure \ref{fig:eps_hist}), we found that there should be a small number of cases in which all of the debris is ejected from the binary, and preferentially-so by the secondary (and this preference was confirmed by the simulations). However, we note that the required energy cutoff given by Equation \eqref{Ncrit1} is linearly dependent on the separation of the binary. Furthermore, the histograms in Figure \ref{fig:eps_hist} show that there is a very sharp cutoff in the number of stars that satisfy $\epsilon_c \gtrsim few\times\epsilon_b$, which shows that even for marginal increases in the binary separation (i.e., by a factor of two) there should be a drastic falloff in the number of totally-ejected streams. Conversely a decrease in the binary separation should dramatically increase the number of entirely-unbound streams, not only because the necessary energy of the center of mass decreases, but also because we expect a larger spread in the energies of the disrupted stars owing to the higher speeds of the SMBHs.

Even though entirely-unbound streams do not result in accretion, they still may generate observational signatures: as the stream exits the sphere of influence of the binary, it will interact with the preexisting gas and dust in the circumnuclear medium (CNM). As investigated by \citet{guillochon16} (see also \citealt{chen16}), this interaction generates a drag on the stream, decelerating the outward motion of the debris, reorienting and tangling the stream due to its radially-dependent density profile \citep{coughlin16b}, and depositing energy into the ambient medium, these features mimicking those of a supernova remnant. Furthermore, because the center of mass of the stream is on a hyperbolic orbit, the total amount of energy deposited into the CNM and the eventual stopping distance of the stream can exceed those from an ordinary TDE (i.e., one produced by an isolated SMBH). 

{}{From the four of our simulations that resulted in the total ejection of the stream, we find that the maximum energy (over all four simulations) is $\epsilon_{ej}\simeq 0.001c^2$, yielding an asymptotic velocity of $v_{\infty} \simeq \sqrt{2\epsilon_{ej}} \simeq 0.045c$. Comparatively, the maximum terminal velocity for the disruption of a Solar-like star by an isolated $10^6M_{\odot}$ SMBH is $v_{\infty} \simeq 0.02c$ (this makes the usual assumption that the stellar COM is on a parabolic orbit; \citealt{rees88}), showing that the shift in the energy of the center of mass in these cases can (at least) double the terminal velocity of the most unbound debris. We also find the total amount of energy contained in these entirely-unbound debris streams to be $E_{inj} = \{1.03,0.703,1.49,0.348\}\times10^{51}$ erg, which is a factor of $\sim 10$ times the amount one would expect from the debris streams ejected by isolated SMBHs \citep{guillochon16}. This additional factor of ten comes both from the fact that the \emph{total} mass of the star is ejected (as opposed to half), and the energies are augmented by roughly the binding energy of the binary. }

{}{In a more general sense, the evolution of the unbound portions of these streams as they impact the CNM should differ substantially from those arising from isolated SMBHs. Specifically, as is evident from Figures \ref{fig:panels_multi}, the streams are tangled into large, sweeping arcs, which extend over hundreds of degrees from end to end, as they escape from the binary system. In contrast, the angle subtended by the unbound portion of the debris stream generated by the disruption of a Solar-like star by a $10^6M_{\odot}$ SMBH is $\theta_{arc} \simeq \sqrt{2}/10 \simeq 0.1$. Since the drag force on the stream is proportional to its area, which scales with its length, this shows that the timescale over which the stream slows appreciably is approximately a factor of 100 shorter in this case.} 

{}{Finally, while the disruption of a star by an isolated SMBH generates one unbound debris stream that expands from the radius of disruption, many of our simulations are characterized by concentric ejections of loops of material over time. Thus, the unbound debris remnants produced from the disruptions by binaries may not necessarily have one, well-defined remnant. Furthermore, if two unbound loops separated in time are relatively coplanar, one might expect that the debris stream ejected at a later time could traverse the already-cleared path made by a preceding stream, culminating in the second stream encountering the stalled, initial remnant. In this case, the interaction of the second stream with the original remnant may cause a sudden rebrightening.}

In addition to entirely unbound streams, some tidally-disrupted streams can also be completely bound to the binary, which requires an energy less than the negative of equation \eqref{Ncrit1}. It is apparent from Figure \ref{fig:eps_hist} that this situation is much easier to achieve than the entirely-unbound case, which results from the fact that most disrupted stars are placed on unstable, but temporarily-bound orbits before being disrupted. 

\begin{figure} 
   \centering
   \includegraphics[width=0.47\textwidth]{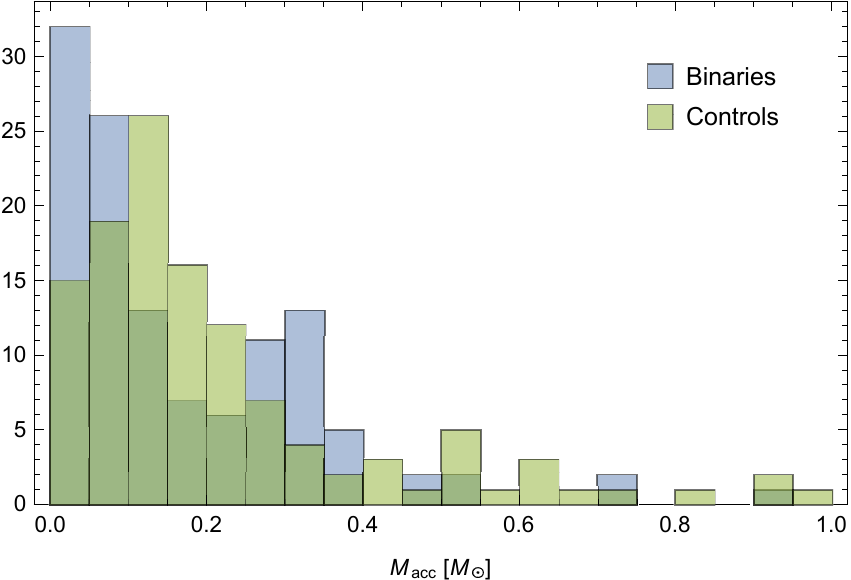} 
   \caption{A histogram of the total amount of mass accreted by the binary (blue bars) and the control SMBH (green bars) at the terminus of each simulation (which amounts to roughly 2 orbits for most cases, but is somewhat less for a few runs). The value of this number is a lower limit on the total amount of disrupted debris that is bound to each system.}
   \label{fig:macc_hist}
\end{figure}

Figure \ref{fig:macc_hist} shows a histogram the total amount of mass accreted at the end of each simulation, corresponding to a time of roughly two orbits for most cases. The blue bars are the binary runs, while the green bars are from the controls. This plot demonstrates that, while roughly a quarter of the simulations have not yet accreted much material, a substantial fraction have already consumed greater than 20\% of the disrupted stellar debris by 5.8 months. Comparatively, it would take roughly nine months for an isolated, $10^6M_{\odot}$ SMBH to accrete 0.2$M_{\odot}$ of disrupted debris under the impulse approximation (assuming the disrupted star was solar-like and that all of the mass that returns to the point of disruption is immediately accreted; \citealt{coughlin14}). In addition, four binary runs and a dozen controls already accreted more than half all of the available debris, thereby exceeding the theoretical limit one would normally invoke for the disruption of a star by an isolated SMBH. 

\section{Summary and Conclusions}
\label{sec:conclusions}
In this paper we have taken a two-step approach to analyzing the tidal disruption of stars by supermassive black hole binaries. First, we performed a large number of restricted three-body integrations of a point mass in a binary potential. By injecting stars (considered point masses) from random orientations into a binary of fixed separation, we were able to determine -- from a statistical standpoint -- the properties of the stellar orbits at the time of disruption, defined as the moment at which the star crossed the tidal sphere of either SMBH. Importantly, our results also demonstrated that the energies and angular momenta of the center of mass of the disrupted stars can differ drastically from those of the originally-assumed parabolic orbit, a discrepancy that arises from the time-dependent nature of the binary potential.

We then simulated the hydrodynamical evolution of 120 different TDEs, with the initial conditions for those TDEs randomly drawn from our sample of three-body integrations. Specifically, we determined -- from the restricted three body results -- the position and velocity of the center of mass of 120 to-be-disrupted stars when they were five tidal radii from the disrupting hole. Those positions and velocities were then used to initiate SPH simulations with the code {\sc phantom} \citep{price10}, with which we followed the hydrodynamical evolution of the TDEs for two binary orbital periods. In our analysis we opted to focus on a circular binary with fixed properties, being a primary mass of $10^6M_{\odot}$, a secondary mass of $2\times10^5M_{\odot}$ (a mass ratio of 0.2), and a separation of 100 tidal radii of the primary (or roughly $10^{-4}$ pc). We also ran 120 control simulations in which we removed the non-disrupting hole, ultimately in an effort to distinguish the effects of changing the energy and angular momentum of the stellar center of mass from the dynamical effects of the binary itself. 

Our restricted three body integrations yielded a number of interesting results concerning the statistical nature of tidal disruption events by SMBH binaries, the first being that the rate of TDEs (assuming an isotropic initial distribution of stars on parabolic orbits with uniformly distributed angular momenta) is nearly independent of the mass ratio of the binary (see Figure \ref{fig:tderate}), that rate being equal to approximately 0.02 (i.e., 2\% of stars that encounter the binary will be tidally disrupted). We also showed that the likelihood of being disrupted by the primary, $\lambda_p$, is greater than that of the secondary, and is very well approximated by the linear relation $\lambda_p(q) = 0.96-0.46q$, $q = M_2/M_1$ being the mass ratio of the binary. The probability distribution functions (PDFs) of the impact parameter $\beta = r_t/r_p$ ($r_t$ being the tidal radius of the disrupting hole, $r_p$ the pericenter distance of the disrupted star), the time taken to be disrupted $T$, and the number of additional encounters $N$ (an additional encounter being any time a star came within 3 $r_t$ of either hole \emph{without} being disrupted), were found to be well-fit by the power-laws $f_\beta = \beta^{-2}$, $f_T \propto T^{-2.5}$, and $f_N \propto N^{-3.5}$ (see Figures \ref{fig:beta_pdf}, \ref{fig:t_pdf}, and \ref{fig:Nenc}). Finally, the PDFs describing the energy and angular momentum of the stellar center of mass at the time of disruption were widely distributed (see Figures \ref{fig:ell_hist} and \ref{fig:eps_hist}), showing that the time-dependence of the binary potential can dramatically alter the dynamics of the ensuing TDE -- even \emph{before} the star is actually disrupted.

The hydrodynamical simulations demonstrated that the second SMBH can have large effects on the evolution of the tidally-disrupted debris. As confirmed by Figures \ref{fig:inplane_tile} -- \ref{fig:panels_multi}, the distribution of the material rapidly (i.e., after less than one binary orbit post-disruption) becomes very chaotic, spreading to large radii in both the plane of the binary and out of the plane. Interestingly, in some cases the debris conformed to an approximately-spherical configuration around the binary, as can be seen in the bottom-right panel of Figure \ref{fig:outofplane_tile}. Furthermore, a subset of simulations in which the secondary was the disrupting SMBH generated entirely-unbound debris streams, creating no fallback whatsoever. 

The accretion rates following these disruptions also showed an intense amount of variability on short and long timescales, the accretion rates themselves shown in Figures \ref{fig:mdots_1} -- \ref{fig:mdots_2} (see also the figures in Appendix \ref{sec:sup}). Interestingly, a number of the disruptions generated quasi-periodic behavior, with dips in the accretion rates occurring on the order of the binary orbital period (the top-right panel of Figure \ref{fig:mdots_2} illustrates one particularly good example of this phenomenon). As we argued in Section \ref{sec:periodicity}, this periodicity likely arises from the motion of the \emph{primary} SMBH, and should therefore occur on timescales comparable to half of the binary period. This interpretation is in contrast to those put forth by \citet{ricarte16}, who argued that the secondary SMBH could -- at least in the instances in which the disrupted debris is confined approximately to the plane of the binary -- intersect the tidally-disrupted debris stream and thereby reduce the accretion onto the primary once per binary orbit. While this situation could certainly occur, and it does seem to be the case that at least some of our simulations show evidence for periodic behavior on the timescale of one binary orbital period, the \href{http://w.astro.berkeley.edu/~eric_coughlin/movies.html}{movie}{ showing the formation and evolution of one accretion disk} and our Figure \ref{fig:power}, which shows the power spectrum for one simulation, support the interpretation that the motion of the primary induces the majority of the periodicity. However, for smaller mass ratios where the motion of the primary becomes negligible, the periodic interruption proposed by \citet{liu14} and \citet{ricarte16} may be the dominant contributor to the variation in the fallback rate.

Many disruptions resulted in small-scale (i.e., extending to only a few percent of the binary separation) accretion disks surrounding one or both of the SMBHs, some of which are evident in Figure \ref{fig:panels_multi}. A closeup of one such accretion disk, shown in Figure \ref{fig:disk_closeup}, illustrates that these disks can have very complex geometries, including a non-uniform, elliptical geometry, and, especially from the right-hand panel of Figure \ref{fig:panels_multi}, highly inclined tilts and warps. We have found that, at least for the case appropriate to Figure \ref{fig:panels_multi}, these features arise from the time-dependent and highly-chaotic origin of the accretion disk; in particular, the disk itself is actually composed of two disks, each formed at discrete times in the evolution of the binary. The misalignment angle between the inner and outer disk is also a function of time, the viscous interaction between the two dragging the inner disk into coalignment with the outer one (see the movie \href{http://w.astro.berkeley.edu/~eric_coughlin/movies.html}{here}).  

It is apparent from Figures \ref{fig:inplane_tile} -- \ref{fig:mdots_2} that many of the simulations still show a high degree of variability in both the morphology of the accretion flows and the accretion rates themselves. We ran one simulation for a total of six orbits (amounting to 1.7 years) to investigate the long-term evolution of one such system, and we found that, even after this amount of time, there was still a display of chaotic behavior in both the appearance of the distribution of disrupted material and the accretion rates (see Figures \ref{fig:lt_panels} and \ref{fig:lt_mdots}). While the accretion rate of the secondary SMBH (the disrupting hole for this case) fell off at a rate that, on average, mimicked the $t^{-5/3}$ decay, the primary accretion rate remained approximately flat and did not resemble anything indicative of a TDE. However, we note that this particular simulation was rather extreme in terms of the chaotic distribution of debris, and other, more ``tame'' TDEs may tend to follow a more regular pattern at earlier times (indeed, this is arguably evident from some of the panels in Figures \ref{fig:mdots_1} -- \ref{fig:mdots_2}). 

{}{Our simulations are generally characterized by the formation of small-scale, gravitationally-bound clumps, resulting from the gravitational instability of the stream \citep{coughlin15}; these are not visible in Figures \ref{fig:inplane_tile} and \ref{fig:panels_multi} because of their incredibly small scale ($\sim R_{\odot}$), and they are correspondingly smoothed over the resolution element of the figures. This instability is expected on analytical grounds if the impact parameter is not sufficiently high, though the growth rate of the instability is stunted by the tidal field of the binary \citep{coughlin16b}. Specifically, the decay of the background density profile of the stream resulting from its tidal deformation causes the overdensities to grow as power-laws, instead of exponentials, in time, meaning that they are very difficult to resolve and their properties are, in general, dependent on the resolution of the simulation. Clumps bound to the binary are generally sheared apart upon returning to the vicinity of the binary, as their densities are significantly lower than that of the original star, and thus add to the chaotic nature of the distribution of the debris around the binary. In the unbound segment of the stream, the clumps recede from the binary and could, conceivably, escape from the host galaxy after interacting with the surrounding circumnuclear medium (in a manner that could differ from that suspected from unbound debris streams that do not harbor clumps; \citealt{guillochon16}). However, recombinations \citep{kasen10} or magnetic fields \citep{guillochon16b} could provide significant pressure support to resist the collapse into small-scale clumps.}

The simulations we performed involved the disruption of a single type of progenitor (being solar-like and following a polytropic, $\gamma = 5/3$ density distribution) by a circular binary of fixed properties. While this approach obviously leaves a large number of parameters to explore in this problem, we have here performed the first extensive parameter-space sweep involving the properties of the orbit of the star (e.g., the point of closest approach of the star to the disrupting SMBH, the orientation of the orbit with respect to the plane of the binary, etc.). Even though the accretion curves generated from these simulations display a wide range of properties, we have found that many exhibit periodic dips that vary on the order of the binary orbital period, and Figure \ref{fig:power} -- a power spectrum from one particular simulation -- confirms this statement on more quantitative grounds. This finding points to the promising conclusion that upcoming wide-field surveys (e.g., LSST; \citealt{ivezic08}) may be able to start to determine, at the very least, the orbital parameters of SMBH binaries that generate tidal disruption events.

\section*{Acknowledgments}
{}{We thank the referee, James Guillochon, for useful and insightful comments and suggestions.} Support for this work was provided by NASA through the Einstein Fellowship Program, grant PF6-170150, NSF grants AST 1313021 and 1411879, and NASA grants NNX14AB42G and NNX16AI40G. Chris Nixon was supported by the Science and Technology Facilities Council (grant number ST/M005917/1). Research in theoretical astrophysics at Leicester is supported by an STFC Consolidated Grant. This work utilized the Janus supercomputer, which is supported by the National Science Foundation (award number CNS-0821794) and the University of Colorado Boulder. The Janus supercomputer is a joint effort of the University of Colorado Boulder, the University of Colorado Denver and the National Center for Atmospheric Research.

\appendix{}
\section{Binary equations}
\label{sec:bineqs}
In this appendix we write down and derive, for the reference of the reader, the equations describing the evolution of the SMBH binary and the motion of the star (considered a point mass) in the gravitational potential of the binary.

We will let the binary occupy the $x$-$y$ plane, with the $x$-axis parallel to the original (i.e., at a time of zero) separation vector between the two black holes. The black holes will have masses of $M_1$ and $M_2$, with $M_1 \ge M_2$, and the (assumed-stationary) center of mass (COM) of the binary will be the origin of the coordinate system. With this setup, the evolution of the binary is described by the following equations:

\begin{equation}
R_1 = \frac{M_2}{M}r, \label{R1eq}
\end{equation}
\begin{equation}
R_2 = \frac{M_1}{M}r,
\end{equation}
\begin{equation}
r = \frac{a\left(1-e^2\right)}{1+e\cos\phi}, \label{req}
\end{equation}
\begin{equation}
\frac{a^{3/2}}{\sqrt{GM}}\frac{\left(1-e^2\right)^{3/2}}{(1+e\cos\phi)^2}\frac{d\phi}{dt} = 1. \label{phieq}
\end{equation}
Here $R_1$ is the separation of $M_1$ from the COM, $R_2$ is the separation of $M_2$ from the COM, $M \equiv M_1+M_2$ is the total mass of the binary, $r \equiv |R_2-R_1|$ is the magnitude of the displacement vector between the two masses, $\phi$ is the angle made between $r$ and the $x$-axis, $a \equiv (r_{max}+r_{min})/2$ is the semi-major axis of the binary (where $r_{max}$ and $r_{min}$ are the maximum and minimum values of the separation between the masses, respectively), and $e = (r_{max}-r_{min})/(r_{max}+r_{min})$ is the eccentricity of the binary.

We will define the positive-$x$ direction as pointing from $M_1$ to $M_2$ at $t = 0$, while the positive-$y$ direction will be orthogonal to the $x$-axis moving in a counterclockwise sense. The $z$-axis is then constructed in a right-handed manner. We will also restrict ourselves to circular binaries, for which $e = 0$, and in this case the $x$ and $y$ positions of the black holes as functions of time are analytically found to be

\begin{equation}
X_1 = -(1-m)\cos{\tau}, \quad Y_{1} = -(1-m)\sin\tau,
\end{equation}
\begin{equation}
X_2 = m\cos\tau, \quad Y_2 = m\sin\tau.
\end{equation}
Here $X_i$ and $Y_i$ are the $x$ and $y$ positions of the black holes normalized by the semi-major axis, $m \equiv M_1/M$ is the ratio of the primary black hole mass to the total mass, and $\tau \equiv t\sqrt{GM}/a^{3/2}$ is time normalized by the orbital time of the binary (modulo a factor of $2\pi$). 

Now consider a star -- treated as a point mass -- that initially encounters the binary from a very large distance. At any instant in time, the total gravitational potential affecting the star is

\begin{equation}
\Phi = -\frac{GM_1}{\mathscr{D}_1}-\frac{GM_2}{\mathscr{D}_2}, \label{Phieq}
\end{equation}
where $\mathscr{D}_1$ and $\mathscr{D}_2$ represent the magnitudes of the separations between the star and the black holes of masses $M_1$ and $M_2$, respectively. We can parameterize these distances in terms of the $x$, $y$, and $z$ coordinates of the star measured from the binary COM; doing so and letting $x$, $y$, and $z$ be normalized by the binary semimajor axis gives

\begin{equation}
\mathscr{D}_1^2 = a^2\left(\left(x-X_1\right)^2+\left(y-Y_1\right)^2+z^2\right),
\end{equation}
\begin{equation}
\mathscr{D}_2^2 = a^2\left(\left(x-X_2\right)^2+\left(y-Y_2\right)^2+z^2\right).
\end{equation}
The Lagrangian of the star moving under the influence of the binary is then

\begin{equation}
\mathscr{L} = \frac{GM}{a}\left\{\frac{1}{2}\left(\dot{x}^2+\dot{y}^2+\dot{z}^2\right)+\Phi_1+\Phi_2\right\}, \label{L}
\end{equation}
where

\begin{equation}
\Phi_1 = \frac{m}{\sqrt{\left(x+(1-m)\cos\tau\right)^2+\left(y+(1-m)\sin\tau\right)^2+z^2}}
\end{equation}
and

\begin{equation}
\Phi_2 = \frac{1-m}{\sqrt{\left(x-m\cos\tau\right)^2+\left(y-m\sin\tau\right)^2+z^2}}. \label{Phi2}
\end{equation}

From the Lagrangian we can construct the Euler-Lagrange equations that govern the motion of the star in the binary potential, given by

\begin{equation}
\frac{d}{dt}\left(\frac{\partial\mathscr{L}}{\partial\dot{x}_i}\right)-\frac{\partial\mathscr{L}}{\partial{x_i}} = 0, \label{EL}
\end{equation}
where $x_i = \{x,y,z\}$.

\section{Supplemental Figures}
\label{sec:sup}
In this section we present, for completeness and for the reference of the reader, the accretion and fallback rates for the additional simulations that were not shown in Section \ref{sec:rates}. In all of these figures the units, color-codings, etc.~are identical to those used in Figures \ref{fig:mdots_1} and \ref{fig:mdots_2}, with the black and green curves being the respective accretion and fallback rates onto the primary, the red and blue curves being the accretion and fallback rates onto the secondary, the grey curves being the accretion rate from the control simulations, and the cyan and magenta lines are the Eddington limits for the primary and secondary. All accretion rates are measured in units of Solar masses per year, and the top-left corner of each panel indicates the disrupting hole, P being the primary, S the secondary.

\begin{figure*} 
   \centering
   \includegraphics[width=\textwidth]{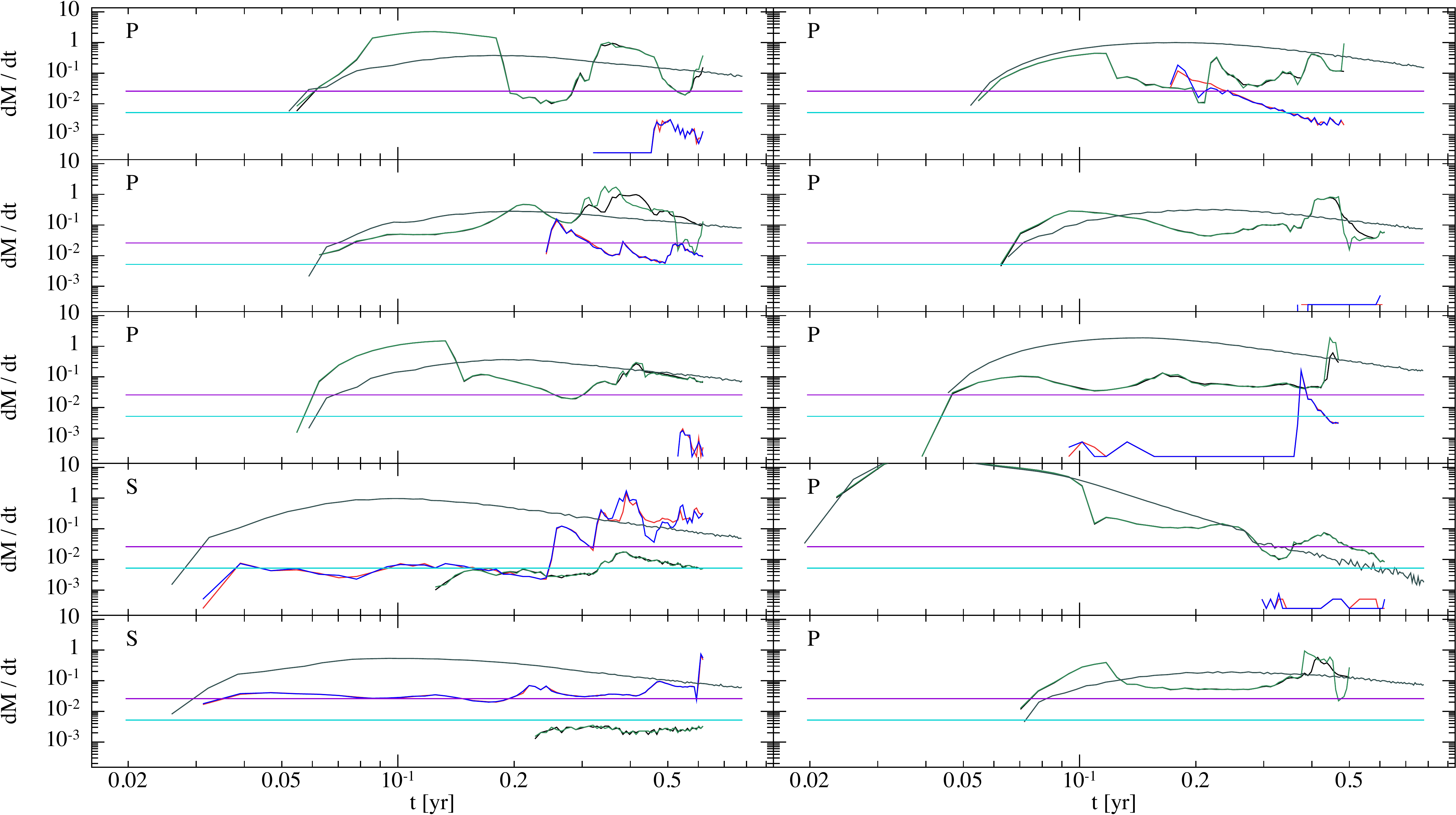} 
   \includegraphics[width=\textwidth]{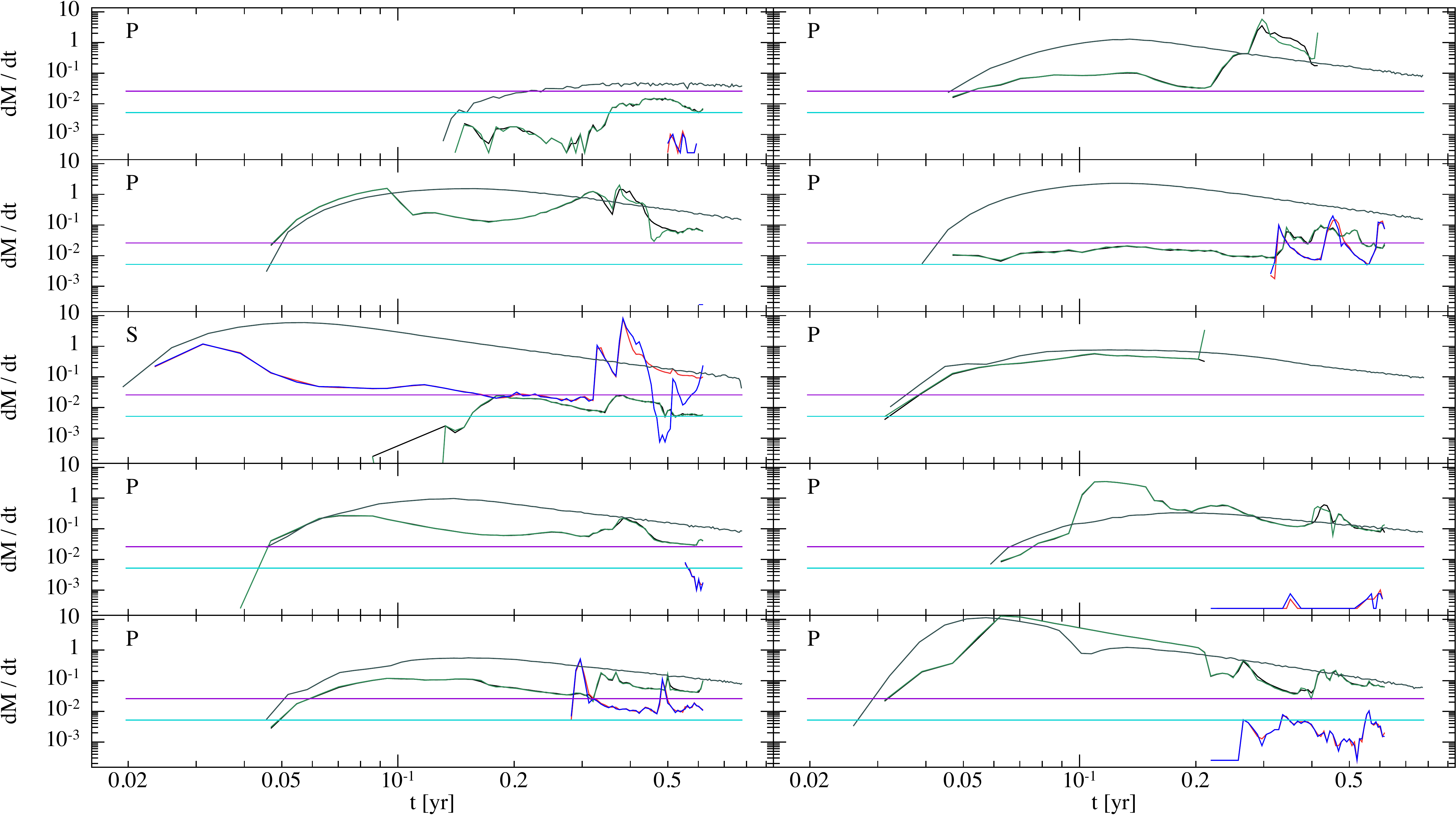} 
   \caption{The accretion rate, on a log-log scale, of the primary (black curve), the fallback rate of the primary (green curve), the accretion rate of the secondary (red curve), the fallback rate of the secondary (blue curve), and the accretion rate of the control (gray curve) for 20 different simulations. The disrupting SMBH is indicated by a ``P'' for primary and ``S'' for secondary in the top-left corner of each panel, accretion and fallback rates are measured in units of Solar masses per year, and time is measured in years (one binary orbit is roughly 0.3 years). The magenta and cyan lines represent the Eddington limit of the primary and secondary, respectively, assuming an accretion efficiency of 10\%. Simulations with no accretion whatsoever were complete ejections of the stream.   \newline\vspace{4in}}
   \label{fig:mdots_3}
\end{figure*}

\begin{figure*} 
   \centering
   \includegraphics[width=\textwidth]{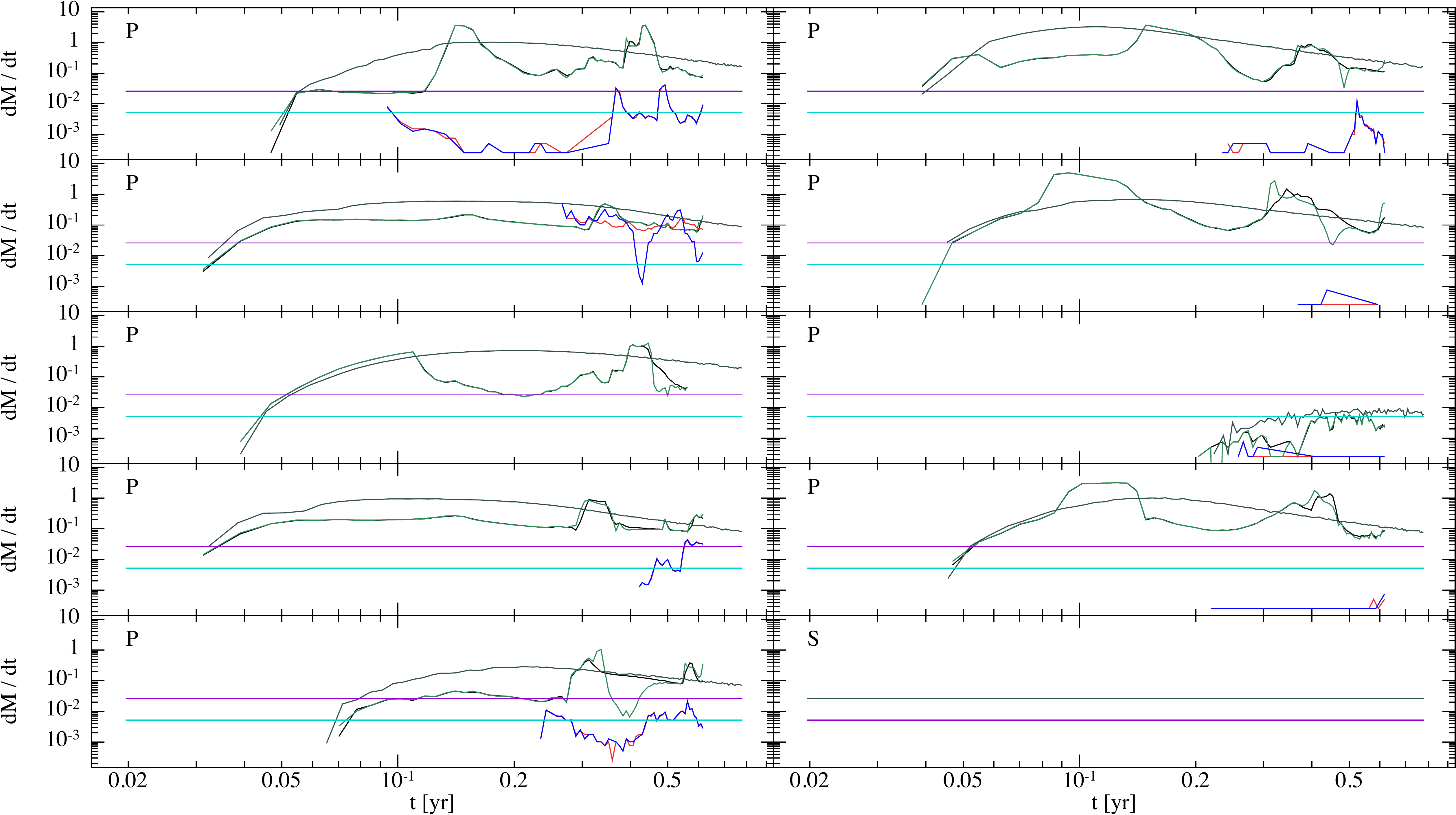} 
   \includegraphics[width=\textwidth]{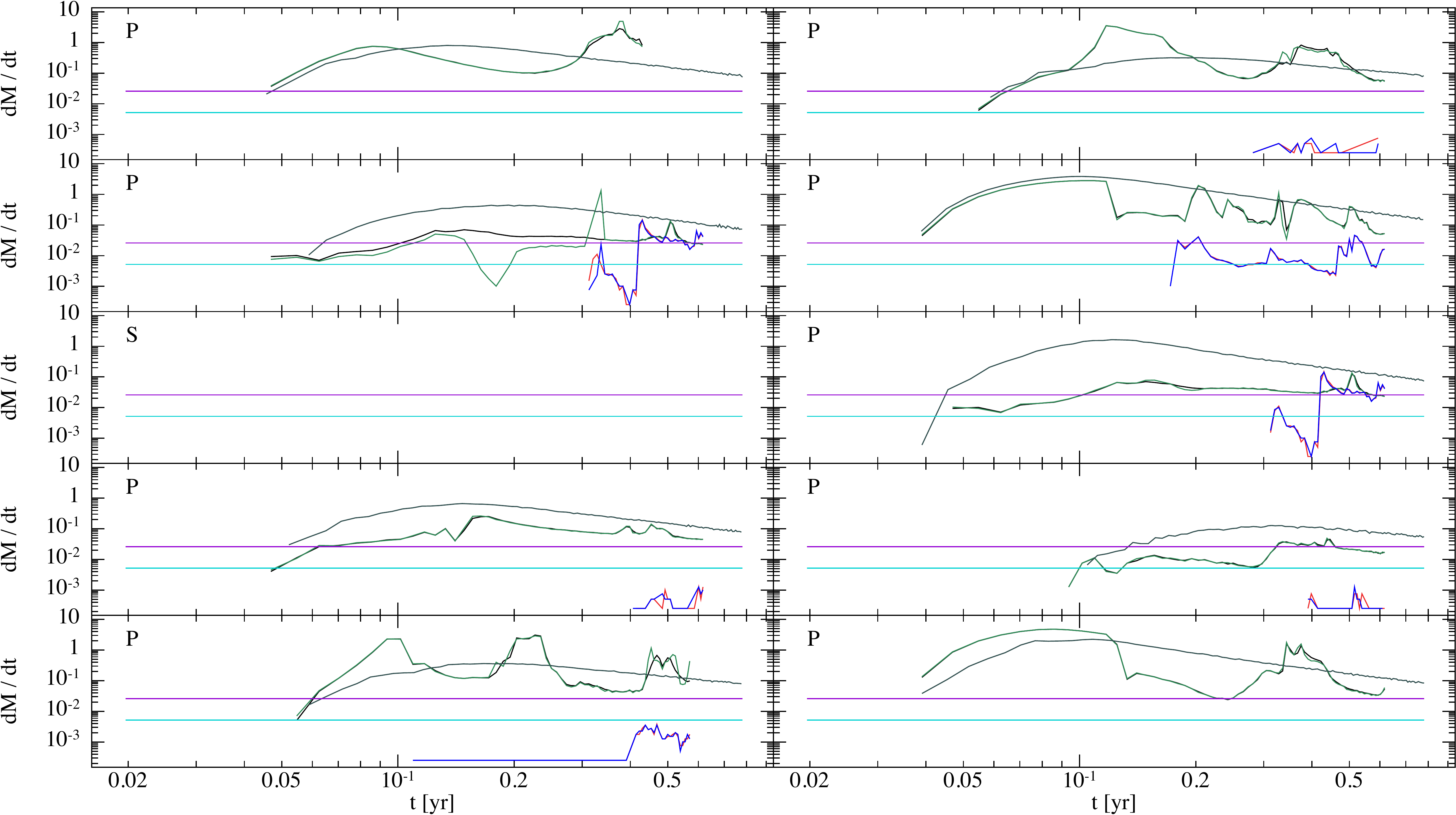} 
   \caption{The accretion rate, on a log-log scale, of the primary (black curve), the fallback rate of the primary (green curve), the accretion rate of the secondary (red curve), the fallback rate of the secondary (blue curve), and the accretion rate of the control (gray curve) for 20 different simulations. The disrupting SMBH is indicated by a ``P'' for primary and ``S'' for secondary in the top-left corner of each panel, accretion and fallback rates are measured in units of Solar masses per year, and time is measured in years (one binary orbit is roughly 0.3 years). The magenta and cyan lines represent the Eddington limit of the primary and secondary, respectively, assuming an accretion efficiency of 10\%. Simulations with no accretion whatsoever were complete ejections of the stream.}
   \label{fig:mdots_4}
\end{figure*}

\begin{figure*} 
   \centering
   \includegraphics[width=\textwidth]{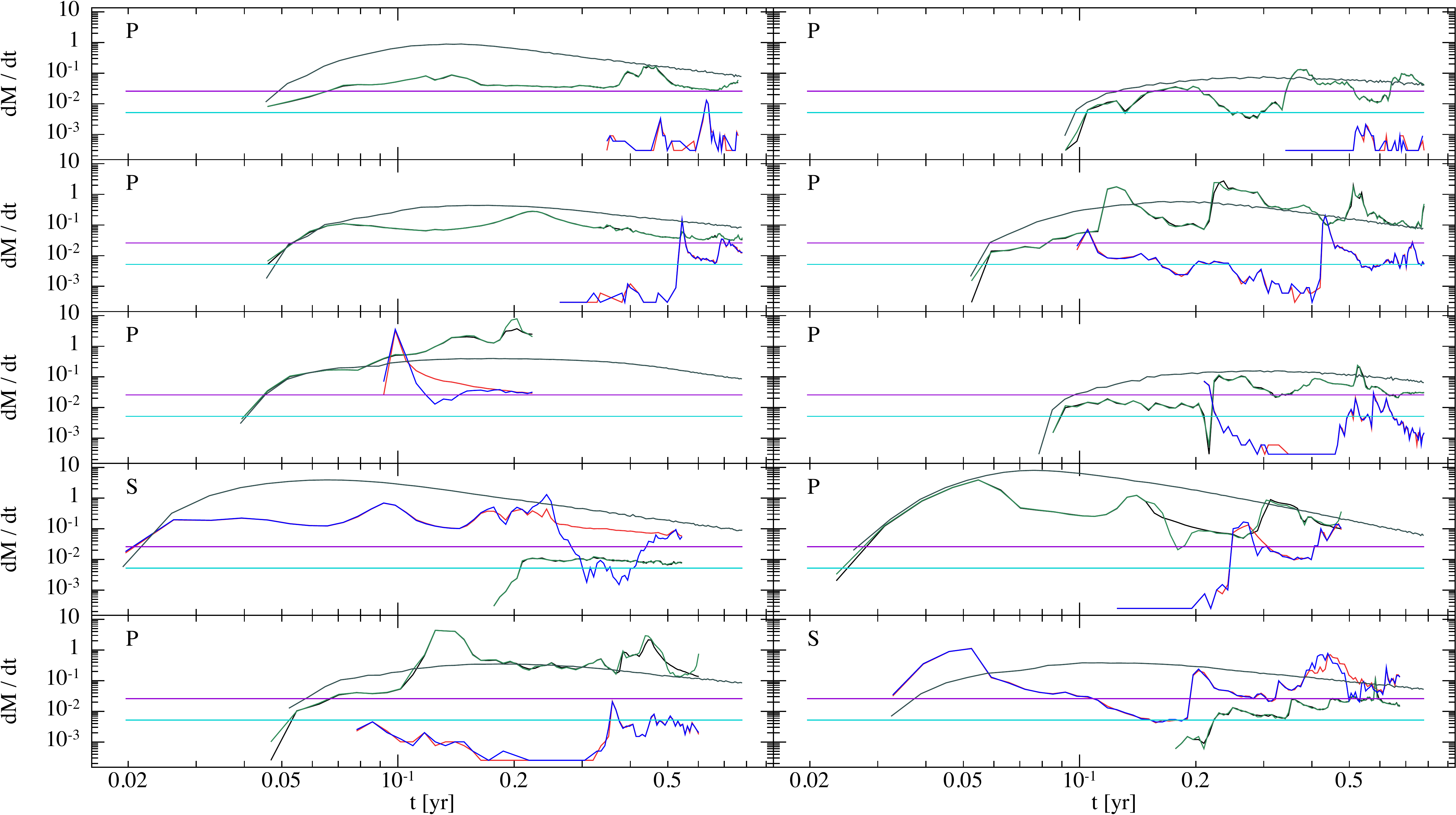} 
   \includegraphics[width=\textwidth]{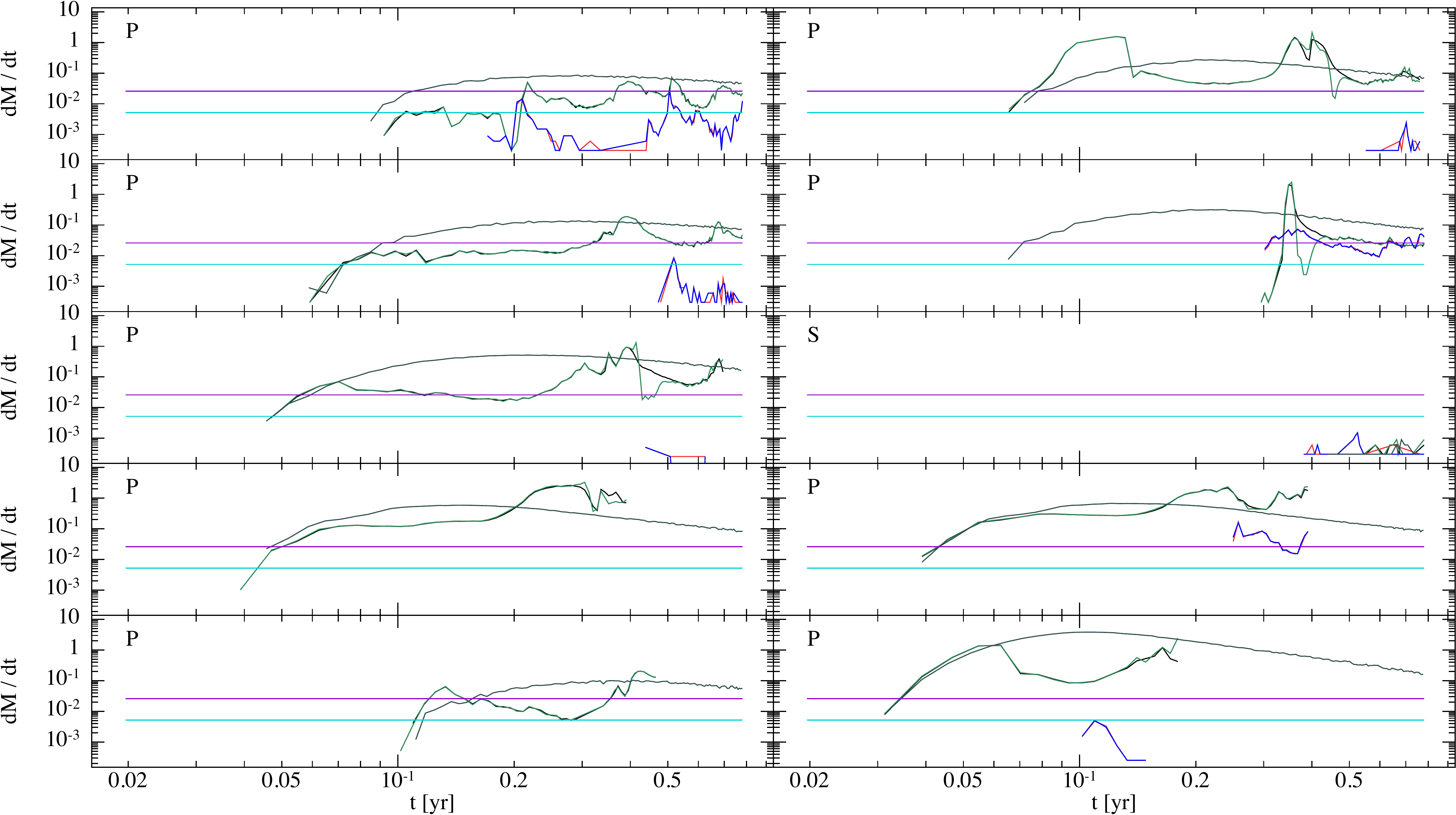} 
   \caption{The accretion rate, on a log-log scale, of the primary (black curve), the fallback rate of the primary (green curve), the accretion rate of the secondary (red curve), the fallback rate of the secondary (blue curve), and the accretion rate of the control (gray curve) for 20 different simulations. The disrupting SMBH is indicated by a ``P'' for primary and ``S'' for secondary in the top-left corner of each panel, accretion and fallback rates are measured in units of Solar masses per year, and time is measured in years (one binary orbit is roughly 0.3 years). The magenta and cyan lines represent the Eddington limit of the primary and secondary, respectively, assuming an accretion efficiency of 10\%. Simulations with no accretion whatsoever were complete ejections of the stream.}
   \label{fig:mdots_5}
\end{figure*}
\clearpage
\begin{figure*} 
   \centering
   \includegraphics[width=\textwidth]{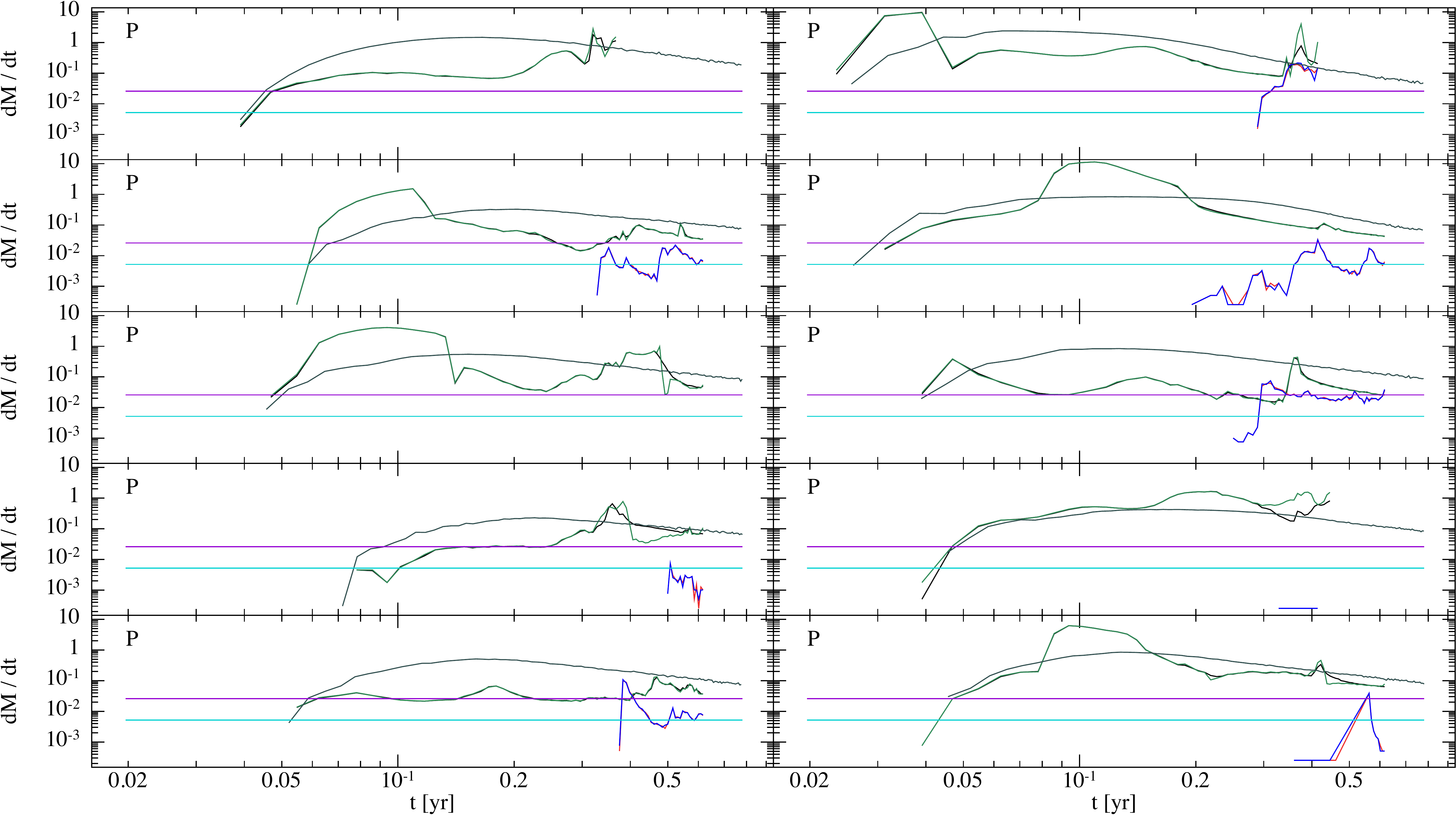} 
   \includegraphics[width=\textwidth]{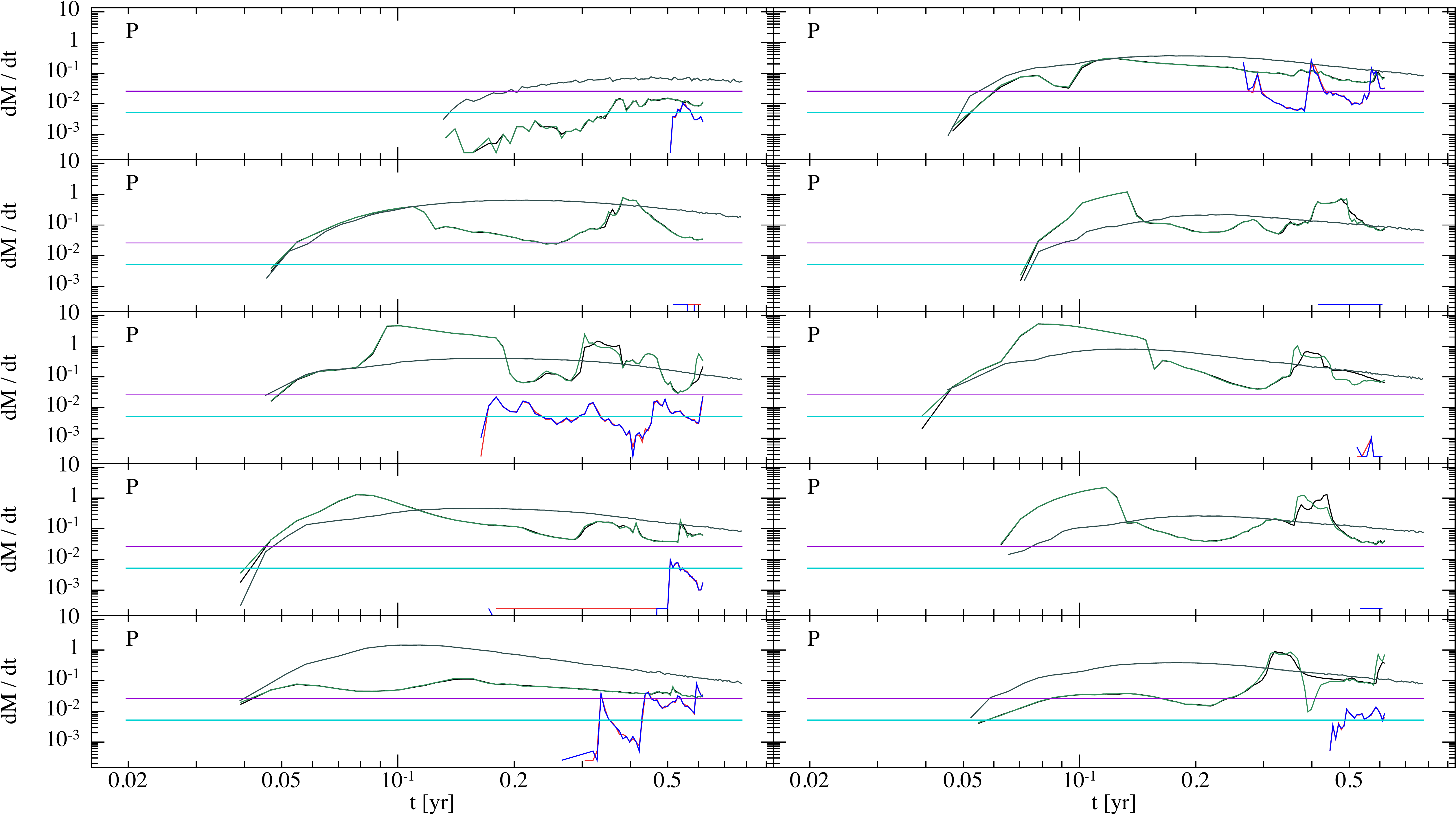} 
   \caption{The accretion rate, on a log-log scale, of the primary (black curve), the fallback rate of the primary (green curve), the accretion rate of the secondary (red curve), the fallback rate of the secondary (blue curve), and the accretion rate of the control (gray curve) for 20 different simulations. The disrupting SMBH is indicated by a ``P'' for primary and ``S'' for secondary in the top-left corner of each panel, accretion and fallback rates are measured in units of Solar masses per year, and time is measured in years (one binary orbit is roughly 0.3 years). The magenta and cyan lines represent the Eddington limit of the primary and secondary, respectively, assuming an accretion efficiency of 10\%. Simulations with no accretion whatsoever were complete ejections of the stream.}
   \label{fig:mdots_6}
\end{figure*}
\clearpage

\bibliographystyle{mnras}
\bibliography{tde_binary_references}

\begin{thebibliography}{}
\makeatletter
\relax
\def\mn@urlcharsother{\let\do\@makeother \do\$\do\&\do\#\do\^\do\_\do\%\do\~}
\def\mn@doi{\begingroup\mn@urlcharsother \@ifnextchar [ {\mn@doi@}
  {\mn@doi@[]}}
\def\mn@doi@[#1]#2{\def\@tempa{#1}\ifx\@tempa\@empty \href
  {http://dx.doi.org/#2} {doi:#2}\else \href {http://dx.doi.org/#2} {#1}\fi
  \endgroup}
\def\mn@eprint#1#2{\mn@eprint@#1:#2::\@nil}
\def\mn@eprint@arXiv#1{\href {http://arxiv.org/abs/#1} {{\tt arXiv:#1}}}
\def\mn@eprint@dblp#1{\href {http://dblp.uni-trier.de/rec/bibtex/#1.xml}
  {dblp:#1}}
\def\mn@eprint@#1:#2:#3:#4\@nil{\def\@tempa {#1}\def\@tempb {#2}\def\@tempc
  {#3}\ifx \@tempc \@empty \let \@tempc \@tempb \let \@tempb \@tempa \fi \ifx
  \@tempb \@empty \def\@tempb {arXiv}\fi \@ifundefined
  {mn@eprint@\@tempb}{\@tempb:\@tempc}{\expandafter \expandafter \csname
  mn@eprint@\@tempb\endcsname \expandafter{\@tempc}}}

\bibitem[\protect\citeauthoryear{{Antonini}, {Lombardi}  \&
  {Merritt}}{{Antonini} et~al.}{2011}]{antonini11}
{Antonini} F.,  {Lombardi} Jr. J.~C.,   {Merritt} D.,  2011, \mn@doi [\apj]
  {10.1088/0004-637X/731/2/128}, \href
  {http://adsabs.harvard.edu/abs/2011ApJ...731..128A} {731, 128}

\bibitem[\protect\citeauthoryear{{Bahcall} \& {Wolf}}{{Bahcall} \&
  {Wolf}}{1976}]{bahcall76}
{Bahcall} J.~N.,  {Wolf} R.~A.,  1976, \mn@doi [\apj] {10.1086/154711}, \href
  {http://adsabs.harvard.edu/abs/1976ApJ...209..214B} {209, 214}

\bibitem[\protect\citeauthoryear{{Barnes} \& {Hut}}{{Barnes} \&
  {Hut}}{1986}]{barnes86}
{Barnes} J.,  {Hut} P.,  1986, \mn@doi [\nat] {10.1038/324446a0}, \href
  {http://adsabs.harvard.edu/abs/1986Natur.324..446B} {324, 446}

\bibitem[\protect\citeauthoryear{{Begelman}, {Blandford}  \& {Rees}}{{Begelman}
  et~al.}{1980}]{begelman80}
{Begelman} M.~C.,  {Blandford} R.~D.,   {Rees} M.~J.,  1980, \mn@doi [\nat]
  {10.1038/287307a0}, \href {http://adsabs.harvard.edu/abs/1980Natur.287..307B}
  {287, 307}

\bibitem[\protect\citeauthoryear{{Bonnerot}, {Rossi}  \& {Lodato}}{{Bonnerot}
  et~al.}{2016a}]{bonnerot16b}
{Bonnerot} C.,  {Rossi} E.~M.,   {Lodato} G.,  2016a, \mn@doi [\mnras]
  {10.1093/mnras/stw2547}, \href
  {http://adsabs.harvard.edu/abs/2016MNRAS.tmp.1535B} {}

\bibitem[\protect\citeauthoryear{{Bonnerot}, {Rossi}, {Lodato}  \&
  {Price}}{{Bonnerot} et~al.}{2016b}]{bonnerot16}
{Bonnerot} C.,  {Rossi} E.~M.,  {Lodato} G.,   {Price} D.~J.,  2016b, \mn@doi
  [\mnras] {10.1093/mnras/stv2411}, \href
  {http://adsabs.harvard.edu/abs/2016MNRAS.455.2253B} {455, 2253}

\bibitem[\protect\citeauthoryear{{Cannizzo}, {Lee}  \& {Goodman}}{{Cannizzo}
  et~al.}{1990}]{cannizzo90}
{Cannizzo} J.~K.,  {Lee} H.~M.,   {Goodman} J.,  1990, \mn@doi [\apj]
  {10.1086/168442}, \href {http://adsabs.harvard.edu/abs/1990ApJ...351...38C}
  {351, 38}

\bibitem[\protect\citeauthoryear{{Carter} \& {Luminet}}{{Carter} \&
  {Luminet}}{1983}]{carter83}
{Carter} B.,  {Luminet} J.-P.,  1983, \aap, \href
  {http://adsabs.harvard.edu/abs/1983A%26A...121...97C} {121, 97}

\bibitem[\protect\citeauthoryear{{Chen}, {Liu}  \& {Magorrian}}{{Chen}
  et~al.}{2008}]{chen08}
{Chen} X.,  {Liu} F.~K.,   {Magorrian} J.,  2008, \mn@doi [\apj]
  {10.1086/527412}, \href {http://adsabs.harvard.edu/abs/2008ApJ...676...54C}
  {676, 54}

\bibitem[\protect\citeauthoryear{{Chen}, {Madau}, {Sesana}  \& {Liu}}{{Chen}
  et~al.}{2009}]{chen09}
{Chen} X.,  {Madau} P.,  {Sesana} A.,   {Liu} F.~K.,  2009, \mn@doi [\apjl]
  {10.1088/0004-637X/697/2/L149}, \href
  {http://adsabs.harvard.edu/abs/2009ApJ...697L.149C} {697, L149}

\bibitem[\protect\citeauthoryear{{Chen}, {Sesana}, {Madau}  \& {Liu}}{{Chen}
  et~al.}{2011}]{chen11}
{Chen} X.,  {Sesana} A.,  {Madau} P.,   {Liu} F.~K.,  2011, \mn@doi [\apj]
  {10.1088/0004-637X/729/1/13}, \href
  {http://adsabs.harvard.edu/abs/2011ApJ...729...13C} {729, 13}

\bibitem[\protect\citeauthoryear{{Chen}, {G{\'o}mez-Vargas}  \&
  {Guillochon}}{{Chen} et~al.}{2016}]{chen16}
{Chen} X.,  {G{\'o}mez-Vargas} G.~A.,   {Guillochon} J.,  2016, \mn@doi
  [\mnras] {10.1093/mnras/stw437}, \href
  {http://adsabs.harvard.edu/abs/2016MNRAS.458.3314C} {458, 3314}

\bibitem[\protect\citeauthoryear{{Coughlin} \& {Begelman}}{{Coughlin} \&
  {Begelman}}{2014}]{coughlin14}
{Coughlin} E.~R.,  {Begelman} M.~C.,  2014, \mn@doi [\apj]
  {10.1088/0004-637X/781/2/82}, \href
  {http://adsabs.harvard.edu/abs/2014ApJ...781...82C} {781, 82}

\bibitem[\protect\citeauthoryear{{Coughlin} \& {Nixon}}{{Coughlin} \&
  {Nixon}}{2015}]{coughlin15}
{Coughlin} E.~R.,  {Nixon} C.,  2015, \mn@doi [\apjl]
  {10.1088/2041-8205/808/1/L11}, \href
  {http://adsabs.harvard.edu/abs/2015ApJ...808L..11C} {808, L11}

\bibitem[\protect\citeauthoryear{{Coughlin}, {Nixon}, {Begelman}, {Armitage}
  \& {Price}}{{Coughlin} et~al.}{2016a}]{coughlin16a}
{Coughlin} E.~R.,  {Nixon} C.,  {Begelman} M.~C.,  {Armitage} P.~J.,   {Price}
  D.~J.,  2016a, \mn@doi [\mnras] {10.1093/mnras/stv2511}, \href
  {http://adsabs.harvard.edu/abs/2016MNRAS.455.3612C} {455, 3612}

\bibitem[\protect\citeauthoryear{{Coughlin}, {Nixon}, {Begelman}  \&
  {Armitage}}{{Coughlin} et~al.}{2016b}]{coughlin16b}
{Coughlin} E.~R.,  {Nixon} C.,  {Begelman} M.~C.,   {Armitage} P.~J.,  2016b,
  \mn@doi [\mnras] {10.1093/mnras/stw770}, \href
  {http://adsabs.harvard.edu/abs/2016MNRAS.459.3089C} {459, 3089}

\bibitem[\protect\citeauthoryear{{Cullen} \& {Dehnen}}{{Cullen} \&
  {Dehnen}}{2010}]{cullen10}
{Cullen} L.,  {Dehnen} W.,  2010, \mn@doi [\mnras]
  {10.1111/j.1365-2966.2010.17158.x}, \href
  {http://adsabs.harvard.edu/abs/2010MNRAS.408..669C} {408, 669}

\bibitem[\protect\citeauthoryear{{Dai}, {Escala}  \& {Coppi}}{{Dai}
  et~al.}{2013}]{dai13}
{Dai} L.,  {Escala} A.,   {Coppi} P.,  2013, \mn@doi [\apjl]
  {10.1088/2041-8205/775/1/L9}, \href
  {http://adsabs.harvard.edu/abs/2013ApJ...775L...9D} {775, L9}

\bibitem[\protect\citeauthoryear{{Dai}, {McKinney}  \& {Miller}}{{Dai}
  et~al.}{2015}]{dai15}
{Dai} L.,  {McKinney} J.~C.,   {Miller} M.~C.,  2015, \mn@doi [\apjl]
  {10.1088/2041-8205/812/2/L39}, \href
  {http://adsabs.harvard.edu/abs/2015ApJ...812L..39D} {812, L39}

\bibitem[\protect\citeauthoryear{{Diener}, {Kosovichev}, {Kotok}, {Novikov}  \&
  {Pethick}}{{Diener} et~al.}{1995}]{diener95}
{Diener} P.,  {Kosovichev} A.~G.,  {Kotok} E.~V.,  {Novikov} I.~D.,   {Pethick}
  C.~J.,  1995, \mn@doi [\mnras] {10.1093/mnras/275.2.498}, \href
  {http://adsabs.harvard.edu/abs/1995MNRAS.275..498D} {275, 498}

\bibitem[\protect\citeauthoryear{{Do{\u g}an}, {Nixon}, {King}  \&
  {Price}}{{Do{\u g}an} et~al.}{2015}]{dogan15}
{Do{\u g}an} S.,  {Nixon} C.,  {King} A.,   {Price} D.~J.,  2015, \mn@doi
  [\mnras] {10.1093/mnras/stv347}, \href
  {http://adsabs.harvard.edu/abs/2015MNRAS.449.1251D} {449, 1251}

\bibitem[\protect\citeauthoryear{{Eggleton}}{{Eggleton}}{1983}]{eggleton83}
{Eggleton} P.~P.,  1983, \mn@doi [\apj] {10.1086/160960}, \href
  {http://adsabs.harvard.edu/abs/1983ApJ...268..368E} {268, 368}

\bibitem[\protect\citeauthoryear{{Evans} \& {Kochanek}}{{Evans} \&
  {Kochanek}}{1989}]{evans89}
{Evans} C.~R.,  {Kochanek} C.~S.,  1989, \mn@doi [\apjl] {10.1086/185567},
  \href {http://esoads.eso.org/abs/1989ApJ...346L..13E} {346, L13}

\bibitem[\protect\citeauthoryear{{Fragner} \& {Nelson}}{{Fragner} \&
  {Nelson}}{2010}]{fragner10}
{Fragner} M.~M.,  {Nelson} R.~P.,  2010, \mn@doi [\aap]
  {10.1051/0004-6361/200913088}, \href
  {http://adsabs.harvard.edu/abs/2010A%26A...511A..77F} {511, A77}

\bibitem[\protect\citeauthoryear{{Frank} \& {Rees}}{{Frank} \&
  {Rees}}{1976}]{frank76}
{Frank} J.,  {Rees} M.~J.,  1976, \mn@doi [\mnras] {10.1093/mnras/176.3.633},
  \href {http://adsabs.harvard.edu/abs/1976MNRAS.176..633F} {176, 633}

\bibitem[\protect\citeauthoryear{{Gafton} \& {Rosswog}}{{Gafton} \&
  {Rosswog}}{2011}]{gafton11}
{Gafton} E.,  {Rosswog} S.,  2011, \mn@doi [\mnras]
  {10.1111/j.1365-2966.2011.19528.x}, \href
  {http://adsabs.harvard.edu/abs/2011MNRAS.418..770G} {418, 770}

\bibitem[\protect\citeauthoryear{{Graham} \& {Scott}}{{Graham} \&
  {Scott}}{2013}]{graham13}
{Graham} A.~W.,  {Scott} N.,  2013, \mn@doi [\apj]
  {10.1088/0004-637X/764/2/151}, \href
  {http://adsabs.harvard.edu/abs/2013ApJ...764..151G} {764, 151}

\bibitem[\protect\citeauthoryear{{Guillochon} \& {McCourt}}{{Guillochon} \&
  {McCourt}}{2016}]{guillochon16b}
{Guillochon} J.,  {McCourt} M.,  2016, preprint, \href
  {http://adsabs.harvard.edu/abs/2016arXiv160908160G} {} (\mn@eprint {arXiv}
  {1609.08160})

\bibitem[\protect\citeauthoryear{{Guillochon} \& {Ramirez-Ruiz}}{{Guillochon}
  \& {Ramirez-Ruiz}}{2013}]{guillochon13}
{Guillochon} J.,  {Ramirez-Ruiz} E.,  2013, \mn@doi [\apj]
  {10.1088/0004-637X/767/1/25}, \href
  {http://adsabs.harvard.edu/abs/2013ApJ...767...25G} {767, 25}

\bibitem[\protect\citeauthoryear{{Guillochon} \& {Ramirez-Ruiz}}{{Guillochon}
  \& {Ramirez-Ruiz}}{2015}]{guillochon15}
{Guillochon} J.,  {Ramirez-Ruiz} E.,  2015, \mn@doi [\apj]
  {10.1088/0004-637X/809/2/166}, \href
  {http://adsabs.harvard.edu/abs/2015ApJ...809..166G} {809, 166}

\bibitem[\protect\citeauthoryear{{Guillochon}, {Loeb}, {MacLeod}  \&
  {Ramirez-Ruiz}}{{Guillochon} et~al.}{2014}]{guillochon14}
{Guillochon} J.,  {Loeb} A.,  {MacLeod} M.,   {Ramirez-Ruiz} E.,  2014, \mn@doi
  [\apjl] {10.1088/2041-8205/786/2/L12}, \href
  {http://adsabs.harvard.edu/abs/2014ApJ...786L..12G} {786, L12}

\bibitem[\protect\citeauthoryear{{Guillochon}, {McCourt}, {Chen}, {Johnson}  \&
  {Berger}}{{Guillochon} et~al.}{2016}]{guillochon16}
{Guillochon} J.,  {McCourt} M.,  {Chen} X.,  {Johnson} M.~D.,   {Berger} E.,
  2016, \mn@doi [\apj] {10.3847/0004-637X/822/1/48}, \href
  {http://adsabs.harvard.edu/abs/2016ApJ...822...48G} {822, 48}

\bibitem[\protect\citeauthoryear{{G{\"u}ltekin} et~al.,}{{G{\"u}ltekin}
  et~al.}{2009}]{gultekin09}
{G{\"u}ltekin} K.,  et~al., 2009, \mn@doi [\apj] {10.1088/0004-637X/698/1/198},
  \href {http://adsabs.harvard.edu/abs/2009ApJ...698..198G} {698, 198}

\bibitem[\protect\citeauthoryear{{Hansen}, {Kawaler}  \& {Trimble}}{{Hansen}
  et~al.}{2004}]{hansen04}
{Hansen} C.~J.,  {Kawaler} S.~D.,   {Trimble} V.,  2004, {Stellar interiors :
  physical principles, structure, and evolution}

\bibitem[\protect\citeauthoryear{{Hayasaki} \& {Loeb}}{{Hayasaki} \&
  {Loeb}}{2015}]{hayasaki16b}
{Hayasaki} K.,  {Loeb} A.,  2015, preprint, \href
  {http://esoads.eso.org/abs/2015arXiv151005760H} {} (\mn@eprint {arXiv}
  {1510.05760})

\bibitem[\protect\citeauthoryear{{Hayasaki}, {Stone}  \& {Loeb}}{{Hayasaki}
  et~al.}{2013}]{hayasaki13}
{Hayasaki} K.,  {Stone} N.,   {Loeb} A.,  2013, \mn@doi [\mnras]
  {10.1093/mnras/stt871}, \href
  {http://adsabs.harvard.edu/abs/2013MNRAS.434..909H} {434, 909}

\bibitem[\protect\citeauthoryear{{Hayasaki}, {Stone}  \& {Loeb}}{{Hayasaki}
  et~al.}{2016}]{hayasaki16}
{Hayasaki} K.,  {Stone} N.,   {Loeb} A.,  2016, \mn@doi [\mnras]
  {10.1093/mnras/stw1387}, \href
  {http://adsabs.harvard.edu/abs/2016MNRAS.461.3760H} {461, 3760}

\bibitem[\protect\citeauthoryear{{Hills}}{{Hills}}{1975}]{hills75}
{Hills} J.~G.,  1975, \mn@doi [\nat] {10.1038/254295a0}, \href
  {http://adsabs.harvard.edu/abs/1975Natur.254..295H} {254, 295}

\bibitem[\protect\citeauthoryear{{Hills}}{{Hills}}{1988}]{hills88}
{Hills} J.~G.,  1988, \mn@doi [\nat] {10.1038/331687a0}, \href
  {http://adsabs.harvard.edu/abs/1988Natur.331..687H} {331, 687}

\bibitem[\protect\citeauthoryear{{Ivanov}, {Polnarev}  \& {Saha}}{{Ivanov}
  et~al.}{2005}]{ivanov05}
{Ivanov} P.~B.,  {Polnarev} A.~G.,   {Saha} P.,  2005, \mn@doi [\mnras]
  {10.1111/j.1365-2966.2005.08843.x}, \href
  {http://adsabs.harvard.edu/abs/2005MNRAS.358.1361I} {358, 1361}

\bibitem[\protect\citeauthoryear{{Ivezic}, {Tyson}, {Abel}, {Acosta},
  {Allsman}, {AlSayyad}, {Anderson}  \& {Andrew}}{{Ivezic}
  et~al.}{2008}]{ivezic08}
{Ivezic} Z.,  {Tyson} J.~A.,  {Abel} B.,  {Acosta} E.,  {Allsman} R.,
  {AlSayyad} Y.,  {Anderson} S.~F.,   {Andrew} J.,  2008, preprint, \href
  {http://adsabs.harvard.edu/abs/2008arXiv0805.2366I} {} (\mn@eprint {arXiv}
  {0805.2366})

\bibitem[\protect\citeauthoryear{{Jiang}, {Guillochon}  \& {Loeb}}{{Jiang}
  et~al.}{2016}]{jiang16}
{Jiang} Y.-F.,  {Guillochon} J.,   {Loeb} A.,  2016, \mn@doi [\apj]
  {10.3847/0004-637X/830/2/125}, \href
  {http://adsabs.harvard.edu/abs/2016ApJ...830..125J} {830, 125}

\bibitem[\protect\citeauthoryear{{Kasen} \& {Ramirez-Ruiz}}{{Kasen} \&
  {Ramirez-Ruiz}}{2010}]{kasen10}
{Kasen} D.,  {Ramirez-Ruiz} E.,  2010, \mn@doi [\apj]
  {10.1088/0004-637X/714/1/155}, \href
  {http://adsabs.harvard.edu/abs/2010ApJ...714..155K} {714, 155}

\bibitem[\protect\citeauthoryear{{Kelley}, {Tchekhovskoy}  \&
  {Narayan}}{{Kelley} et~al.}{2014}]{kelley14}
{Kelley} L.~Z.,  {Tchekhovskoy} A.,   {Narayan} R.,  2014, \mn@doi [\mnras]
  {10.1093/mnras/stu2041}, \href
  {http://adsabs.harvard.edu/abs/2014MNRAS.445.3919K} {445, 3919}

\bibitem[\protect\citeauthoryear{{Kim}, {Park}  \& {Lee}}{{Kim}
  et~al.}{1999}]{kim99}
{Kim} S.~S.,  {Park} M.-G.,   {Lee} H.~M.,  1999, \mn@doi [\apj]
  {10.1086/307394}, \href {http://adsabs.harvard.edu/abs/1999ApJ...519..647K}
  {519, 647}

\bibitem[\protect\citeauthoryear{{Kochanek}}{{Kochanek}}{1994}]{kochanek94}
{Kochanek} C.~S.,  1994, \mn@doi [\apj] {10.1086/173745}, \href
  {http://adsabs.harvard.edu/abs/1994ApJ...422..508K} {422, 508}

\bibitem[\protect\citeauthoryear{{Komossa}}{{Komossa}}{2015}]{komossa15}
{Komossa} S.,  2015, \mn@doi [Journal of High Energy Astrophysics]
  {10.1016/j.jheap.2015.04.006}, \href
  {http://adsabs.harvard.edu/abs/2015JHEAp...7..148K} {7, 148}

\bibitem[\protect\citeauthoryear{{Kozai}}{{Kozai}}{1962}]{kozai62}
{Kozai} Y.,  1962, \mn@doi [\aj] {10.1086/108790}, \href
  {http://adsabs.harvard.edu/abs/1962AJ.....67..591K} {67, 591}

\bibitem[\protect\citeauthoryear{{Kroupa}}{{Kroupa}}{2001}]{kroupa01}
{Kroupa} P.,  2001, \mn@doi [\mnras] {10.1046/j.1365-8711.2001.04022.x}, \href
  {http://adsabs.harvard.edu/abs/2001MNRAS.322..231K} {322, 231}

\bibitem[\protect\citeauthoryear{{Larwood}, {Nelson}, {Papaloizou}  \&
  {Terquem}}{{Larwood} et~al.}{1996}]{larwood96}
{Larwood} J.~D.,  {Nelson} R.~P.,  {Papaloizou} J.~C.~B.,   {Terquem} C.,
  1996, \mn@doi [\mnras] {10.1093/mnras/282.2.597}, \href
  {http://adsabs.harvard.edu/abs/1996MNRAS.282..597L} {282, 597}

\bibitem[\protect\citeauthoryear{{Li}, {Naoz}, {Kocsis}  \& {Loeb}}{{Li}
  et~al.}{2015}]{li15}
{Li} G.,  {Naoz} S.,  {Kocsis} B.,   {Loeb} A.,  2015, \mn@doi [\mnras]
  {10.1093/mnras/stv1031}, \href
  {http://adsabs.harvard.edu/abs/2015MNRAS.451.1341L} {451, 1341}

\bibitem[\protect\citeauthoryear{{Lidov}}{{Lidov}}{1962}]{lidov62}
{Lidov} M.~L.,  1962, \mn@doi [\planss] {10.1016/0032-0633(62)90129-0}, \href
  {http://adsabs.harvard.edu/abs/1962P%26SS....9..719L} {9, 719}

\bibitem[\protect\citeauthoryear{{Lightman} \& {Shapiro}}{{Lightman} \&
  {Shapiro}}{1977}]{lightman77}
{Lightman} A.~P.,  {Shapiro} S.~L.,  1977, \mn@doi [\apj] {10.1086/154925},
  \href {http://adsabs.harvard.edu/abs/1977ApJ...211..244L} {211, 244}

\bibitem[\protect\citeauthoryear{{Liu}, {Li}  \& {Chen}}{{Liu}
  et~al.}{2009}]{liu11}
{Liu} F.~K.,  {Li} S.,   {Chen} X.,  2009, \mn@doi [\apjl]
  {10.1088/0004-637X/706/1/L133}, \href
  {http://adsabs.harvard.edu/abs/2009ApJ...706L.133L} {706, L133}

\bibitem[\protect\citeauthoryear{{Liu}, {Li}  \& {Komossa}}{{Liu}
  et~al.}{2014}]{liu14}
{Liu} F.~K.,  {Li} S.,   {Komossa} S.,  2014, \mn@doi [\apj]
  {10.1088/0004-637X/786/2/103}, \href
  {http://adsabs.harvard.edu/abs/2014ApJ...786..103L} {786, 103}

\bibitem[\protect\citeauthoryear{{Lodato} \& {Price}}{{Lodato} \&
  {Price}}{2010}]{lodato10}
{Lodato} G.,  {Price} D.~J.,  2010, \mn@doi [\mnras]
  {10.1111/j.1365-2966.2010.16526.x}, \href
  {http://adsabs.harvard.edu/abs/2010MNRAS.405.1212L} {405, 1212}

\bibitem[\protect\citeauthoryear{{Lodato} \& {Rossi}}{{Lodato} \&
  {Rossi}}{2011}]{lodato11}
{Lodato} G.,  {Rossi} E.~M.,  2011, \mn@doi [\mnras]
  {10.1111/j.1365-2966.2010.17448.x}, \href
  {http://adsabs.harvard.edu/abs/2011MNRAS.410..359L} {410, 359}

\bibitem[\protect\citeauthoryear{{Lodato}, {King}  \& {Pringle}}{{Lodato}
  et~al.}{2009}]{lodato09}
{Lodato} G.,  {King} A.~R.,   {Pringle} J.~E.,  2009, \mn@doi [\mnras]
  {10.1111/j.1365-2966.2008.14049.x}, \href
  {http://adsabs.harvard.edu/abs/2009MNRAS.392..332L} {392, 332}

\bibitem[\protect\citeauthoryear{{Loeb} \& {Ulmer}}{{Loeb} \&
  {Ulmer}}{1997}]{loeb97}
{Loeb} A.,  {Ulmer} A.,  1997, \apj, \href
  {http://adsabs.harvard.edu/abs/1997ApJ...489..573L} {489, 573}

\bibitem[\protect\citeauthoryear{{Magorrian} \& {Tremaine}}{{Magorrian} \&
  {Tremaine}}{1999}]{magorrian99}
{Magorrian} J.,  {Tremaine} S.,  1999, \mn@doi [\mnras]
  {10.1046/j.1365-8711.1999.02853.x}, \href
  {http://adsabs.harvard.edu/abs/1999MNRAS.309..447M} {309, 447}

\bibitem[\protect\citeauthoryear{{Manukian}, {Guillochon}, {Ramirez-Ruiz}  \&
  {O'Leary}}{{Manukian} et~al.}{2013}]{manukian13}
{Manukian} H.,  {Guillochon} J.,  {Ramirez-Ruiz} E.,   {O'Leary} R.~M.,  2013,
  \mn@doi [\apjl] {10.1088/2041-8205/771/2/L28}, \href
  {http://adsabs.harvard.edu/abs/2013ApJ...771L..28M} {771, L28}

\bibitem[\protect\citeauthoryear{{Martin}, {Nixon}, {Armitage}, {Lubow}  \&
  {Price}}{{Martin} et~al.}{2014a}]{martin14a}
{Martin} R.~G.,  {Nixon} C.,  {Armitage} P.~J.,  {Lubow} S.~H.,   {Price}
  D.~J.,  2014a, \mn@doi [\apjl] {10.1088/2041-8205/790/2/L34}, \href
  {http://adsabs.harvard.edu/abs/2014ApJ...790L..34M} {790, L34}

\bibitem[\protect\citeauthoryear{{Martin}, {Nixon}, {Lubow}, {Armitage},
  {Price}, {Do{\u g}an}  \& {King}}{{Martin} et~al.}{2014b}]{martin14b}
{Martin} R.~G.,  {Nixon} C.,  {Lubow} S.~H.,  {Armitage} P.~J.,  {Price} D.~J.,
   {Do{\u g}an} S.,   {King} A.,  2014b, \mn@doi [\apjl]
  {10.1088/2041-8205/792/2/L33}, \href
  {http://adsabs.harvard.edu/abs/2014ApJ...792L..33M} {792, L33}

\bibitem[\protect\citeauthoryear{{Martin}, {Lubow}, {Nixon}  \&
  {Armitage}}{{Martin} et~al.}{2016}]{martin16}
{Martin} R.~G.,  {Lubow} S.~H.,  {Nixon} C.,   {Armitage} P.~J.,  2016, \mn@doi
  [\mnras] {10.1093/mnras/stw605}, \href
  {http://adsabs.harvard.edu/abs/2016MNRAS.458.4345M} {458, 4345}

\bibitem[\protect\citeauthoryear{{M{\"u}ller-S{\'a}nchez}, {Comerford},
  {Nevin}, {Barrows}, {Cooper}  \& {Greene}}{{M{\"u}ller-S{\'a}nchez}
  et~al.}{2015}]{muller15}
{M{\"u}ller-S{\'a}nchez} F.,  {Comerford} J.~M.,  {Nevin} R.,  {Barrows} R.~S.,
   {Cooper} M.~C.,   {Greene} J.~E.,  2015, \mn@doi [\apj]
  {10.1088/0004-637X/813/2/103}, \href
  {http://adsabs.harvard.edu/abs/2015ApJ...813..103M} {813, 103}

\bibitem[\protect\citeauthoryear{{Nealon}, {Price}  \& {Nixon}}{{Nealon}
  et~al.}{2015}]{nealon15}
{Nealon} R.,  {Price} D.~J.,   {Nixon} C.~J.,  2015, \mn@doi [\mnras]
  {10.1093/mnras/stv014}, \href
  {http://adsabs.harvard.edu/abs/2015MNRAS.448.1526N} {448, 1526}

\bibitem[\protect\citeauthoryear{{Nixon} \& {King}}{{Nixon} \&
  {King}}{2016}]{nixon16}
{Nixon} C.,  {King} A.,  2016, in {Haardt} F.,  {Gorini} V.,  {Moschella} U.,
  {Treves} A.,   {Colpi} M.,  eds,  Lecture Notes in Physics, Berlin Springer
  Verlag Vol. 905, Lecture Notes in Physics, Berlin Springer Verlag. p.~45
  (\mn@eprint {arXiv} {1505.07827}), \mn@doi{10.1007/978-3-319-19416-5_2}

\bibitem[\protect\citeauthoryear{{Nixon}, {King}, {Price}  \& {Frank}}{{Nixon}
  et~al.}{2012}]{nixon12}
{Nixon} C.,  {King} A.,  {Price} D.,   {Frank} J.,  2012, \mn@doi [\apjl]
  {10.1088/2041-8205/757/2/L24}, \href
  {http://adsabs.harvard.edu/abs/2012ApJ...757L..24N} {757, L24}

\bibitem[\protect\citeauthoryear{{Nixon}, {King}  \& {Price}}{{Nixon}
  et~al.}{2013}]{nixon13}
{Nixon} C.,  {King} A.,   {Price} D.,  2013, \mn@doi [\mnras]
  {10.1093/mnras/stt1136}, \href
  {http://adsabs.harvard.edu/abs/2013MNRAS.434.1946N} {434, 1946}

\bibitem[\protect\citeauthoryear{{Peebles}}{{Peebles}}{1972}]{peebles72}
{Peebles} P.~J.~E.,  1972, \mn@doi [\apj] {10.1086/151797}, \href
  {http://adsabs.harvard.edu/abs/1972ApJ...178..371P} {178, 371}

\bibitem[\protect\citeauthoryear{{Peters} \& {Mathews}}{{Peters} \&
  {Mathews}}{1963}]{peters63}
{Peters} P.~C.,  {Mathews} J.,  1963, \mn@doi [Physical Review]
  {10.1103/PhysRev.131.435}, \href
  {http://adsabs.harvard.edu/abs/1963PhRv..131..435P} {131, 435}

\bibitem[\protect\citeauthoryear{{Phinney}}{{Phinney}}{1989}]{phinney89}
{Phinney} E.~S.,  1989, in {Morris} M.,  ed.,  IAU Symposium Vol. 136, The
  Center of the Galaxy. p.~543

\bibitem[\protect\citeauthoryear{{Polnarev} \& {Rees}}{{Polnarev} \&
  {Rees}}{1994}]{polnarev94}
{Polnarev} A.~G.,  {Rees} M.~J.,  1994, \aap, \href
  {http://adsabs.harvard.edu/abs/1994A%26A...283..301P} {283, 301}

\bibitem[\protect\citeauthoryear{{Price} \& {Federrath}}{{Price} \&
  {Federrath}}{2010}]{price10}
{Price} D.~J.,  {Federrath} C.,  2010, \mn@doi [\mnras]
  {10.1111/j.1365-2966.2010.16810.x}, \href
  {http://adsabs.harvard.edu/abs/2010MNRAS.406.1659P} {406, 1659}

\bibitem[\protect\citeauthoryear{{Ramirez-Ruiz} \& {Rosswog}}{{Ramirez-Ruiz} \&
  {Rosswog}}{2009}]{ramirez-ruiz09}
{Ramirez-Ruiz} E.,  {Rosswog} S.,  2009, \mn@doi [\apjl]
  {10.1088/0004-637X/697/2/L77}, \href
  {http://adsabs.harvard.edu/abs/2009ApJ...697L..77R} {697, L77}

\bibitem[\protect\citeauthoryear{{Rees}}{{Rees}}{1988}]{rees88}
{Rees} M.~J.,  1988, \mn@doi [\nat] {10.1038/333523a0}, \href
  {http://adsabs.harvard.edu/abs/1988Natur.333..523R} {333, 523}

\bibitem[\protect\citeauthoryear{{Ricarte}, {Natarajan}, {Dai}  \&
  {Coppi}}{{Ricarte} et~al.}{2016}]{ricarte16}
{Ricarte} A.,  {Natarajan} P.,  {Dai} L.,   {Coppi} P.,  2016, \mn@doi [\mnras]
  {10.1093/mnras/stw355}, \href {http://esoads.eso.org/abs/2016MNRAS.458.1712R}
  {458, 1712}

\bibitem[\protect\citeauthoryear{{Rosswog}, {Ramirez-Ruiz}  \& {Hix}}{{Rosswog}
  et~al.}{2009}]{rosswog09}
{Rosswog} S.,  {Ramirez-Ruiz} E.,   {Hix} W.~R.,  2009, \mn@doi [\apj]
  {10.1088/0004-637X/695/1/404}, \href
  {http://adsabs.harvard.edu/abs/2009ApJ...695..404R} {695, 404}

\bibitem[\protect\citeauthoryear{{Shapiro} \& {Lightman}}{{Shapiro} \&
  {Lightman}}{1976}]{shapiro76}
{Shapiro} S.~L.,  {Lightman} A.~P.,  1976, \mn@doi [\nat] {10.1038/262743a0},
  \href {http://adsabs.harvard.edu/abs/1976Natur.262..743S} {262, 743}

\bibitem[\protect\citeauthoryear{{Shen}, {Nakar}  \& {Piran}}{{Shen}
  et~al.}{2016}]{shen16}
{Shen} R.-F.,  {Nakar} E.,   {Piran} T.,  2016, \mn@doi [\mnras]
  {10.1093/mnras/stw645}, \href
  {http://adsabs.harvard.edu/abs/2016MNRAS.tmp..433S} {}

\bibitem[\protect\citeauthoryear{{Shiokawa}, {Krolik}, {Cheng}, {Piran}  \&
  {Noble}}{{Shiokawa} et~al.}{2015}]{shiokawa15}
{Shiokawa} H.,  {Krolik} J.~H.,  {Cheng} R.~M.,  {Piran} T.,   {Noble} S.~C.,
  2015, \mn@doi [\apj] {10.1088/0004-637X/804/2/85}, \href
  {http://adsabs.harvard.edu/abs/2015ApJ...804...85S} {804, 85}

\bibitem[\protect\citeauthoryear{{Stone} \& {Loeb}}{{Stone} \&
  {Loeb}}{2012}]{stone12}
{Stone} N.,  {Loeb} A.,  2012, \mn@doi [Physical Review Letters]
  {10.1103/PhysRevLett.108.061302}, \href
  {http://adsabs.harvard.edu/abs/2012PhRvL.108f1302S} {108, 061302}

\bibitem[\protect\citeauthoryear{{Stone} \& {Metzger}}{{Stone} \&
  {Metzger}}{2016}]{stone16}
{Stone} N.~C.,  {Metzger} B.~D.,  2016, \mn@doi [\mnras]
  {10.1093/mnras/stv2281}, \href
  {http://adsabs.harvard.edu/abs/2016MNRAS.455..859S} {455, 859}

\bibitem[\protect\citeauthoryear{{Strubbe} \& {Quataert}}{{Strubbe} \&
  {Quataert}}{2009}]{strubbe09}
{Strubbe} L.~E.,  {Quataert} E.,  2009, \mn@doi [\mnras]
  {10.1111/j.1365-2966.2009.15599.x}, \href
  {http://adsabs.harvard.edu/abs/2009MNRAS.400.2070S} {400, 2070}

\bibitem[\protect\citeauthoryear{{Strubbe} \& {Quataert}}{{Strubbe} \&
  {Quataert}}{2011}]{strubbe11}
{Strubbe} L.~E.,  {Quataert} E.,  2011, \mn@doi [\mnras]
  {10.1111/j.1365-2966.2011.18686.x}, \href
  {http://adsabs.harvard.edu/abs/2011MNRAS.415..168S} {415, 168}

\bibitem[\protect\citeauthoryear{{Ulmer}}{{Ulmer}}{1999}]{ulmer99}
{Ulmer} A.,  1999, \mn@doi [\apj] {10.1086/306909}, \href
  {http://adsabs.harvard.edu/abs/1999ApJ...514..180U} {514, 180}

\bibitem[\protect\citeauthoryear{{Wang} \& {Merritt}}{{Wang} \&
  {Merritt}}{2004}]{wang04}
{Wang} J.,  {Merritt} D.,  2004, \mn@doi [\apj] {10.1086/379767}, \href
  {http://adsabs.harvard.edu/abs/2004ApJ...600..149W} {600, 149}

\bibitem[\protect\citeauthoryear{{Yu} \& {Tremaine}}{{Yu} \&
  {Tremaine}}{2001}]{yu01}
{Yu} Q.,  {Tremaine} S.,  2001, \mn@doi [\aj] {10.1086/319401}, \href
  {http://adsabs.harvard.edu/abs/2001AJ....121.1736Y} {121, 1736}

\makeatother
\end{thebibliography}

\bsp	
\label{lastpage}
\end{document}